\documentclass[a4paper,aps,11pt]{article}

\usepackage{amsmath,amssymb}
\usepackage{pifont}
\usepackage{latexsym}
\usepackage{graphicx}
\usepackage{dcolumn}
\usepackage{bm}
\usepackage{psfrag}
\usepackage{subfigure}
\usepackage{tabularx}
\usepackage{verbatim}

\usepackage{placeins}
\usepackage{grffile}
\usepackage{color}
\usepackage[usenames,dvipsnames,svgnames,table]{xcolor}
\usepackage{hyperref} 
\hypersetup{backref,pdfpagemode=FullScreen,colorlinks=true,linkcolor=NavyBlue,citecolor=BrickRed}
 \usepackage[top=3cm, bottom=3cm, left=2.8cm, right=2.8cm] {geometry}
\usepackage{times}

\def\spinup{\uparrow}
\def\spindown{\downarrow}

\def\vk{{\bf k}}

\def\vr{{\bf r}}
\def\vrp{{\bf r}^\prime}

\def\vR{{\bf R}}

\def\vx{{\bf x}}

\def\mp{m^\prime}

\def\dx2{$d_{x^2-y^2}$}
\def\dz2{$d_{z^2}$}

\def\beq{\begin{equation}}
\def\eeq{\end{equation}}
\def\be{\begin{equation}}
\def\ee{\end{equation}}

\def\epsk{\epsilon_{{\bf k}}}

\def\cG0{{\cal G}_0}

\def\spinup{\uparrow}
\def\spindown{\downarrow}

\def\vk{{\bf k}}

\def\vr{{\bf r}}

\def\a{\alpha}

\def\d{\delta}
\def\D{\Delta}

\def\eps{\epsilon}

\def\l{\lambda}

\def\s{\sigma}

\def\uc2{$U_{c2}$}
\def\uc1{$U_{c1}$}

\def\bea{\begin{eqnarray}}
\def\eea{\end{eqnarray}}
\def \bal{\begin{align}}
\def \eal{\end{align}} 
\def\#{\!\!}
\def\@{\!\!\!\!}

\def\+{\dagger}

\def\up{\spinup}
\def\down{\spindown}

\begin{document}
\begin{titlepage}
\begin{center}

\vspace{3.5cm}

{\LARGE \bf Modeling many-body physics with Slave-Spin Mean-Field: Mott and Hund's physics in Fe-superconductors}

\vspace{2.5cm}

\hspace{0.5cm} {\Large Luca de' Medici and Massimo Capone}

\vspace{2.5cm}

Lectures prepared for the

\vspace{1.cm}

{\Large XIX Training Course in the Physics of Strongly Correlated Systems}

\vspace{0.5cm}

{Vietri sul mare, 12-16 October, 2015}
\end{center}

\vspace{1.7cm}

\abstract{Slave-spin mean-field theory, a light and accurate technique to model electronic correlations in Fermi-liquid phases of multi-orbital materials, is pedagogically exposed in this chapter, with a focus on its recent successful application to the physics of Iron-based superconductors. Beside introducing electronic correlations and recalling the Fermi-liquid phenomenology, the manuscript accompanies the step-by-step explanation of the slave-spin technique with a set of useful complements providing analytical insight into Mott and Hund's physics, which are at the heart of the physics of strongly correlated materials. Some original research material is also exposed, such as the Hund-induced shrinking of the Hubbard bands flanking the gap of the half-filled Mott insulator, and the low-energy description of the "orbital-decoupling" mechanism.} 

\tableofcontents
\end{titlepage}
\newpage 


\section{The theoretical description of iron-based Superconductors}

The discovery of superconductivity with critical temperatures exceeding 50K in the wide family of iron-based superconductors\cite{Kamihara_pnictides1} has strongly impacted on  the quest for the understanding of the physical origin of high-temperature superconductivity. Until this discovery, high-temperature superconductor has been for long time a synonymous of ``cuprate", a chemical label which stands for the family of doped copper oxides which have been discovered by Bedonorz and M\"uller in 1986\cite{bednorzmuller}.

The critical temperature is not the only aspect in which the iron-based family seems to trail the copper-based relatives, as their whole phase diagram can be seen as a sort of ``light" version of the celebrated doping-temperature diagram of the cuprates. In both cases a superconducting phase is established by doping a stoichiometric (parent compound) with magnetic ordering, and the two families share the existence of layers composed by the transition metal atom (Fe and Cu, respectively) even if the ligand atom in the iron-based material is not oxygen but rather a pnictogen or chalcogen atom, hence the popular label of pnictides and chalcogenides.

The first crucial difference lies in the nature of the magnetic phase from which superconductivity originates. The undoped cuprates are indeed antiferromagnetic Mott insulators, i.e., they are insulators because the strong Coulomb interaction localizes the carriers in a partially filled band, while the iron-based parents remain metallic despite the magnetic ordering in a spin-density wave pattern. This significant difference puts in jeopardy the possibility to develop a unified theoretical framework for the two classes of materials.

As a matter of fact the role of strong correlations is one of the few (almost) undisputed features of the cuprates, and it is not limited to make the parent compounds Mott insulators, but it is widely believed to be the  main driving force behind the phase diagram obtained as a function of doping, including the superconducting state itself.

This suggestion, put forward by P.W. Anderson in the early days of the cuprate era\cite{Anderson_highTc_RVB}, has deeply influenced the research in the field and drove a huge effort to understand the basic models for strongly correlated electrons, whose paradigm is the single-band Hubbard model. The combined use of a variety of techniques has confirmed that a two-dimensional Hubbard model provides at least the backbone for the theoretical description of the cuprates as doped Mott insulators.

The above mentioned chemical and structural similarities between cuprates and iron-based materials coupled with the analogy of their physics obviously led at least a part of the community to pursue the analogy also in more microscopic terms, therefore proposing that electronic correlations had to play a central role also in the iron-based material. Yet, this point of view is immediately challenged by the lack of Mott insulators in the iron-based family, which casts doubts on the strength of the electronic correlations and on their possible role in driving or affecting the superconducting phase. 

This, and many other seemingly conflicting evidences led to the polarization of the community in two main camps, which we may label as weak-coupling and strong-coupling for simplicity. The first approach is corroborated by the success of Density Functional Theory (DFT) to reproduce many aspects of the bandstructure and the Fermi surface, which is typically a fingerprint of minor correlation effects.  Within this approach magnetism is due to Fermi-surface nesting of otherwise metallic mobile carriers  and it has an itinerant character, while superconductivity arises due to the exchange of bosons of magnetic nature\cite{Mazin_Splusminus}. The opposite approach assumes instead that the materials are close to Mott localization and they describe the magnetism as the ordering of almost localized magnetic moments. According to this picture the carriers are poorly mobile and the main interactions can be modeled in terms of a frustrated Heisenberg model with nearest- and next-nearest-neighbor superexchange interactions\cite{Si_Abrahams_J1J2}. In the final part of this chapter we will show that a sort of intermediate scenario where both localized and delocalized carriers are simultaneously present can overcome this dualism and account for most of the properties of the iron-based phase diagram.

Before discussing the general theoretical framework of our contribution and its application to the iron-based superconductors, we briefly introduce two main physical points which motivate the outline of this chapter.

{\bf{Multi-orbital Electronic Structure.}} One clear and significant difference between the two main families of high-temperature superconductors is the nature of the relevant electronic states. While in the cuprates only one band, mainly derived from copper d$_{x^2-y^2}$ and oxygen orbitals, crosses the Fermi surface and contributes to the low-energy physics, including superconductivity and magnetism, in the iron-based materials all the iron d orbital contribute to the low-energy states, leading to multiple Fermi surface sheets and to an inherently multiorbital fermiology.

{\bf{Bandstructure Renormalization and Fermi-liquid behavior.}} The other aspect we would like to focus on is that experiments clearly measure well defined energy bands, even if renormalized with respect to DFT. At the same time, however, these metallic states are relatively incoherent and fragile with respect to thermal excitation, in contrast with weakly interacting metals.  Therefore electronic correlations do not, for a reason or another, completely destroy the itinerant nature of the electrons even if they appear to play some role in reducing the coherence of carriers. Yet, the system remains a Fermi-liquid, even when the renormalization is quite strong.

These two basic observations lead us to focus on a class of theoretical approaches which allow to study correlated metals at a reasonable computational effort. The use of approximate methods which allow for a relatively fast exploration of phase diagrams is a crucial request for multi-orbital systems such as those required to describe iron-based superconductors. Indeed the large Hilbert space implied by the inclusion of five d orbitals severely increases the computational cost and the complexity of all the numerical or seminumerical methods one can use to solve correlated systems. This class of cheap theoretical approaches is given by slave-particle approaches, that we describe in Sec. \ref{sec:slaves}. In particular our method of choice is the slave-spin method, which scales in a favourable way with the number of orbitals and it is therefore suitable for our goal. Importantly, this method is able to capture the quasiparticle renormalization leading to strongly correlated Fermi liquids that we aim to describe.

Of course our simplified approach cannot access all the extremely rich physics of strongly correlated materials. The main limitation is that it can only describe Fermi-liquid metallic phases, and mainly their renormalized quasiparticle structures. Only very roughly it describes the development of high-energy excitations such as the Hubbard bands or their precursors in the metallic state. In particular we have no way to properly describe a Mott insulator within this mean-field approach. On the other hand, we are able to follow the metallic solution all the way to the Mott transition. 

We also limit to the paramagnetic solution without any symmetry breaking, explicitly inhibiting any magnetic ordering or superconductivity. However the modeling of the evolution of the correlation effects as a function of the interactions in the normal phase is a key information to understand how the many possible instabilities may arise at low temperature. 

The chapter is organized as follows. In Sec. \ref{sec:intro} we introduce the basic concepts of strong electron correlation. First we give a brief reminder of the band theory of solids followed by a basic introductions to the concepts of electronic correlation, Mott insulators and the Mott transition, and Femi liquids. Section \ref{sec:Hubbard} presents the Hubbard model and explore its range of application and the very definition of a strongly correlated material. In section \ref{sec:slaves} we introduce the basic concepts behind the slave-particle approaches to strongly correlated systems, and in \ref{sec:SSMF} the slave-spin representation and its mean field approximation, with the main results on the single-band Hubbard model. In Section \ref{sec:Hund} we generalize this approach to multi-orbital systems and discuss the new energy scale intervening in these systems: Hund's coupling. Section \ref{sec:ironbased} is dedicated to the application of all the concepts and techniques introduced in the previous section to the physics of the Iron-based superconductors and the main findings are illustrated. Finally in Sec. \ref{sec:conclusions} the authors' concluding remarks and perspectives are given.

\section{Strong electronic correlations}
\label{sec:intro}

\subsection{The band theory of solids: A Brief reminder}
\label{subsec:bandtheory}
Crystalline solids are made by atoms that bind and form regular arrays or lattices. The chemical bonding and the equilibrium crystal structure of the lattice arise from the competition between the kinetic energy gained by the outermost electrons of each atom in delocalizing through the crystal, contrasted by the mutual repulsion of the ions left behind. The same delocalized outermost electrons (the ``valence electrons") are responsible for most low-energy properties of the solids as they are easily excited by external fields or by thermal excitation.

Our present understanding and modeling of the electronic properties of crystalline solids, such as electrical conductivity, magnetic and optical responses for example, is founded on the ``band theory" of solids\cite{ashcroft}. Within this approach the many-body state of the huge number of electrons in the solid is constructed by populating single-particle energy levels forming a set of bands separated by energy gaps. 
This huge simplification is the consequence of a crucial approximation where the electron-electron Coulomb repulsion is replaced by an average effective potential experienced by each electron, thereby turning a problem of interacting electrons into a sum of independent single-electron problems. Let us briefly discuss how this can be realized in practice.

\subsubsection{An effective Independent-particle Theory}
In the Born-Oppenheimer approximation, in which the electronic and ionic problems are decoupled, and the ionic positions are treated as fixed parameters for the electronic problem - the many-body hamiltonian for the electrons reads:
\be\label{eq:many-bodyH}
\hat H=-\sum_i \frac{\hbar^2\nabla_i^2}{2m_e}-\sum_{i\a}\frac{Z_\a e^2}{| \vR_\a-\vr_i|} + \frac{1}{2}\sum_{i\neq i'}\frac{e^2}{| \vr_i-\vr_{i'}|},
\ee
where $\vr_i$ is the coordinate of electron i, $\vR_\a$ the coordinate of ion $\a$, $Z_\a$ the atomic numbers and $m_e$ and $e$ the mass and the charge of the electron, respectively. 

Eq. (\ref{eq:many-bodyH}) defines an interacting problem which is essentially impossible to solve exactly for a realistic number of atoms. A drastic, yet surprisingly successful approximation is realized by approximating the last term with and an effective single-particle potential experienced by every single electron. With this approximation  the Hamiltonian becomes separable into the sum of  identical single particle hamiltonians, one for each electron, $\hat H=\sum_i \hat h(\vr_i)$, with
\be
\hat h(\vr_i) \equiv \frac{\hbar^2\nabla_i^2}{2m_e}+v_{eff}(\vr_i),
\ee
where $v_{eff}(\vr_i)$ includes the two last terms in Eq. (\ref{eq:many-bodyH}): the second, the coulomb potential due to the nuclei, treated exactly,  and the third, the electronic repulsion treated in the independent electron approximation. 

The next basic step is simply to solve the time-independent single-particle Schr\"odinger equation to obtain the single-particle eigenfunctions and eigenstates \be\label{eq:single_particle_Schreq}
\hat h(\vr)\phi_{a}(\vr)=\eps_{a}\phi_{a}(\vr).
\ee

For perfect crystals, where the ions sit in the positions of a Bravais lattice, as we will be considering throughout this course, $v_{eff}(\vr)$ is a function with the periodicity of the lattice, and the solutions to Eq. (\ref{eq:single_particle_Schreq}) will be the so-called Bloch functions of the form $\phi_a(\vr) \equiv \phi_{\vk n}(\vr)=e^{i\vk\cdot \vr}u_{n\vk}(\vr)$, where $u_{n\vk}(\vr)$ is also a function with the periodicity of the lattice. These solutions are labeled by their wave number vector $\vk$ that can take discrete values for a finite lattice with periodic boundary conditions, becoming infinitely dense for the volume of the crystal going to infinity. In practice $\vk$ can be considered a continuous variable in macroscopic systems. On the other hand the periodicity of the lattice results in a strict equivalence between wave vectors in the first Brillouin zone of the reciprocal lattice and in the other zones (in each zone the number of allowed $\vk$ values equals the number to the sites of the lattice), so that one usually restricts the treatment of the problem to the first zone. 

In the presence of the periodic potential the eigenvalues of Bloch functions distribute in energy in distinct sets (labeled by n) within which their energy varies continuously with k following a dispersion relation $\eps_n(\vk)\equiv \eps_{n\vk}$. Each of these sets spans a finite energy range, a ``band" of energy, separated among each others by forbidden regions of energies, the ``gaps". 

The many-body eigenfunction $\psi(\vx_1,\vx_2,\ldots)$ (where $\vx_i\equiv\{\vr_i,S^z_i\}$ indicates both position and spin coordinate of particle $i$) solving the full Hamiltonian $H$ in the single-particle approximation will then be a product of one-body wave functions $\phi_a(\vx_i)$.


Quantum mechanics constraints the fulll many body wave function for many indistinguishable fermions, like electrons are, to be antisymmetric under the exchange of any pair of particles.This means that, in the simples case of two electrons, solutions of the form $\phi_a(\vx_1)\phi_b(\vx_2)$ are not good physical states, and the actual state must be  the antisymmetric combination 
\be\label{eq:2p_SlatDet}
\psi(\vx_1,\vx_2)=\frac{1}{\sqrt{2}}\left[\phi_a(\vx_1)\phi_b(\vx_2)-\phi_b(\vx_1)\phi_a(\vx_2)\right].
\ee
This in turn implies that $\phi_a(\vx)$ and $\phi_b(\vx)$ must be different ($a, b,\ldots$ indicate collectively all the quantum numbers characterizing a given single-particle eigenstate, thus including the spin state), or the function would vanish, and is the reason why fermions obey the Pauli exclusion principle, that states, in its most familiar formulation, that two fermions can not occupy the same quantum state.

The generalization of this construction to N particles is obtained using Slater determinants, i.e.:
\be\label{eq:Slater_det}
\psi(\vx_1,\vx_2,\ldots,\vx_N)=\frac{1}{\sqrt{N!}}\left|
\begin{array}{cccc}
     \phi_{a_1}(\vx_1) & \phi_{a_1}(\vx_2) & \ldots & \phi_{a_1}(\vx_N) \\
     \phi_{a_2}(\vx_1)& \phi_{a_2}(\vx_2) & \ldots & \phi_{a_2}(\vx_N) \\
      \vdots & \vdots & \vdots & \vdots \\
     \phi_{a_N}(\vx_1) & \phi_{a_N}(\vx_2)& \ldots & \phi_{a_3}(\vx_N)
     \end{array}
\right|,
\ee
which indeed ensures that the exchange of the coordinates of any two electrons results in an overall minus sign, since it implies an exchange of two columns in the determinant. Also any two identical wave functions $\phi_{a_i}(\vx)=\phi_{a_j}(\vx)$ result in two identical lines and thus to the vanishing of the determinant, thus enforcing the Pauli principle. The energy of such wave functions is obviously $E=\sum_i \eps_{a_i}$. 

\subsubsection{Metals and Insulators in the band theory of solids}

The most celebrated success of the  band theory is explaining the difference between metals and insulators\cite{ashcroft}. 
Indeed the N-electron ground state of a solid is constructed using, in the Slater determinant Eq. (\ref{eq:Slater_det}), the N single-particle wave functions  corresponding to the N lowest eigenvalues. In practice due to the spin degeneracy, we have to take only the $N/2$ lowest-energy states, each occupied twice by the two electrons with different spin directions $\{\uparrow,\downarrow\}$. The energy of highest state which is actually occupied defines the Fermi energy  $\eps_F$. Therefore, the electronic structure is obtained simply ``filling" the levels from the bottom of the spectrum with the total number of electrons corresponding to the configuration of the individual atoms. As a result, we will end up with some completely filled bands and a (some) final one(s), called the valence band(s), which can be either partially filled or exactly filled. In the latter case, adding one more electron or exciting one of the existing electrons would have an energy cost equal to the energy gap between the valence and the next energy band, often called conduction band. In this condition the system is said to have a gap at the Fermi level.

A system with a filled valence band and empty conduction band can be shown to be inert to an applied electric field and it is therefore insulating. This can be somewhat intuitively understood by realizing that in order to create an electric current the system must be excited in a state with a finite momentum, while the equilibrium state carries no current. If the system has a gap, the excitation has a large energy cost, and the system remains in its ground state as long as the field does not reach the large values necessary to directly promote carriers from the valence to the conduction band, leading to the Landau-Zener breakdown of the insulator. On the other hand, in partially filled bands (for which the distinction between "valence" and "conduction" bands becomes meaningless, and both denominations are customarily used) finite-momentum excitations lie at a negligible distance from the Fermi energy and electrons can be excited at zero cost. Therefore band theory predicts that materials with filled bands are insulators whereas those with partially filled bands are metals.

 Since each atom composing the solid brings in an integer number of electrons, in a perfect crystal the total number of electrons is a multiple of the number of lattice sites. The spin degeneracy then implies that only materials having an even number of electrons per unit cell can ``fill" entirely a band.
Therefore an even number of electrons per unit cell is a necessary --but not sufficient-- condition to have a gap at the Fermi level, in band theory. On the other hand systems with an odd number of electrons per unit cell will always end up with partially filled bands. Indeed all elemental insulators that one finds by browsing the periodic table have an even number of electrons per unit cell. If the band gap in an insulator is smaller than $k_BT$ (where $k_B$ is the Boltzmann constant and $T$ the absolute temperature) at ambient temperature, thermal processes can excite electrons across the gap opening a conduction channels. These systems are called semiconductors. Another general prediction of band theory is that each time the population of a band insulator is modified either by chemical doping or by electrostatic gating, either the conduction or the valence band necessarily become partially populated and the system becomes a metal.

\subsubsection{Fermi velocity and band mass}
Within the framework of the band theory, all the electrons in a solid behave as independent particles, but the parabolic dispersion of electron in the vacuum $\epsilon_k = \hbar^2k^2/m_e$  is replaced by a series of band dispersions $\eps_{n\vk}$ determined by the periodic potential of the ionic lattice and by the effective potential arising from the electron-electron interactions. 

The band dispersion is therefore largely responsible of the electronic properties of a solid. In particular the electrons near to the Fermi energy are responsible for most physical responses of a given material. It is particularly useful to introduce effective quantities relating the behavior of band electrons with electrons in free space. 

By linearizing the dispersion close to the Fermi energy we can define a band Fermi velocity  
\be
\label{eq:vfermi}
{\bf v}_{n\vk}=\frac{1}{\hbar}\nabla_\vk \eps_{n\vk}.
\ee
This quantity will be obviously different from that of a non-interacting electron in vacuum $|{\bf v}_{n\vk}|=\hbar k/{m_e}$ , and we can use it to define a band effective mass through the relation 
$|{\bf v}_{n\vk}|=\hbar k/{m_b}$, i.e., the same expression of free electrons where the bare mass $m_e$ is replaced by an effective mass $m_b$, which measures the intertia of the bands electrons. Notice that this definition is unambiguous only close to band minima, where the dispersion is quadratic and the Fermi velocity is actually proportional to the momentum, while in general cases different definitions can be used according to the context. 

The band mass is however a very useful parameter which characterizes how some experimental quantities differ in band theory from the free-electron value: for instance the low-temperature electronic specific heat in metals is typically altered by a factor $\sim m_b/m_e$ with respect to a Fermi gas and similar behavior is found for response functions such as the spin susceptibility. 

\subsection{Electron-electron Interactions and Correlations}
\label{subsec:correlations}

Band theory is so successful in general that is regarded as a sort of  ``standard model" of the solid state. While many qualitative predictions can be based on simple analysis of the chemistry of materials, namely on the electron count and the consequent filling of the bands, 
in order to obtain a quantitative theory able to capture the differences between different materials, the choice of the effective one-electron potential $v_{eff}(\vr)$ becomes crucial but is is also highly non trivial. The most popular and successful method to derive $v_{eff}(\vr)$ starting from the actual Coulomb interactions is Kohn-Sham Density-Functional Theory (DFT), where effective Kohn-Sham single-particle orbitals and the effective potential are self-consistently determined\cite{kohn_sham}. DFT requires in turn a choice of the so-called exchange-correlation functional, which defines possible "flavours", among which the local density approximation (LDA) and generalized gradient approximation (GGA) play a prominent role in actual implementations.  Alternative single-particle potentials can be defined, for example through a Hartree-Fock variational principle which defines an effective single-particle description alternative to Kohn-Sham DFT. 
As a matter of principle, all the methods that derive describe the electronic structure of a solid in terms of a collection of well defined bands are based on the definition of some kind or explicit of implicit effective potential $v_{eff}(\vr)$.  

There are however important cases in which even the most powerful band theory calculations are grossly inaccurate, or fail even at the most qualitative level, misrepresenting as good metals materials which are experimentally insulating. This happens when  the effects of electron-electron interactions are so relevant that the independent-electron approximation of band theory fails to correctly account for the physics of the system. Broadly speaking, the valence electrons of these materials are said to be ``strongly correlated", a label which is often extended to the materials. 

For an oversimplified yet transparent picture of how strong correlations imply a failure of the independent-electron scenario we can  consider a simple two-electron problem with two ``lattice sites", which can be thought as ions of an H$^2$ molecule. We assume that the electrons can only occupy wave functions localized around one of the two sites, say $\phi_L(\vr)$ and $\phi_R(\vr)$. Within the independent-electron approximation the wave function of the electron will be a linear combination of the two basis states with normalized coefficients: $\phi_a(\vx)=c_{aL}\phi_L(\vx) + c_{aR}\phi_R(\vx)$. The 2-particle function will then be the antisymmetrized sum Eq. (\ref{eq:2p_SlatDet}) of two products of the kind:
\bea\label{eq:corr_2part}
\phi_a(\vx_1)\phi_b(\vx_2)&=& c_{aL}c_{bL}\phi_L(\vx_1)\phi_L(\vx_2)+c_{aR}c_{bR}\phi_R(\vx_1)\phi_R(\vx_2)\nonumber \\
&+&c_{aL}c_{bR}\phi_L(\vx_1)\phi_R(\vx_2)+c_{aR}c_{bL}\phi_R(\vx_1)\phi_L(\vx_2))
\eea

If the relative effect of the repulsive electron-electron interaction (which depends on the specific shape and distance of the basis functions, on possible external screening, etc.) is important, it is easy to understand that in the exact solution the first two terms in Eq. (\ref{eq:corr_2part}) in which the two electrons are close to one another and thus maximize the energy due to mutual repulsion must have a smaller weight than the last two.
However direct inspection of the coefficient leads to realize that there is \emph{no way} for the independent electron approximation to account for this effect, since there is no way to depress the relative contribution of the first two terms in the sum by acting on the coefficient of the single-particle solutions. 

The reason for the failure of the independent-electron picture becomes now clear. If the interactions are strong the electrons pay a large energy cost if they are on the same site (or in general if they are close) or, in other words, they gain energy in \emph{avoiding} each other. As a consequence  the many-electron wave function is depressed when one or more electronic coordinates are at small reciprocal distances, i.e., when $\vr_i -\vr_j$ is small with respect to the effective range of interactions. Therefore the behavior of each electron depends on the state of the other and a single-particle description breaks down.  This is what we define as a  \emph{correlated} behavior, which can not be encoded in a Slater determinant, or equivalently in a band-theory description.

\subsection{The Mott transition and Mott insulators}

If we imagine to continuously vary the degree of correlation in our system, for example changing the inter-site distance in our two-site example, we expect an evolution from a ``weakly interacting" regime (small distances) where the single-particle description is a fair approximation of the exact solution and a ``strongly interacting" regime, where the correlation between the electrons dominates and the independent electron picture breaks down. 

Let us consider an idealized system with a single band and a number of electrons equal to the number of lattice sites. In the weakly interacting regime, where a single-particle picture holds, the solution of our system describes a metal with a half-filled band due to the odd number of electrons per idealized atom. When the interaction is increased, which can be realized increasing the lattice spacing as in our two-site example, the electrons will become more and more correlated in the sense we described above. 

This process where the correlation between the electrons increases is characterized by the competition  between the kinetic energy, which favors delocalized band states and the interaction, which disfavors configurations with two electrons on the same site, which determines constraints to the motion of the electrons. When the energetic cost of a doubly occupied site exceeds the possible energy gain, the best compromise to minimize the energy is simply to localize the electrons, one on each site. This avoids to pay an energy of order $U$, even if the kinetic energy is obviously sacrificed. Therefore the systems becomes an insulator, despite the electron count would imply a metal according to the band theory of solids. This quantum state, where the electrons are localized by the mutual Coulomb repulsion, is called a Mott insulator after N. Mott, who introduced the concept in a pioneering work\cite{mott49}. The transition from a metal to a Mott insulator which occurs as a function of the ratio between Coulomb repulsion and hopping is called a Mott transition. It is important to notice that Mott localization can happen only when the number of electrons equals the number of sites, or in multi orbital systems, for an integer number of electrons per atom. If the system is doped away from half-filling, electronic correlation can have a major effect, but they can not drive a Mott insulating state in the absence of symmetry breaking. Therefore doping of a Mott insulator in principle leads to a metallization. As a matter of fact, one can drive an insulator-to-metal transition by doping a Mott insulator with either holes or electrons.

Needless to say, the very existence of Mott insulators represents a clear failure of the band theory of solids. Indeed Mott insulators have been experimentally discovered already in 1937\cite{DeBoer-Mott_Ins_1937}, and a number of these materials have been identified in the years including related charge-transfer insulators\cite{ZaanenSawatzkyAllen}, whose prominent members are the parent compounds of high-temperature superconductors. The physics of high-temperature superconductivity is indeed widely believed to be the consequence of the doping of a (charge-transfer) Mott insulator.

It must also be mentioned that Mott insulators are usually characterized by long-range ordering of the spin (and orbital) degrees of freedom. It is indeed easy to show that a half-filled Hubbard model is unstable towards two-sublattice antiferromagnetism. In the case where the hopping is limited to nearest-neighbors sites only, the Hubbard model has indeed an antiferromagnetic ground state for every value of $U$, while longer-range hoppings lead to a critical value of the interaction. The long-range ordered phase can survive for a finite range of doping thereby partially hiding the intrinsic Mott transition. In this work we will focus only on paramagnetic solutions, where the long-range magnetic ordering is inhibited. This choice allows to follow the intrinsic effects of electron-electron correlations, disentangling them from the appearance of low-energy broken symmetry phases.


\subsection{Fermi-liquids and effective mass}

The very existence of Mott insulators and the Mott transition are qualitative and unequivocal fingerprints of strong correlations leading to a complete breakdown of any independent-electron approximation. Yet, many materials display strong correlations but remain metallic, either because their interactions are not sufficient to drive Mott localization or simply because the electron count is not commensurate with the number of sites. 
	Interestingly, the behavior of many correlated metals is not qualitatively different from weakly interacting metals. As a matter of fact the low temperature behavior of many observables has the same temperature dependence of free electrons only with different parameters, as described for example by Drude-Sommerfeld theory, even in very strongly correlated materials on the brink of Mott localization.

	The robustness of the metallic behavior in interacting fermionic systems is described by the Landau theory of normal Fermi liquids\cite{landau1,landau2,nozieres}. The basic assumption of this phenomenological theory is that the low-energy and low-temperature elementary excitation of a system of interacting fermions are in one-to-one correspondence with the excitations of a system of non-interacting particles, as if one could adiabatically switch on the interaction without changing the character of the excitations.

	More concretely, the excitations of a non-interacting Fermi gas are given by variations of the occupations of the different single-particle eigenstates and they are associated with the creation and/or destruction of actual fermions. Analogously, the excitations of the interacting systems correspond to ``quasiparticle states", excitations with fermionic statistics which carry the same quantum numbers of the original fermions. Obviously, these states are not real eigenstates of the interacting systems and the crucial idea is that the above correspondence is only valid for weak excitations at low-energy. In general a quasiparticle excitation has a finite lifetime, as opposed to non-interacting particles. The Landau theory is based on the fact that the inverse of the lifetime vanishes quadratically  both as a function of the distance from the fermi Energy and as a function of temperature
	\be
	\label{scatteringrate}
	\frac{1}{\tau} = a(\epsilon - \epsilon_F)^2 + b T^2
	\ee

	Notice that no assumption is made about the strength of the interactions, that can well be large, as long as it does not lead to a complete breakdown of the metallic state. On the other hand, the residual interactions between quasiparticles will be small. 
	The whole Fermi liquid theory essentially amounts to phenomenologically describe a system of strongly interacting particles in terms of a low-energy picture of renormalized quasiparticles with small residual interactions. The effect of the original interactions is therefore included both in the residual interactions and in effective parameters characterizing the quasiparticle states. Among these, a central role is played by the effective mass, which emerges in a similar way as in band theory in terms of an effective renormalized dispersion $\eps^*_{n\vk}$ --which now arises also from interaction effects-- and a renormalized Fermi velocity 
${\bf v}^*_{n\vk}=\frac{1}{\hbar}\nabla_\vk \eps^*_{n\vk}$. The effective mass $m^*$  is now defined by the relation
\be
\vert{{\bf v}}^*_{n\vk}\vert=\frac{\hbar k}{m^*}
\ee

The physical meaning of the effective mass is the same of the band effective mass, namely it is an attempt to compare the motion of the interacting electron with that of non-interacting electrons. In this context, where we focus on the interaction effects, it is natural to compare the effective mass with the band mass rather than with the free-electron mass, and the ratio $m^*/m_b$ can be used as an estimate of the degree of correlation of a given system.

This effective mass describes the reduction of the mobility of the electrons as a consequence of the interactions that we discussed above. The whole process of Mott localization that we described in the above section can be mirrored in a progressive enhancement of the effective mass from the non-interacting value to a divergent value which corresponds to the localization of the electron. The effective mass in strongly interacting materials can range from a  few times the bare mass to huge values of the order of hundred times the bare value, which correspond to the so-called heavy fermion materials\cite{Stewart-HeavyFermions_RMP}. 

As long as the system remains in a Fermi-liquid state, i.e. it does not turn into a Mott insulator, into a state with some kind of ordering (magnetic, orbital, \ldots) or into a more exotic non-Fermi liquid metal, the effective mass embodies the main effect of correlations and it can be used to understand the influence of interactions on a given system. It should be mentioned that also other interaction mechanism, as e. g. any kind of electron-boson interaction, also lead to an analogous  enhancement of the effective mass.

The Landau Fermi-liquid theory is a phenomenological approach. However it can be founded on a diagrammatic perturbative expansion of the interaction. The 
condition (\ref{scatteringrate}) is verified when the imaginary part of the self-energy $\Sigma(\omega)$ vanishes like $\omega^2$ for small frequencies and like $T^2$ for small temperatures. Focusing on the zero-temperature behavior, up to order $\omega$ the self-energy is purely real and it can be expanded as
\be 
\left.\Sigma (\vk,\omega) \simeq \Sigma'(\vk,0)+ \omega \frac{\partial\Sigma'(\vk,\omega)}{\partial\omega} \right\rvert_{0}, 
\ee
where $\Sigma'(\vk,\omega)$ is the real part of the self-energy. The Green's function then becomes

\begin{align}
G(\vk,\omega) &\simeq \frac{1}{\omega - \epsilon_\vk + \mu - \Sigma'(0) -\omega \left.\frac{\partial\Sigma'(\vk,\omega)}{\partial\omega}\right\vert_0}= 
\frac{1}{\omega\left(\left.1- \frac{\partial\Sigma'(\vk,\omega)}{\partial\omega}\right\vert_0\right) - \epsilon'_\vk + \mu }\nonumber\\
& = \frac{1}{\omega/Z_\vk - \epsilon'_\vk + \mu } = \frac{Z_\vk}{\omega - Z_\vk(\epsilon'_\vk - \mu)},
\end{align}
where we defined a shifted dispersion $\epsilon'_\vk = \epsilon_\vk + \Sigma'(\vk,0)$ and the quasiparticle weight 

\be
Z_\vk= \frac{1}  {1- \left.\frac{\partial\Sigma'(\vk,\omega)}{\partial\omega}\right\vert_0}.
\ee

The final result shows that the renormalized Green's function features a pole with a weight $Z_\vk$ (hence the name of quasiparticle weight) at a frequency $\omega = Z_\vk(\epsilon'_\vk - \mu)$, which plays the role of a renormalized quasiparticle energy. If we further expand the self-energy and the dispersion around the Fermi momentum, we obtain the expression for the effective mass

\be
\label{mefffl}
\frac{m_e}{m*} = \frac {1-\left.\frac{\partial\Sigma'(\vk,0)}{\partial k}\right\vert_{k_F}} {1- \left.\frac{\partial\Sigma'(\vk_F,\omega)}{\partial\omega}\right\vert_0}.
\ee

We notice that if the self-energy does not depend on momentum,  as it will happen in the mean-field approaches that we will discuss in the present chapter and for the Dynamical Mean-Field theory where the spatial dependence of the observables is frozen, the numerator of Eq. (\ref{mefffl}) becomes 1 and we find

\be
\frac{m*}{m_e} = \frac{1}{Z},
\ee

so that the effective mass is just given by the inverse of the quasiparticle weight. An increased effect of correlation will be then reflected in the reduction of $Z$ and a metal-insulator transition will occur when $Z$ vanishes or, equivalently, the effective mass diverges.

The Landau theory of Fermi liquids focuses on the low-energy excitations of a metallic system and it does not make any assumption for the high-energy and high-temperature behavior, which could be accessible through the knowledge of the self-energy at all frequencies. 
The methods and the results presented in this chapter indeed focus on strongly correlated Fermi liquids which can be described within the above formalism, and will not address possible violations of the Fermi liquid as well as the development of the high-energy Hubbard bands. This choice is motivated and justified by the fact that the metallic phases of iron-based superconductor appear experimentally as regular Fermi liquids with only minor exceptions.



\section{The Hubbard model}\label{sec:Hubbard}

In this section we introduce the celebrated Hubbard model, which is almost universally considered the paradigmatic model to study the physics of strong correlations. The model has been introduced independently by Hubbard, Kanamori and Gutzwiller to study itinerant magnetism, but later became a general reference system to study correlated metals and their metal-insulator transitions.
As we will see, the model can be derived as a simplification of the many-body Hamiltonian describing electrons in a solid, which allows to treat explicitly an  essential part of electron-electron correlation.
In second quantization the many-body hamiltonian for the electrons eq. (\ref{eq:many-bodyH}) reads $\hat H=\hat H_0+\hat H_{int}$

\be\label{eq:2quant_H0}
\hat H_0= \sum_\s \int d\vr \Psi^\dag_\s(\vr) \left[ -\frac{\hbar^2\nabla^2}{2m_e}+V(\vr) \right ]\Psi_\s(\vr)
\ee
where $V(\vr)=-\sum_{\a}\frac{Z_\a e^2}{| \vR_\a-\vr|}$, and
\be\label{eq:2quant_Hint}
\hat H_{int}= \sum_\s \sum_{\s'} \int d\vr d\vr' \Psi^\dag_\s(\vr) \Psi^\dag_{\s'}(\vr') \left[ \frac{1}{2}\frac{e^2}{| \vr-\vr'|} \right ]\Psi_{\s'}(\vr') \Psi_\s(\vr)
\ee
where $\Psi_\s(\vr)$ is the field operator creating a particle with spin $\s$ at point \vr.
$H_0$ is diagonalized by bloch wavefuctions $\phi_{\vk n}(\vr)$. These are delocalized through the solid and to implement any sort of approximation involving the interaction term it is convenient to re-express them using a complete orthonormal basis of localized functions, like a basis of Wannier orbitals\cite{ashcroft}, for example (but not uniquely\cite{Marzari_RMP}) defined by: 
\be\label{eq:wannier}
w_n(\vr-\vR)=\frac{1}{\sqrt{N}} \sum_{\vk \in BZ} e^{-i\vk \cdot \vR} \phi_{\vk n}(\vr),
\ee
where a Wannier orbital is associated to each lattice position $\vR$. 
This choice is justified and particularly useful in strongly correlated materials because, as discussed in section \ref{sec:chemistry}, strong correlations emerge when the conduction electron density is very concentrated around the ionic positions. The field operator can be then expressed in this basis: 
\be
\Psi^\dag_\s(\vr)=\sum_{i n}w^*_{n}(\vr-\vR_i)d^\dag_{i n\s}
\ee
(where we used the index to label the site with position $R_i$). Now $d^{\dagger}_{in\sigma}$ creates an electron with spin $\sigma$ in the n-th Wannier orbital associated to site $R_i$. In this basis the Hamiltonian reads
\footnote{Here, and throughout this manuscript, we use light (i.e. non-bold) characters for the site indices $i,j,\ldots$. Indeed each of these can be seen formally as a scalar discrete index labeling all the sites. It is however more natural to think about them as vectors of 3 discrete indices, on a 3-dimensional lattice, indeed.}
\be
\hat H = \sum_{ijmm'\s} t^{m m'}_{ij} d^\dag_{im\s}d_{jm'\s}+\frac{1}{2}\sum_{ijkl}\sum_{mm'nn'}\sum_{\s\s'} V^{mm'nn'}_{ijkl} d^\dag_{im\s}d^\dag_{jm'\s}d_{kn'\s'}d_{ln\s}
\ee
where 
\be
 t^{m m'}_{ij}= \int d\vr\, w^*_m(\vr-\vR_i) \left[ \frac{-\hbar^2\nabla^2}{2m_e}+V(\vr) \right ]w_{m'}(\vr-\vR_j)
\ee
and
\be\label{eq:interaction_integrals}
V^{mm'nn'}_{ijkl}= \int d\vr d\vr' w^*_m(\vr-\vR_i)w^*_{m'}(\vr'-\vR_j) \frac{e^2}{| \vr-\vr'|} w_{n}(\vr'-\vR_k)w_{n'}(\vr-\vR_l)
\ee

Up to here the only approximation performed is the Born-Oppenheimer one and we have merely rewritten the many-body problem into a Wannier basis, which emphasizes the real space coordinates of the ions. 
A typical treatment of the many-body Hamiltonian uses band theory as a starting point, i.e. uses the basis of Bloch functions that diagonalize the effective problem in which $V_{eff}(\vr)$ is used in eq. (\ref{eq:2quant_H0}) instead of $V(\vr)$ (as explained in section \ref{sec:intro}), and the corresponding basis of Wannier orbitals. This means including then the static electrostatic potential due to the electrons in $H_0$ and will obviously require to subtract it during the treatment of eq. (\ref{eq:2quant_Hint}) to avoid a "double counting" of this term. 

The convenience of the Wannier basis in strongly correlated materials stems out of the fact that when the basis orbitals are very localized in space the hopping integral $t^{mm'}_{ij}$ decays rapidly with the distance between the sites $i$ and $j$, and thus the number of independent parameters for the one-body part of the Hamiltonian is reduced, and the corresponding Hamiltonian matrix rather sparse.

This is true also for the interaction parameters $V^{mm'nn'}_{ijkl}$ which indeed decay faster with the distance because of the spatial decay of the Coulomb interaction and because they involve products of four wave functions instead of two. 

Even with a wise choice of the set of basis functions the problem at hand remains formidably complex and simplifications are needed. Two key physical facts are to be invoked here. First, when analyzing the properties of correlated materials at the temperatures and for most of the probes relevant to experiments\footnote{With the notable exception of high-energy probes like e.g. X-rays that involve states vary far from the Fermi energy.} only a subset of bands is relevant. Indeed those coming from deep shells of very low energy are completely filled and can be considered inert to weak perturbations, like empty bands far above the Fermi energy. It is thus reasonably safe to limit the explicit treatment of many-body effect to a subset of bands (i.e. limiting the orbitals m in the sum ) near the Fermi energy. Second, these degrees of freedom can be eliminated modulo the inclusion of the screening effects on the interaction between the remaining electrons. This amounts to replacing in the integrals eq.(\ref{eq:interaction_integrals}) the coulomb interaction $e^2/(|\vr-\vr'|)$ with the screened interaction $W(|\vr-\vr'|)$ that decays much faster with distance\footnote{The simplest analytical form is the Yukawa potential: $W(|\vr-\vr'|)=e^2\frac{e^{-\lambda|\vr-\vr'|}}{|\vr-\vr'|}$, where $\lambda$ is the screening length.} It is to be noted that only the screening from the eliminated degrees of freedom has to be included, since the explicit treatment of the interactions in the retained subset of bands will account for the rest of it.

This simplification allows then a clear hierarchy of the interaction integrals, those for which all the Wannier functions are on the same site being much larger than those involving other sites. One customarily (but indeed not always) retains only the former. i.e. the local interaction
\be\label{eq:Hubbard_interaction}
V^{mm'nn'}_{ijkl}=U^{mm'nn'}\d_{ij}\d_{ik}\d_{il}
\ee
to obtain what is called a (multi-orbital) Hubbard model, where the interactions are local, but involve electrons in different orbitals. When we consider just one orbital on every site we obtain the single-band Hubbard model, reading:
\be\label{eq:Hubbard_oneband}
\hat H = \sum_{ij\s} t_{ij} d^\dag_{i\s}d_{j\s}+U\sum_i n_{i\up}n_{i\down},
\ee
where we introduced the number operators $n_{i\s}\equiv d^\dag_{i\s}d_{i\s}$

The Hubbard model is clearly a huge simplification of the original problem, and in many realistic systems we can even speak of an oversimplification\footnote{Despite its simplicity, this is not necessarily too crude an approximation, as it is believed to describe, at least qualitatively, the main physics of the copper-oxygen planes of the cuprates, and it can be simulated with fermionic ultra cold atoms  trapped in optical lattices.}. Yet, it retains the basic aspects of the strong correlation physics and it includes the essential physics behind Mott localization and formation of bad metals with large effective masses. For this reason the model remains very hard to solve and it is unsolved except in one or infinite spatial dimensions or in extreme and trivial limits  $t_{ij}=0$ or $U=0$.).

\subsection{\emph{Complement}: single-band Hubbard model at particle-hole symmetry}\label{compl:1band_PH}

Here we show that, for the single-band Hubbard model on a bipartite lattice (i.e. a lattice that can be divided in two sublattices A and B such that the electrons hop only from one sublattice to the other, the simplest example being a square lattice with hopping only between nearest neighbours), the choice $\mu=U/2$ fixes the filling to one electron per site. This corresponds to particle-hole symmetry in this model, and this result is thus independent of the value of U and of the temperature, and holds irrespectively of the phase in which the system is (metallic, insulating,..).

Indeed if we apply to eq. (\ref{eq:Hubbard_oneband}) the particle-hole transformation:
\bea
d^\dag_{i\s}&\rightarrow& d_{i\bar \s}\nonumber\\
d_{i\s}&\rightarrow& d^\dag_{i\bar \s}
\eea
the Hamiltonian becomes
\be
\hat H-\mu \hat N=-\sum_{ij\s} t_{ij} d^\dag_{i\s}d_{j\s}+U\sum_i n_{i\up}n_{i\down}+(\mu-U)\sum_{i\s}n_{i\s}+U-2\mu,
\ee
in which $t_{ij}=t_{ji}$ (hermiticicity of the Hamiltonian) was used when renaming the site indices.
One can then apply a gauge transformation of the fermionic fields only on one sublattice,

\bea
d^\dag_{i\s}&\rightarrow& -d_{i\s}\nonumber\\
d_{i\s}&\rightarrow& -d^\dag_{i\s}\nonumber\\
 i &\in& sublattice\; A
\eea 

to obtain
\be
\hat H-\mu \hat N=\sum_{ij\s} t_{ij} d^\dag_{i\s}d_{j\s}+U\sum_i n_{i\up}n_{i\down}+(\mu-U)\sum_{i\s}n_{i\s}+U-2\mu,
\ee
a Hamiltonian identical to $\hat H-\mu \hat N$ with $\hat H$ as in eq. (\ref{eq:Hubbard_oneband}) if the chemical potential has the value $\mu=U/2$.

Thus $\mu=U/2$ enforces the particle-hole symmetry and hence $\langle n_{i\up}+n_{i\down}\rangle =1$, i.e. half-filling.

It can be convenient to rewrite the Hubbard Hamiltonian in a manifestly invariant form by rescaling the chemical potential. Indeed  completely equivalent forms for the interaction to eq. (\ref{eq:Hubbard_oneband}) are both
\be\label{eq:Hubbard_PH1}
U\sum_i (n_{i\up}-\frac{1}{2})(n_{i\down}-\frac{1}{2})
\ee
and 
\be\label{eq:Hubbard_PH2}
\frac{U}{2}\sum_i(n_{i\up}+n_{i\down}-1)^2=\frac{U}{2}\sum_i\left(\sum_{\sigma}(n^d_{i\s}-\frac{1}{2})\right)^2
\ee
in which the particle-hole symmetry is evident. Using these forms of interaction the chemical potential yielding half-filling will then be $\mu=0$. By developing the above forms\footnote{Here one uses the fact that for fermions $n_{i\s}^2=n_{i\s}$.} one can convince himself that this corresponds indeed to the standard form eq. (\ref{eq:Hubbard_oneband}) with $\mu=U/2$.

\subsection{Which materials are likely to be strongly correlated?}\label{sec:chemistry}
In the simple example of Sec. \ref{subsec:correlations}, we have shown that ``strong correlations" appear when some short-range Coulomb interaction exceeds the energy gain in freely delocalizing the carriers, thus making configurations with two electrons on the same atom unfavourable. In this section we give some basic guidelines to understand how the balance between these two energy terms changes in different materials, and in particular we identify which compounds are more likely to display strong electronic correlations.

Within the tight-binding parametrization leading to the Hubbard modeling described in Sec. \ref{sec:Hubbard} the delocalizing energy is given by the hopping matrix elements $t^{m m'}_{ij}$, which have to be compared with the interaction parameters $U^{mm'nn'}$. A material is expected to be more or less correlated according to the ratio between the Hubbard $U$ and some combination of the hopping matrix elements.

As we anticipated, we focus on the valence electrons, and we expect that the kinetic energy is mainly determined by the largest hoppings, which naturally correspond to nearest-neighboring ions due to the rapid decay of the overlap between wavefunctions when the distance increases. Roughly speaking, the hopping is determined by the overlap between valence wavefunctions associated with neighbouring lattice sites. This quantity is expected to depend on the spatial extension of the relevant wavefunctions, which naturally depends on their atomic quantum numbers $n$, $l$ and $m_l$ and on the value of the lattice spacing, which also depends on the chemistry of the materials and on the atomic species involved. 

On the other hand, the Coulomb interaction is a local quantity, which depends much less on the chemistry of the material and the orbitals involved. Of course this last statement has to be taken with a grain of salt, as $U$ changes in the various materials, mainly because of the different screening, but certainly the leading effect in determining the degree of correlation in a materials is the variation of the hopping matrix elements. 

With these guidelines, we can understand why certain families of materials are more likely to display strong correlation effects than others. As we already mentioned, the key factor is given by the symmetry of the orbitals which mainly contribute to the valence bands. the hopping matrix elements are controlled by the lattice spacing and the spatial extension of the wavefunctions. 
As a matter of fact the critical parameter to determine the degree of correlation of a materials is the latter, which depends directly on the nature of the involved orbitals, while the lattice spacing has a weaker and much less obvious dependence on the atomic number and it is influenced by the core electrons.

On the other hand, the  spatial extension of the valence wavefunctions essentially increases as the atomic number grows, as we know from basic quantum mechanics. The reason for this is that each wavefunction must be orthogonal to all the previous ones, which is realized by increasing the number of nodes and by shifting the charge further away from the nucleus. Based on this considerations, it would be hard to find a general trend for the evolution of the hopping (hence of the degree of correlation) as a function of the atomic number of the atoms involved. There is however a crucial aspect to take into account. As it is well known the orbital quantum number $l$ can assume integer values smaller than the principal quantum number $n$. Therefore, at each value of $n$ the orbitals with $l=n-1$ are authomatically orthogonal to all the "previous" orbitals because they are the first with that value of $l$. This implies that these orbitals do not need to push the charge far from the nucleus, and they are more localized around the lattice sites. This leads to relatively small value of the hopping. Since the Hubbard $U$ is not equally affected, this results in an increased level of correlation.

Hence, we can expect strong correlation effects are particularly strong in materials where the valence bands arise from 3d orbitals, the first orbital with $l=2$ and the 4f orbitals, the first with $l=3$. This expectation is actually met, and the prototype materials for strong correlation physics are characterized by the presence of transition-metal atoms or rare earths, the two groups which correspond to the above quantum numbers. As a matter of fact the 4f are more correlated than the 3d, and display extreme correlation effects (e.g. the huge effective masses of heavy-fermion compounds), while the 3d happen to be exactly in the "interesting" range, and they can either be strongly correlated metals or Mott insulators according to different chemical and structural factors. For the same reason 3d materials can be driven across the Mott metal insulator transition either by doping or by pressure and chemical substitution and they represent the ideal playground to study and understand the properties of strongly correlated materials. Besides the academic interest, these materials also display a wealth of spectacular phenomena, including high-temperature superconductivity since both iron and copper belong to the 3d transition metal series.

\section{Slave-particle approaches}
\label{sec:slaves}
In the previous sections of this chapter we have emphasized that the physics of electron-electron correlations driving the Mott transition can not be described within {\it{any}} single-particle picture. This is true also for the Hartree-Fock approximation, which amounts to search for the optimal Slater determinant for a given problem according to the variational principle, which obviously corresponds to an effective single-particle picture. Since the Hartree-Fock method is completely equivalent to a mean-field decoupling of the interaction, any attempt to theoretically account for strong correlation physics requires to go beyond the standard mean field.

The need to understand the effects of correlations, primarily triggered by the discovery of high-temperature superconductivity, led to the development of methods able to treat correlated systems, ranging from purely numerical methods such as quantum Monte Carlo simulations or the Density-Matrix Renormalization Group to analytical approximate approaches like the functional renormalization group, to mention only a few members of a very long list\cite{Dagotto_RMP,imada_mit_review,LeeNagaosaWen_RMP,Metzner_func_ren_group_RMP}.

In this contribution we will focus on a class of methods which are based on the introduction of auxiliary ``slave" particles. These approaches allow to construct a new kind of mean-field theories specifically designed to deal with strongly correlated systems and to describe the Mott-Hubbard transition and the appearance of heavy quasiparticles with large effective masses. The first and most popular of these approaches is the slave-boson method, which we briefly describe here to give a general perspective on this class of methods.

Within the slave-boson approach the physical fermion, which is associated with the annihilation operators $c_{i\sigma}$ (here we limit to a single band model for simplicity) is represented in terms of a pseudofermion variable $f_{i\sigma}$ which has fermionic statistics and a boson variable $b_{i\sigma}$, where here the spin index of the boson operator does refer to an internal spin degree of freedom, but it should be viewed as a label which establishes the link between the slave boson and the correspondent fermion.
\be 
c_{i\sigma} \rightarrow f_{i\sigma}b_{i\sigma}
\ee

In this way the original Hiibert space is enlarged to include also the auxiliary bosonic degree of freedom. The nickname of ``slave" particles derives from a constraint which will impose that the auxiliary variable contains the same physical information of the fermionic variable. This would obviously be redundant if the model was to be solved exactly. The idea is instead to perform a suitable mean-field after which the pseudofermion variable describes the itinerant quasiparticle fraction of the electron, while the auxiliary boson describes its localized fraction. 

We give some more details of one popular slave-boson approach introduced  by Kotliar and Ruckenstein\cite{kotliar_ruckenstein}, where the slave bosons are introduced as a sort of label attached to each local electronic configuration.

Explicitly, on each lattice site one can have four physical states, the empty state $\vert 0 \rangle$, the states where there is one electron with a given spin $c^{\dagger}_{i\uparrow} \vert 0 \rangle \equiv \vert \uparrow\rangle$ and   $c^{\dagger}_{i\downarrow}  \vert 0 \rangle \equiv \vert \downarrow\rangle$ and the doubly occupied configuration $c^{\dagger}_{i\uparrow} c^{\dagger}_{i\downarrow} \vert 0 \rangle \equiv \vert \uparrow\downarrow\rangle$. On each site we introduce four bosons which ``label" these four states, $e^{\dagger}$, $s^{\dagger}_{\uparrow}$, $s^{\dagger}_{\downarrow}$, $d^{\dagger}$. When the fermions define a given local configuration, the bosons have to follow and the boson with the correct "label" must be created.The original Hilbert space can be mapped onto an extended space of bosons and fermions according to the following correspondence (repeated for every lattice site)

\begin{align}
  \vert 0\rangle  & \Longleftrightarrow  e^{\dagger}\vert\tilde{0}\rangle\\
  c^{\dagger}_{\uparrow}\vert 0\rangle =\vert\uparrow\rangle  & \Longleftrightarrow  f^{\dagger}_{\uparrow}s^{\dagger}_{\uparrow}\vert\tilde{0}\rangle\\
  c^{\dagger}_{\downarrow}\vert 0\rangle =\vert\downarrow\rangle  & \Longleftrightarrow  f^{\dagger}_{\downarrow}s^{\dagger}_{\downarrow}\vert\tilde{0}\rangle\\
  c^{\dagger}_{\uparrow}c^{\dagger}_{\downarrow}\vert 0\rangle =\vert\uparrow\downarrow\rangle  & \Longleftrightarrow   f^{\dagger}_{\uparrow}f^{\dagger}_{\downarrow}d^{\dagger}\vert\tilde{0}\rangle
\end{align}

In the new Hilbert space whenever the fermionic degrees of freedom describe a given local configuration, the bosons have to follow, which means that one (and only one) given boson must exist on every lattice site. This constrained is enforced through a Lagrange multiplier which adds to the constraint which imposes that the physical state on each lattice site is equivalently described by fermionic and bosonic variables. One easily realizes that the creation operator for a physical fermion can be represented as
\be
\label{fermionsb}
c^{\dagger}_{\sigma} \Longleftrightarrow f^{\dagger}_{i\sigma}(s^{\dagger}_{i\sigma}e_i + d^{\dagger}s_{i\bar\sigma})o_i,
\ee
where $o_i$ is any bosonic operator which gives 1 if evaluated on the physical Hilbert space.
Clearly, if the original problem could not be solved exactly, there is no hope to solve the new problem, which is severely more complicated, as it now features a fermion-boson interaction and several constraints. The idea  behind the whole construction is to perform a rather crude mean-field approximation which in turn leads to a non-interacting effective fermionic problem in which the physics of strong correlation is included in mean-field parameters arising from the bosonic degrees of freedom. 

In the bosonic case, the way the mean-field is formulated is by taking a saddle-point approximation for all the bosonic degrees of freedom. As a matter of fact this amounts to replace every bosonic field in a path integral representation with a c-number. This can be pictured as a Bose condensation of the slave bosons. A crucial point is that one can easily see that the original fermionic interaction can be expressed exclusively in terms of bosons, which means that, in the saddle-point approximation it will become a number and the remaining fermionic Hamiltonian will be non-interacting and therefore easily solvable.

We do not enter here in the details of the slave-boson method, but we emphasize that the main result is that the interacting system will be described as an effective non-interacting system with a renonormalization of the kinetic energy which essentially entails all the effects of the interactions. The renormalization is a momentum-independent number, which indeed coincides with the quasiparticle weight $Z$.

The correlated metal is therefore described as a Fermi liquid as long as $Z$ is finite, and the Mott-Hubbard metal-insulator transition is associated with a vanishing of $Z$, which corresponds, within the slave-boson mean-field, to a divergent effective mass.

The approach followed in the slave-spin method, that we present in the next section, is analogous in many ways to the slave-boson mean-field. The respective physical descriptions of the modeled system share many of their main features.


\section{The Slave-spin formalism and its mean-field}\label{sec:SSMF}

In this section we introduce in some details  the slave-spin mean-field method. This approach represents one of the simplest and computationally cheap approaches to study the strong correlation physics arising from large short-ranged interactions. As we anticipated above, this method belongs to the broader field of slave-particle methods, in which auxiliary particles are introduces to represent local degrees of freedom in a strongly correlated system. In the previous section we have sketched the most popular of these approaches, the slave bosons, where local bosons are introduced to represent local electronic configurations. While this method is appealing because one can easily perform a saddle-point approximation of the bosonic degrees of freedom, it is evident that the use of bosonic degrees of freedom to represent the same physics of fermions is hugely redundant. Bosons, with an infinite Hilbert space, are in fact used to describe the same physics as fermions whose Hilbert space comprises only two states.  In the Kotliar-Ruckenstein approach\cite{kotliar_ruckenstein} this implies two constraints that impose that one and only one boson exists at every site, and that its "label" corresponds to the fermionic state.

Within the slave-spin approach, we reduce the redundancy and get rid of one of the two constraints by introducing spin-1/2 auxiliary variables in correspondence with each fermionic operator. 

In this section we introduce the slave-spin approach and its mean field in the version of Refs. \cite{demedici_Slave-spins,Hassan_CSSMF}. Another version of the method, completely equivalent for the purpose of this section has been proposed in Ref. \cite{YuSi_LDA-SlaveSpins_LaFeAsO} and applied on the Fe-superconductors (see references in sec. \ref{sec:ironbased}). Further studies on interesting different developments of this technique are Refs. \cite{Ruegg-Z2-Slave-Spin,Schiro_Quenches_SlaveSpin,Mardani_SlaveSpin,Zhong_SlaveSpin_Path-Integral,Georgescu-Generalized_SlaveParticle}.
As we will see the slave-spin mean field provides us with a simple and fast method to study strongly correlated metals and their disappearance into a Mott transition. Just like in the slave-boson method, the metallic solution will be described as Fermi liquids all the way to the Mott transition. One important aspect of the slave-spin method is that it can be easily generalized to multi-orbital models, as we will discuss in Sec. \ref{sec:Hund}, where it remains numerically cheap and agile compared to similar methods.

In the slave-spin representation, we map the original local Fock space of the problem onto a larger local Fock space
that contains as many fermionic degrees of freedom (named $f_{i\sigma}$)  as the original plus the same number of 
spin-$1/2$ quantum variables, one for each $f_{i\sigma}$. It is worth noting that the auxiliary spins have nothing to do with the physical spin of the electrons, but they are just auxiliary variables which have been chosen to have the commutation relations of spins. In fact a slave-spin variable is introduced for every fermion species, taking into account the fermion spin multiplicity, so that slave-spins are also labeled with a physical-spin index $\sigma$ i.e. $S^z_{i\sigma}$.

We then associate to every state of the original physical space one of the states in this larger space by using the correspondence:
\begin{equation}
\vert n^d_{i\sigma}=1\rangle \Longleftrightarrow \vert n^f_{i\sigma}=1, \, S^z_{i\sigma}=+1/2\rangle ,
\end{equation}
\begin{equation}
\vert n^d_{i\sigma}=0\rangle \Longleftrightarrow  \vert n^f_{i\sigma}=0, \, S^z_{i\sigma}=-1/2\rangle.
\end{equation}
In words, when a local orbital and spin state is occupied then the corresponding slave-spin is "up" and if it is empty the slave-spin is "down". With these one-particle local states one construct the many-particle states as usual.

The enlarged local Fock space contains also unphysical states 
such as $\vert n^f_{i\sigma}=0, S^z_{i\sigma}=+1/2\rangle$ and $\vert n^f_{i\sigma}=1, S^z_{\i \sigma}=-1/2\rangle$. 
These unphysical states are excluded if the following local constraint is enforced at each site and for each $\sigma$:
\begin{equation}\label{constraint}
f^{\dag}_{i\sigma}f_{i\sigma}= S^z_{i\sigma}+\frac{1}{2}.
\end {equation}

The next step is to map the physical operators onto operators that act in the enlarged Fock space.
The electron number operator is equivalently represented by the auxiliary fermions number, i.e.,
$n^d_{i\sigma}\rightarrow n^f_{i\sigma}$, but also by the z component of the slave-spin $n^d_{i\sigma} \rightarrow S^z_{i\sigma}+1/2$. This allows us to rewrite any density-density interaction term in the hamiltonian in terms of the spins only. This is a crucial point, which recalls the fact that in the slave-boson language the interaction can be recast in terms of bosons only and will be crucial to eventually obtain an effective non-interacting Hamiltonian for the fermionic variables. 

For the single-band Hubbard model it is convenient to use the particle-hole symmetric form eq (\ref{eq:Hubbard_PH2}):
\begin{align}
H_{int}[d]=\frac{U}{2}\sum_i\left(\sum_{\sigma}(n^d_{i\s}-\frac{1}{2})\right)^2\longrightarrow H_{int}[S]=\frac{U}{2}\sum_{i}(\sum_{\sigma}S^{z}_{i\sigma})^2
\end{align}

For the non-diagonal operators appearing in the hopping terms we have some freedom of choice, equivalent to the choice of the operators $o_i$ in Eq. (\ref{fermionsb}).
Indeed several operators will have the same action on the physical states while differing when acting on the non-physical ones. Any of these operators is completely equivalent, as long as the constraint Eq. (\ref{constraint}) is enforced exactly, excluding non-physical states. 
For example both the choices $f_{i\sigma}^{\dag}S^+_{i\sigma}$ and $f_{i\sigma}^{\dag}2S^x_{i\sigma}$ will have the same action on the physical state $\vert n^f_{i\sigma}=0, \, S^z_{i\sigma}=-1/2\rangle$, transforming it into $\vert n^f_{i\sigma}=1, \, S^z_{i\sigma}=+1/2\rangle$, like does  $d_{i\sigma}^{\dag}$ on the corresponding original state, i.e. $d_{i\sigma}^{\dag}\vert n^d_{i\sigma}=0\rangle=\vert n^d_{i\sigma}=1\rangle$. 

In the following we derive the most general form of the spin-part of the composite operator. Indeed we can write a generic expression
\begin{equation}
d_{i\sigma} \rightarrow f_{i\sigma}O_{i\sigma},\qquad d_{i\sigma}^{\dag}  \rightarrow f_{i\sigma}^{\dag}O_{i\sigma}^{\dag}
\label{eq:generic}
\end{equation}
in which  $O_{i\sigma}$ is a generic spin-$1/2$ operator, i.e.  a $2\times 2$ complex matrix.
It is easy to determine that the most general form for $O_{i\sigma}${ is
\begin{equation}\label{eq:O}
O_{i\sigma}=\left(\begin{array}{cc} 0 & c_{i\sigma}\\1 & 0 \end{array}\right)=S^-_{i\s}+c_{i\s}S^+_{i\s},
\end{equation}
where $c_{i \sigma}$ is an arbitrary complex number.

Let's check that the operators (\ref{eq:generic}) have, in 
the physical states of the enlarged Fock space, the same effect
as the fermionic operators in the original Fock space, i.e.;
\begin{eqnarray}\label{eq:physical_d}
d_{i\sigma}|n^d_{i\sigma}=0\rangle=0, &&\quad d_{i\sigma}|n^d_{i\sigma}=1\rangle=|n^d_{i\sigma}=0\rangle\nonumber \\
d^{\dag}_{i\sigma}|n^d_{i\sigma}=1\rangle=0,&&\quad d^{\dag}_{i\sigma}|n^d_{i\sigma}=0\rangle=|n^d_{i\sigma}=1\rangle
\end{eqnarray}

The two conditions on the right determine three out of four elements of $O_{i\sigma}$.
\begin{eqnarray}
\!\!\!\!\!\!\!\!\!\!\!\! f^\dag_{i\sigma}O^\dag_{i\sigma}|n^f_{i\sigma} \!\!=\! 0, S^z_{i\sigma} \!\!=\!\! -1/2\rangle \!\!&=&\!\! |n^f_{i\sigma} \!=\! 1, S^z_{i\sigma} \!=\! +1/2\rangle,\\
\!\!\!\!\!\!\!\!\!\!\!\! f_{i\sigma}O_{i\sigma}|n^f_{i\sigma}\!\!=\!\! 1, S^z_{i\sigma}\!\!=\!\! +1/2\rangle \!\!&=&\!\! |n^f_{i\sigma}\!=\! 0, S^z_{i\sigma} \!=\! -1/2\rangle,
\end{eqnarray}
imply 
\begin{eqnarray}
&O^\dag_{i\sigma}|S^z_{i\sigma}=-1/2\rangle=|S^z_{i\sigma}=+1/2\rangle,\\
&O_{i\sigma}|S^z_{i\sigma}=+1/2\rangle=|S^z_{i\sigma}=-1/2\rangle,
\end{eqnarray}
which impose $O_{{i\sigma};11}=0$, $O_{{i\sigma};21}=1$, and $O^{*}_{{i\sigma};22}=0$. 

The two conditions on the left hand side of eqs. (\ref{eq:physical_d}) are instead assured by the fermionic operators $f_{i\sigma}$, i.e.
\begin{eqnarray}
&&f_{i\sigma}O_{i\sigma}|n^f_{i\sigma}=0, S^z_{i\sigma}=-1/2\rangle=0,\\
 &&f^{\dag}_{i\sigma}O^\dag_{i\sigma}|n^f_{i\sigma}=1,S^z_{i\sigma}=+1/2\rangle=0
\end{eqnarray}
for any $O_{i\sigma}$.
This implies that $O_{{i\sigma};12}=c_{i\sigma}$ remains undetermined\footnote{For any choice of $c_{i\sigma}$ the operators (\ref{eq:generic}) duly respect anticommutation relations, in the subspace of physical states.}.

The arbitrariness of the complex number $c_{i\sigma}$ is a sort of gauge of our formulation as long as the constraint is treated exactly.  In any practical implementation some  approximation has to be performed to enforce the constraint in our mean-field treatment. In these approximations the particular choice of the gauge comes into play. 
$c_{i\sigma}$ can indeed be tuned in order to give rise to the most physical approximation scheme, by  
imposing, for instance, that the mean field correctly reproduces solvable limits of the problem, like the non-interacting 
limit. A similar freedom is indeed present also in the Kotliar-Ruckenstein slave-boson mean-field, where the choice of reproducing the non-interacting limit implies that the mean-field coincides with the Gutzwiller approximation\cite{kotliar_ruckenstein}.

We will see that the correct choice of $c_{i\sigma}$ depends on the average occupation of the local 
state, and is such that it reduces to 1 at occupation $1/2$, so that $O_{i\sigma}=2S^x_{i\sigma}$ and 
the prescription for half-filling is recovered.

Thus finally, in the enlarged Fock space the Hamiltonian can be written exactly as:
\be
\hat H-\mu \hat N=\sum_{<ij>\sigma}t_{ij}O^{\dag}_{i\sigma}O_{j\sigma}f^{\dag}_{i\sigma}f_{j\sigma} + \frac{U}{2}\sum_{i}(\sum_{\sigma}S^{z}_{i\sigma})^2 -\mu\sum_{i\sigma}n^f_{i\sigma},
\label{eq:SS_ham_oneband}
\ee
subject to the constraint  (\ref{constraint}).

\subsection{Mean-field approximation}

So far we have only formulated a more involuted and redundant representation for the original Hubbard model, which obviously does not allow us to solve exactly the model. In the same spirit of the slave-boson mean field, we now perform a mean-field approximation, which consists of three main steps

\begin{enumerate}
\item{Decoupling auxiliary fermions and slave-spin degrees of freedom}
\item{Treating the constraint on average, using a site- and spin-independent (since we will investigate uniform non-magnetic phases)
   Lagrange multiplier $\lambda$}
\item{Solving the resulting slave-spin Hamiltonian in a single-site mean-field approximation (a simple Weiss-like mean-field)}
  \end{enumerate}
  
As we will see, this approximation leads to an effective non-interacting Hamiltonian for the pseudofermions similar to the result of the slave-boson mean-field. In the present case, the parameters of the pseudofermion Hamiltonian are determined from the solution of the slave-spin Hamiltonian rather than from the saddle-point approximation for the slave bosons.  Let us now discuss the mean-field solution of the Hubbard model in details. 

\subsubsection{Decoupling fermions from spins}
The main obstacle to solve the new Hamiltonian (\ref{eq:SS_ham_oneband}) lies in the original hopping terms, which has now turned into an interaction between pseudofermions and spins. We thus decouple the mixed spin-fermion hopping term according to
\be
\sum_{<ij>\sigma}t_{ij}O^{\dag}_{i\sigma}O_{j\sigma}f^{\dag}_{i\sigma}f_{j\sigma}\simeq \sum_{<ij>\sigma}t_{ij}\langle O^{\dag}_{i\sigma}O_{j\sigma}\rangle f^{\dag}_{i\sigma}f_{j\sigma}+\sum_{<ij>\sigma}t_{ij}O^{\dag}_{i\sigma}O_{j\sigma}\langle f^{\dag}_{i\sigma}f_{j\sigma}\rangle.
\ee
As a result the Hamiltonian can be rewritten as a sum of a purely fermionic one and a purely spin one: $H=H_f+H_S$. The effect of the original interaction now lies in the fact that the coefficients of the two Hamiltonians depend on expectation values computed on the other.

\subsubsection{Treating the constraint on average}
Despite the separation of Hamiltonians, the problem remains practically intractable because of the constraint which still couples the two degrees of freedom. 
We can write the partition function in the enlarged Fock space as:
\be
{\cal Z}=Tr[e^{-\beta (\hat H-\mu \hat N)}\d(S^z_{i\sigma}+\frac{1}{2}-f^{\dag}_{i\sigma}f_{i\sigma})]
\ee
where the delta function imposes the constraint by projecting out  all contributions from the unphysical states.
Even having decoupled the hamiltonian in step 1, because of the delta function, the sum over the enlarged Fock space cannot be replaced by two factorized sums over the decoupled spaces of spin and fermions. 

But one can relax the constraint by replacing the delta inside the trace with a simple exponential of the form $e^{-\beta\lambda(S^z_{i\sigma}+\frac{1}{2}-f^{\dag}_{i\sigma}f_{i\sigma})}$, and realize that $\lambda$ can be adjusted so that the constraint is respected on average, i.e., it is satisfied by the expectation values of the operators.   
Indeed by minimizing the grand potential $\Omega=-K_BT\log{\cal{Z}}$ with respect to $\lambda$, that is determining what $\lambda$ satisfies the condition $\partial\Omega/\partial\lambda=\frac{1}{\cal Z}Tr[e^{-\beta (\hat H-\mu \hat N)}(S^z_{i\sigma}+\frac{1}{2}-f^{\dag}_{i\sigma}f_{i\sigma})]=0$, one ensures that $\langle S^z_{i\sigma}\rangle +\frac{1}{2}=\langle f^{\dag}_{i\sigma}f_{i\sigma} \rangle$.
Thus $\lambda$ acts as a Lagrange multiplier for the constrained minimization of $\Omega$ by respect to the mean-field parameters.

After the first two steps, the total Hamiltonian can be written as the sum of the following two effective 
Hamiltonians:
\begin{align}
H_{f}=&\sum_{<ij>,\sigma}t_{ij}(Q_{ij}f^{\dag}_{i\sigma}f_{j\sigma}+ H.c.)
-(\mu+\lambda)\sum_{i}n^f_{i} \label{eq:Hf},   \\
H_{s}=&\sum_{<ij>,\sigma}(J_{ij}O^{\dag}_{i\sigma}O_{j\sigma} + H.c.)
+\lambda\sum_{i,\sigma}(S^{z}_{i\sigma}+\frac{1}{2})
+\frac{U}{2}\sum_{i}(\sum_{\sigma}S^{z}_{i\sigma})^2\label{eq:Hs} .
\end{align}
 The parameters $Q_{ij}$ (hopping renormalization factor), $J_{ij}$ (slave-spin exchange constant) and $\lambda_{i}$ in these expression are  determined from the following  coupled self-consistency equations:
\begin{equation}
Q_{ij}=\langle O^{\dag}_{i\sigma}O_{j\sigma}\rangle_{s},
\label{eq:Qij}
\end{equation}
\begin{equation}
J_{ij}=t_{ij}\langle f^{\dag}_{i\sigma}f_{j\sigma}\rangle_{f},
\label{eq:Jij}
\end{equation}
\begin{equation}
\langle n^f_{i\sigma}\rangle_{f}=\langle S^{z}_{i\sigma}\rangle_{s}+\frac{1}{2},
\label{eq:constraint}
\end{equation}
where $\langle\rangle_{f,s}$ indicates the effective Hamiltonian and the corresponding Fock space used for the calculation of the 
averages. 

\subsubsection{Mean-field approximation for the slave-spin Hamiltonian}
We are thus left with two Hamiltonians, one for the pseudofermions and the other for the slave spins, which must be solved in a self-consistent way so that the conditions (\ref{eq:Qij}), (\ref{eq:Jij}) and (\ref{eq:constraint}) are fullfilled.

The pseudofermionic Hamiltonian is indeed a renormalized independent-fermion Hamiltonian 
for the $f_{i\sigma}$ of which the hopping is renormalized by the expectation values of some spin operators.
However, the spin Hamiltoninan retains essentially the full complexity of the original model. We have thus to perform a further approximation, in 
this case the single-site mean-field, on the spin Hamiltonian.

For this, a Hamiltonian with one site only is considered, on which 
the interaction is treated exactly, and is embedded in the effective ("Weiss") field 
of its surroundings, i.e. the quantum and thermodynamic average value of the operators on other sites coupled to the chosen site. Assuming 
translational invariance this approximate Hamiltonian is self-consistently used to calculate also the same mean-field average values.

Mathematically this means performing the approximation 
\be
\langle O^{\dag}_{i\sigma}O_{j\sigma}\rangle \simeq \langle O^{\dag}_{i\sigma}\rangle O_{j\sigma}+O^{\dag}_{i\sigma}\langle O_{j\sigma}\rangle
\ee
so that we obtain the following mean-field slave-spin Hamiltonian:
\begin{equation}\label{eq:H_s_MF}
H_{s}=\sum_i H^i_{s}=\sum_{i,\sigma}(h_{i\sigma}O^\dagger_{i\sigma} + H.c.) 
+\lambda\sum_{i,\sigma}(S^{z}_{i\sigma}+\frac{1}{2})
+\frac{U}{2}\sum_{i}(\sum_{\sigma}S^{z}_{i\sigma})^2 .
\end{equation}
where
\begin{equation}
h_{i\sigma}=h_\sigma=\sum_{j}J_{ij}\langle O_{j\sigma}\rangle_s=\langle O_{i\sigma}\rangle_s\sum_{j} t_{ij}\langle f^{\dag}_{i\sigma}f_{j\sigma}\rangle_{f},
\label{eq:h}
\end{equation}
given the definition of $J_{ij}$ and the fact that $\langle O_{i\sigma}\rangle=\langle O_{j\sigma}\rangle$ because of translational invariance.
This last sum is independent on the renormalized parameters in the one-band case and it can be evaluated once and for all, yielding:
\be
\sum_{j} t_{ij}\langle f^{\dag}_{i\sigma}f_{j\sigma}\rangle_{f}=\sum_{\vk} \epsk \langle f^{\dag}_{\vk\sigma}f_{\vk\sigma}\rangle_{f}\equiv \bar \eps,
\ee
which is the kinetic energy per spin for the pseudofermions.

Equations (\ref{eq:Hf}) and (\ref{eq:H_s_MF}), with the definitions eqs. (\ref{eq:Qij}) and (\ref{eq:h}), and the implicit condition (\ref{eq:constraint}) constitute the slave-spin mean-field equations for the solution of the single-band Hubbard model Eq. \ref{eq:Hubbard_oneband}, and have to be solved for the mean-field parameters $h_{i\s}$,$Q_{ij}$,$\lambda$. It is worth noting that the hopping renormalization factor $Q_{ij}$, after the last approximation becomes  $Q_{ij} = \langle O_{i\sigma}\rangle^2$  and it is the same for every pair of sites. As a result it becomes a prefactor which multiplies a kinetic energy operator identical to the non-interacting result. The physical picture emerging is that of a kinetic energy renormalized by a factor, which can be shown to coincide with the quasiparticle weight $Z$, as we show explictly in the  complement \ref{subsec:Z_vs_Q}. This is a common result to the single-site mean-field of slave particle methods and it mirrors the natural expectation that the motion of electrons in the presence of strong interaction is reduced. 

Turning back to our solution, the single-site approximation also implies that the slave-spin hamiltonian is the sum of decoupled one-site hamiltonians so that $h_{i\sigma}=h_\sigma$ in Eq. (\ref{eq:h}) and the sum over i can be dropped in Eq. (\ref{eq:H_s_MF}). 
The practical advantage is now that the slave-spin one-site hamiltonian can be solved exactly as it has a Fock space of finite dimension. Notice that including more orbitals in the modelling will increase the size of this local Hilbert space introducing one spin for every fermionic variable. The growth of this local Hilbert space will be the bottleneck of actual calculations.

In practice these equations are solved iteratively: starting with a guess for $Z$ and $\lambda$ one solves the fermionic hamiltonian in k-space and calculates $h_{i\s}$, which is used in the spin hamiltonian to calculate the new $Z$. Within the cycle $\lambda$ is adjusted so that the average constraint eq. (\ref{eq:constraint}) is satisfied, and the scheme is iterated until convergence of all quantities.

The only part of the scheme yet to be specified is the choice of the complex number  $c_{i\sigma}$ ($c_{i\sigma}=c$ in the present site- and spin-independent mean-field)\cite{Hassan_CSSMF}.
The most natural choice is to use this gauge of our formulation to impose that the non-interacting limit, which is exactly soluble, is exactly reproduced, i.e. that Z(U=0)=1.
Imposing this condition (and taking $c$ real) gives an expression for $c$ that depends only on the density $\langle n_{d\s}\rangle=\langle n_{f\s}\rangle\equiv n_\s$, i.e.
\begin{equation}\label{eq:gauge}
c=\frac{1}{\sqrt{n_\s(1-n_\s)}}-1
\end{equation}
The gauge can be fixed then once and for all for a given wanted density, that is reached by adjusting the chemical potential $\mu$. 
In practice it can be advantageous to adjust $\mu$ and $c$ along with the other parameters during the mean-field iteration scheme just outlined.

With this choice we see from Fig.~\ref{fig:comp_SB} that at the same fixed population $n$ the single-site mean field of the Slave Spins method gives exactly the same results of the Kotliar-Ruckenstein mean-field of the Slave Bosons method, i.e. the Gutzwiller Approximation.

\subsection{\emph{Complement}: derivation of the gauge c for arbitrary filling}\label{subsec:c_single-site}

In the single-site approximation, we can determine the gauge $c$ analytically\cite{demedici_PhD,Hassan_CSSMF}. 

The non-interacting single-site slave spin Hamiltonian $H_{s}$ reads:
\begin{equation}
\hat H_{s}=hO^{\dag} + h^*O +\lambda S^z,
\end{equation}
where $O$ is defined as in Eq. (\ref{eq:O}). The single-site fermionic part of the Hamiltonian (where we have absorbed the the $\lambda/2$) is simply spinless non-interacting fermions.
The physical spin index $\sigma$ is supressed in $H_{s}$ since for $U=0$ upspin and downspin fermions are  decoupled, so that we can diagonalize the hamiltonian for one slave-spin in the $S^z=\pm 1/2$ basis.
Thus the matrix to be diagonalized reads: 
\be
H_{s}=\left(
\begin{array}{cc}
\frac{\lambda}{2} & a \\
 a^* & -\frac{\lambda}{2}\\
\end{array}
\right)
\ee
 and $a=h+ch^*$.
 
The ground state eigenvalue $\epsilon_{GS}$ and the corresponding eigenstate are
\begin{equation}
\epsilon_{GS}=-\sqrt{\frac{\lambda^2}{4}+\vert a^2\vert}\equiv-R
\end{equation} 
\begin{equation}
\vert GS\rangle=\left(\begin{array}{c}\frac{\frac{\lambda}{2}+R}{N}\\\frac{-a*}{N}\end{array}\right)
\end{equation}
with $N=\sqrt{2R(\frac{\lambda}{2}+R)}$.

The expectation value of $S^{z}$ and $O$ in the ground state are 
\begin{equation}
\langle S_z\rangle= \frac{\lambda}{4R}
\end{equation}
and
\begin{equation}
\langle O \rangle=-\frac{ca^*+a}{2R}
\end{equation}
The Lagrange multiplier depends on the density $n=\sum_\s n_\s$ and is adjusted in order to satisfy the constraint equation: 
\begin{equation}
n_\s-\frac{1}{2} = \langle S_z\rangle=\frac{\lambda}{4R}
\label{eq:n}
\end{equation}

We want to tune $c$ in order to match the condition that in the limit $U=0$ the renormalization factor Z must be unity:
\begin{equation}
Z=<O>^2=\frac{\vert c a^{*}+a\vert^2}{4R^2}=1
\label{eq:Z_1}
\end{equation}

We can easily eliminate $\lambda$ from these two conditions, by squaring eq. (\ref{eq:n}).
We are left with the following expression for $c$:
\begin{equation}
\frac{\vert a \vert^2}{\vert c a^*+a\vert^2}=\frac{1}{\vert c e^{-i\arg{a}}+e^{i\arg{a}}\vert}=n_\s-n_\s^2
\end{equation}

If we choose c to be real then h and a are also real.
Then, the expression for $c$ in the closed form is
\be
c=\frac{1}{\sqrt{n_\s(1-n_\s)}}-1.
\ee
Note that this result is independent of $h$.

\subsection{\emph{Complement}: Fermi-liquid quasiparticle weight and mass enhancement.}\label{subsec:Z_vs_Q}

Using the correspondence $d_{i\s} \rightarrow O_{i\s}f_{i\s}$ we can rewrite the physical Green function, within our mean-field formalism as a product of fermionic and spin Green's functions:

\be
G^d_{ij}(t)= G^f_{ij}(t)G^S_{ij}(t)
\ee
with 
\bea
G^d_{ij}(t) &\@=&\@ -i\langle T_t d_{i\s}(t) d^{\dag}_{j\s}(0) \rangle_d \\
G^f_{ij}(t) &\@=&\@ -i\langle T_t f_{i\s}(t) f^\dag_{j\s}(0) \rangle_f \\
G^S_{ij}(t) &\@=&\@ \langle T_t O_{i\s}(t) O^\dag_{j\s}(0) \rangle_S
\eea

a Fourier transform with respect to time transforms the equal time product into a convolution
\be
G^d_\vk(\omega)= \int \frac{d\omega' }{2\pi}\frac{d\vk'}{8\pi^3} G^f_{\vk'}(\omega')G^S_{\vk-\vk'}(\omega-\omega').
\ee

In the metallic phase the slave-spins are long-range ordered, this implies that the correlation function $G^S$ has a finite value at large distances and times which is  given by $Z=\langle O_{i\s}^\dag\rangle \langle O_{j\s} \rangle$, and thus its Fourier transform has a Dirac delta contribution at zero frequency
\be
G^S_\vk(\omega)=Z\d(\omega)\d(\vk)+G^{S,reg}_\vk(\omega)
\ee
which in turn implies that 
\be
G^d_\vk(\omega)= Z G^f_{\vk}(\omega) + \int \frac{d\omega' }{2\pi}\frac{d\vk'}{8\pi^3} G^f_{\vk'}(\omega')G^{S,reg}_{\vk-\vk'}(\omega-\omega').
\ee
It is thus obvious that Z plays the role of a quasiparticle weight.

When comparing the expressions of the particle ($d^\dag_{i\s}$) and quasiparticle ($f^\dag_{i\s}$) Green functions, i.e.:
\be
G^d_\vk(\omega)=\frac{1}{\omega+\mu-\epsk-\Sigma(\omega)}
\ee
and 
\be
G^f_\vk(\omega)=\frac{1}{\omega+\mu+\lambda -Z\epsk}
\ee 
one sees that the electronic self-energy is, in this scheme:
\be
\Sigma(\omega)=\mu-\frac{\mu+\lambda}{Z}+\omega(1-\frac{1}{Z}).
\ee

This establishes that the slave-spin mean-field, in complete analogy with the slave-boson mean-field and the Gutzwiller approximation, is equivalent to a self-energy which features a low-frequency linear behavior which determines the quasiparticle renormalization and a constant term, which determines a shift of the chemical potential. 

Before entering in the main argument of this chapter, which is the application of the the slave-spin mean-field to the multi orbital models for iron-based superconductors, we present some benchmark calculations to assess the ability of the approach to study well-known limits and regimes.

\subsection{The half-filled single-band Hubbard model}

\begin{figure}[h]
\includegraphics[width=12cm]{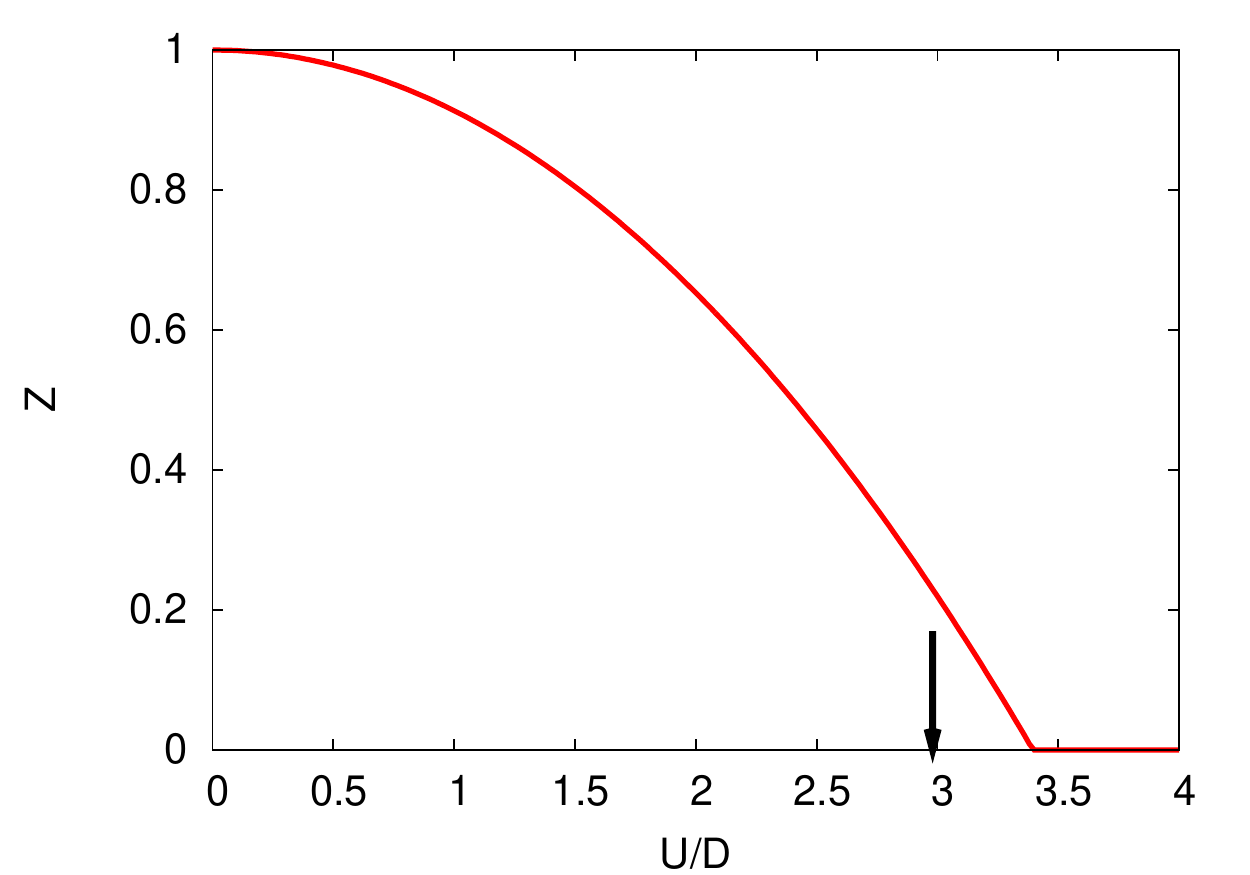}
\caption{\label{fig:Mott_oneband-Spin}
Quasiparticle weight in the half-filled single-band Hubbard model solved with SSMF, for a semi-circular DOS of half-bandwidth D. The arrow show the $U_c$ obtained in DMFT for the same model.}
\end{figure}

We begin our exploration of simple results within SSMF with the half-filled single-band Hubbard model.
In Fig. \ref{fig:Mott_oneband-Spin} we report the quasiparticle weight obtained by solving the slave-spin mean-field equations for a half-filled system with a semicircular density of states with semi-bandwidth D. In this case we can obtain $\lambda =0$ from particle-hole symmetry and solve the remaining equations numerically. The results share the same qualitative and quantitative behavior of popular mean-field approaches to the Mott transition, like the slave-boson mean-field (Brinkmann-Rice transition) and the Gutzwiller approximation. $Z$ decreases regularly from the non-interacting value 1 and vanishes at a critical value  $U_c \simeq 3.39D$ (See complement \ref{comp:perturbative_Uc} for an analytical derivation of this value), slightly larger than the result obtained by means of Dynamical Mean-Field Theory, which captures the local quantum fluctuations beyond mean-field. We recover that a vanishing $Z$ amounts to a divergent effective mass. The main problem of this approach is that it is limited to the strongly correlated metallic region and its description of the Mott insulating state is basically that of the atomic limit.

\subsection{\emph{Complement}: critical U for the Mott transition in the single-orbital Hubbard model}\label{comp:perturbative_Uc}

Here we show how to obtain analytically within the slave-spin mean-field approximation the critical interaction strength $U_c$ for the Mott transition in the half-filled one-band Hubbard model.

Indeed the slave-spin equation equation (\ref{eq:H_s_MF}) can be developed around the insulating solution $h_{i\s}=0$ where $H_s=H_{at}$ with
\be
H_{at}=\frac{U}{2}\left(\sum_\s S^z_\s\right)^2, 
\ee
of with the spectrum can be calculated exactly. One can calculate the effect, as a perturbation, of the "kinetic" part of the slave-spin hamiltonian:
\be
H_{pert}=h\sum_\s 2 S^x_\s.
\ee
where $ O_{i\sigma} = 2 S^x_\s$ because for a half-filled band the gauge $c=1$ (eq. (\ref{eq:gauge}) for n=1/2), and we call $h\equiv h^*_{i\s}+h_{i\s}$. Also $\lambda=0$ at half-filling because of particle-hole symmetry.

The perturbative parameter here is thus $h$, which is small near the metal-insulator transition, if this is second-order.

The unperturbed space is:
\be
\left\{
\begin{array}{cc}
|\up\up\rangle & E=\frac{U}{2}\\
|\up\down\rangle & E=0\\
|\down\up\rangle  & E=0\\
|\down\down\rangle & E=\frac{U}{2}\\
\end{array}
\right.
\ee
The unperturbed ground state being two times degenerate one has to use the degenerate perturbation theory. Furthermore, since the perturbation $H_{pert}$ does not have any nonzero elements in the low-energy subspace one has to use second-order perturbation theory to find the state, in the ground-state subspace, to which the perturbed state tends for $h\rightarrow 0$.

In second-order perturbation one has to diagonalize the matrix $H'\equiv H_{pert}(E_0-H_{at})^{-1}H_{pert}$  in the degenerate subspace to obtain the energy corrections at the leading order. In this case:
\be
H'=\left ( 
\begin{array} {cc}
- \frac{4h^2}{U} & - \frac{4h^2}{U}\\
- \frac{4h^2}{U} & - \frac{4h^2}{U}\\
\end{array}
\right)
\ee
diagonalising which one obtains  $| \phi_0\rangle\equiv (|\up\down\rangle + |\down\up\rangle )/\sqrt{2}$ as the ground state, of energy (to second order in $h$) $E_0^{(II)}=-\frac{8h^2}{U}$.
 
The correction to the ket is however nonzero at first order in $h$ and reads:
 \be
 |\phi_0^{(I)}\rangle=| \phi_0\rangle -\frac{2}{U} 2h (|\up\up\rangle + |\down\down\rangle)/\sqrt{2}.
 \ee
 
 To calculate the critical interaction one simply uses the self-consistency condition $h=2 \bar \eps \langle 2 S_\s^x\rangle$. 
 This condition has to be rigorously satisfied and since we can only calculate here a perturbative development for $\langle 2 S_\s^x\rangle$, 
 which by definition is only exactly valid at the transition, (where $h\rightarrow 0$), the self-consistency condition becomes an equation determining the coupling U for which this condition can be satisfied rigorously, hence the critical coupling $U_c$.
 
The calculation for $\langle 2 S_\s^x\rangle$ on the state $|\phi_0^{(I)}\rangle$ gives:
\be
\langle \phi_0^{(I)} |2 S_\s^x|\phi_0^{(I)}\rangle = -\frac{8h}{U}.
 \ee
 
The self-consistency equation becomes then: 
 \be
 h=2 \bar \eps \langle 2 S_\s^x\rangle=-\frac{16h}{U_c}\bar \eps \quad \Longrightarrow \quad U_c=-16\bar \eps.
 \ee
 For a semi-circular DOS of half-bandwidth $D=1$,  $\bar \eps\simeq -0.2122$, which gives
 \be
 U_c\simeq 3.39
 \ee
 if $D=1$, in agreement with the numerical result fig. \ref{fig:Mott_oneband-Spin}.

\subsection{The Hubbard model for finite doping}

The treatment of the Hubbard within SSMF model can be easily extended to any density away from half-filling. This is indeed obtained by moving the chemical potential away from its particle-hole symmetric value $\mu$ ($=0$ or $=U/2$ depending on which among the equivalent forms of the Hubbard hamiltonian are used - see complement \ref{compl:1band_PH} -  these analytically-derived values hold for particle-hole symmetric DOS), and determining the corresponding value of the Lagrange multiplier $\lambda\neq 0$ at the relevant doping.
As mentioned in sec. \ref{subsec:c_single-site} the gauge eq. (\ref{eq:gauge}) in the slave-spin formalism depends on the final population, so that its value has to be updated in the numerical solution of the mean-field equations.

It can be shown (Fig. \ref{fig:comp_SB}) that this formulation gives exactly the same results than the Kotliar-Ruckenstein slave-boson mean-field\cite{kotliar_ruckenstein}, as illustrated in Fig. \ref{fig:comp_SB}.
\begin{figure}
\includegraphics[width=12cm]{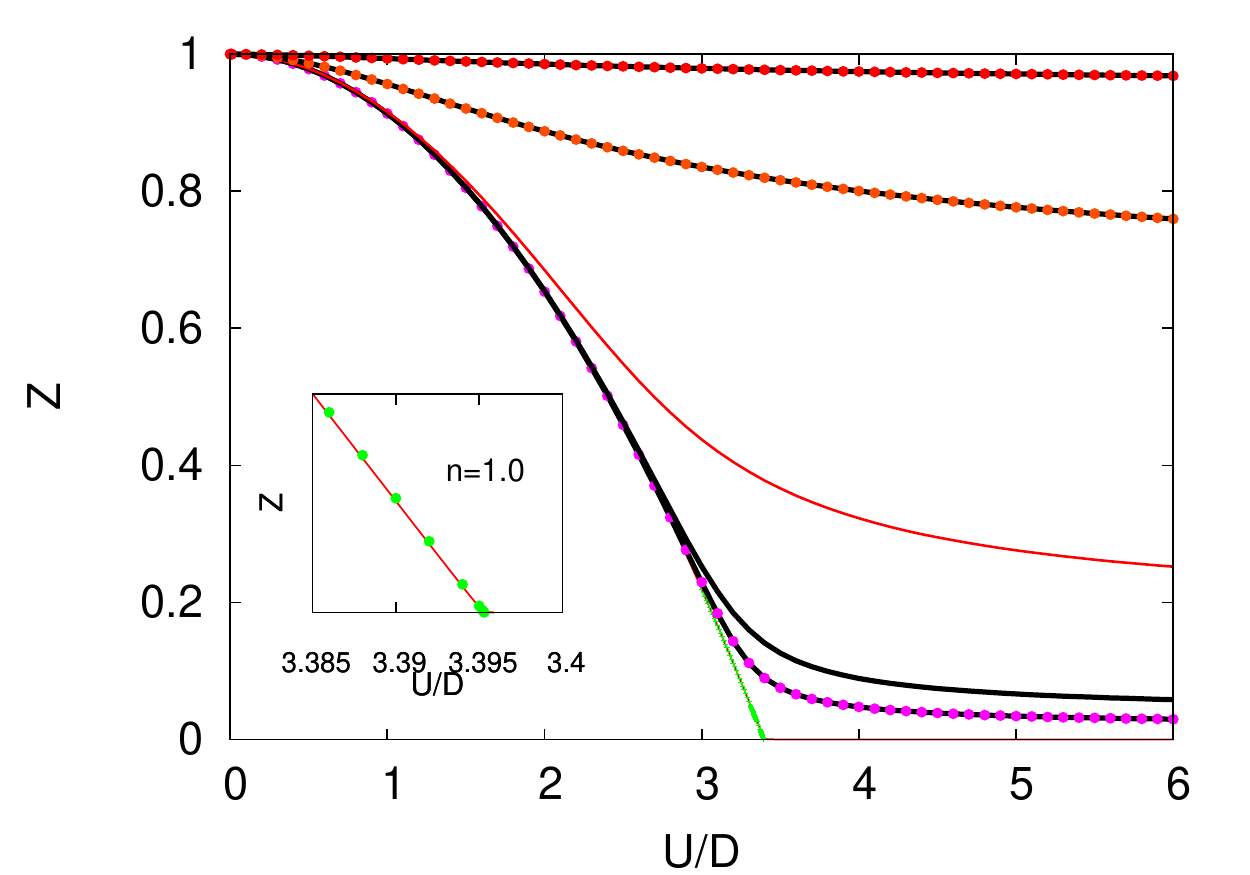}
\caption{ Comparison of Slave-Spin (within the gauge eq. (\ref{eq:gauge}), continuous lines) and Kotliar-Ruckenstein Slave-bosons (dots) single-site mean-fields in the one-band Hubbard model. For any fixed population (from below, the filling is $n=\sum_\s n_\s=1.0, 0.99, 0.98, 0.9, 0.5, 0.1$) the two methods give coincident results. Inset: blow-up of the Mott transition at half-filling, $n=1.0$. From \cite{Hassan_CSSMF}.}
\label{fig:comp_SB}
\end{figure}
The quasiparticle weight is still reduced with increasing interaction but does not vanish. Indeed the Mott transition needs an integer number of electron per site. 
For large values of the doping away from half-filling it tends to the proper value $Z=1$ for a dilute electron system, while at small doping indeed Z is strongly reduced following the trend of the half-filled case, but then it saturates at a small finite value that at large U depends only on the doping. For $U\rightarrow\infty$  the value of Z can be derived analytically\cite{Hassan_CSSMF} and it reproduces the value from the Gutzwiller approximation (equivalent to Kotliar-Ruckenstein slave-boson mean-field\cite{kotliar_ruckenstein})
\be
Z=\frac{2x}{1+x}
\ee
where $x=|1-n|$ is the absolute value of the doping (with $n=\sum_\s n_\s$ the total density).

\section{Multi-orbital correlations: Hund's coupling}\label{sec:Hund}

In this section we introduce one of the main characters which enter into play when the low-energy description of the material requires to include more than one orbital per site.
So far we have considered the simplified case when the low energy model is restricted to one orbital only. In this situation there is obviously only one matrix element corresponding to the local interaction, i.e. the element with $m=m'=n=n'$ in eq. (\ref{eq:Hubbard_interaction}), that we labeled U.

However when multiple orbitals are retained in the basis to expand the electron field, other integrals appear. The number of independent integrals depends on the number of retained orbitals and on the spatial symmetry of the problem. As discussed in Sec. \ref{sec:chemistry} materials where the valence bands arise from 3d orbitals are the reference case for strong correlations. 
In these materials the five orbitals of the 3d shells are typically lifted in energy by a crystal field which has quite often cubic symmetry and separates a pair of $e_g$ orbitals, from a triplet of $t_{2g}$ orbitals. In these cases one can show that there are 3 independent integrals\footnote{All Wannier functions are intended here centered around the site with $R_i=0$.}:

\begin{eqnarray}
V^{mmmm}_{iiii}&\equiv& U= \int d\vr d\vrp\, |w_m(\vr)|^2\, W(\vr,\vrp)\, |w_m(\vrp)|^2 \nonumber \\
V^{m\mp\mp m}_{iiii}&\equiv& U^\prime= \int d\vr d\vrp\, |w_m(\vr)|^2\, W(\vr,\vrp)\, |w_{\mp} (\vrp)|^2 \nonumber \\
V^{m\mp m\mp}_{iiii}&\equiv& J= \int d\vr d\vrp\, w^*_m(\vr)w^*_{\mp} (\vrp) \, W(\vr,\vrp)\, w_m(\vrp)w_{\mp} (\vr)
\label{eq:uupj}
\end{eqnarray}
and all others vanish by symmetry. Thus the interaction used in the multi-orbital Hubbard model in these cases has customarily the form of the "Kanamori" hamiltonian\footnote{The coefficient of the last term, the pair-hopping", is equal to J only if the basis functions are chosen real-valued.}:
\begin{eqnarray}
\hat H_{int}\,=\,U\sum_m n_{m\up}n_{m\down}\,+\,U^\prime\sum_{m\neq\mp} n_{m\up}n_{m'\down}\,
+(U^\prime-J) \sum_{m<\mp,\sigma} n_{m\s}n_{m'\s} + \nonumber \\
-J\,\sum_{m\neq\mp} d^+_{m\spinup}d_{m\spindown}\,d^+_{\mp\spindown}d_{\mp\spinup} \,
+ J\, \sum_{m\neq\mp} d^+_{m\spinup}d^+_{m\spindown}\,d_{\mp\spindown}d_{\mp\spinup} 
\label{eq:ham_kanamori} 
\end{eqnarray}

Cubic symmetry already, for the $e_g$ case, and the requirement of full "rotational" symmetry (i.e. invariance of the hamiltonian under orbital, spin and charge gauge symmetry transformations, separately) for the $t_{2g}$ case impose the further condition
\be
U'=U-2J.
\ee
This form of the interaction Hamiltonian is not exact in symmetries other than cubic or if one considers more orbitals (as it is appropriate for a full 3d shell, for example), and anyway implies the rotational invariance of $W(r-r')$ which is true in free space but only approximate in a solid state environment. It is however customarily used in practice. 

It is quite intuitive that it enforces the fact, transparent from the density-density terms, that the Coulomb repulsion between two electrons is reduced when these occupy two different orbitals. Moreover in this case the Pauli principle does not prevent the electrons to have their spins to be either anti-parallel or parallel along the quantization axis (say $\hat z$). In this case the purely quantum "exchange" integral J further lowers the energy of the parallel configuration. The spin-flip terms account for the same effect along the $\hat x$ and $\hat y$ directions, restoring the spin rotational invariance. 

In an isolated atoms these interaction terms originate the well-known "Hund's rules" that dictate the \emph{aufbau} of an atom (Fig. \ref{fig:Hund}): within the same shell the electrons occupy different orbitals with their spins aligned if possible, and then they start doubly occupying the orbitals. 

The effect of Hund's coupling in magnetic insulators is also well known. What is becoming clear in recent years is that it also plays a crucial role in tuning the correlations in correlated metals (and their tendency to become Mott insulators)\cite{Georges_annrev}, and this physics, explored with the slave-spin mean-field is the subject of the following chapters.

\begin{figure}
\begin{center}
\includegraphics[width=6cm]{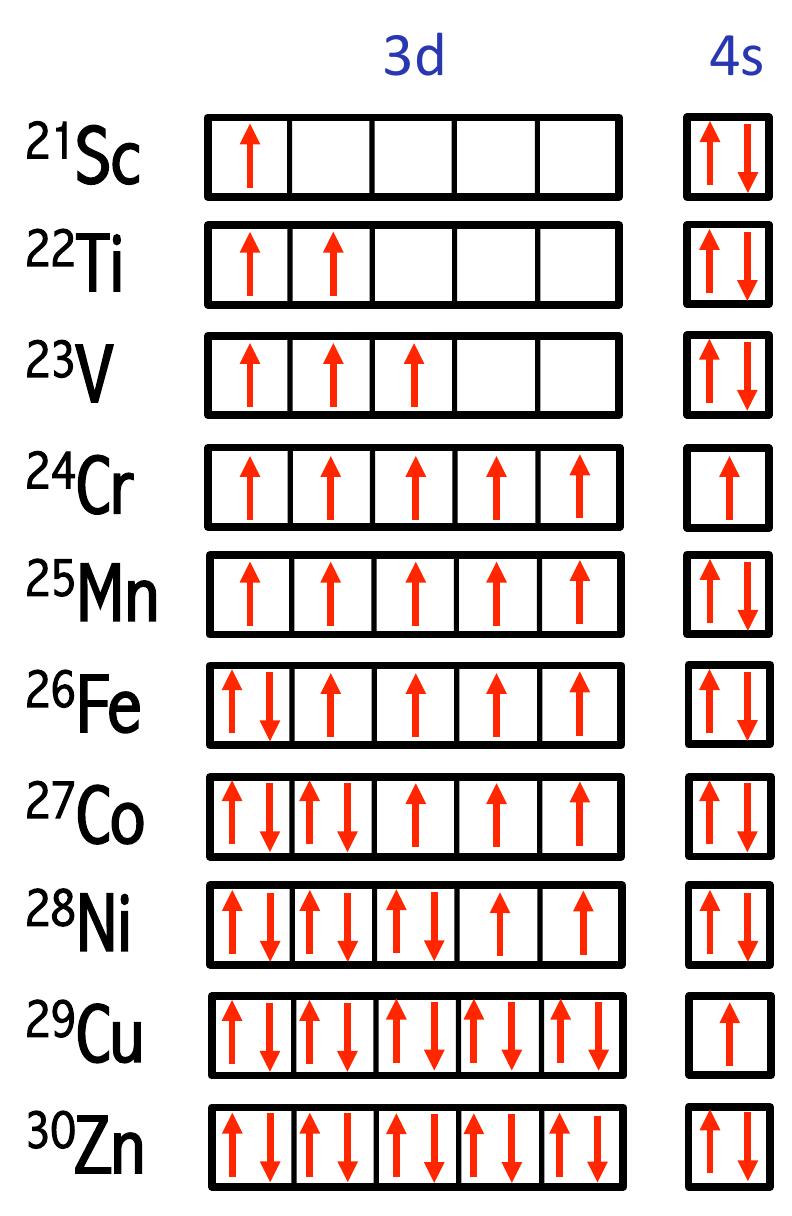}
\caption{Atomic Hund's rules in the 3d series of elements.}
\label{fig:Hund}
\end{center}
\end{figure}

In many practical applications only the three density-density terms of the Kanamori interaction are retained (dubbed interaction \emph{\`a la} "Ising"). 
This introduces somewhat quantitative, but rarely qualitative changes to the results and that is what we will do for the rest of the manuscript.

\subsection{\emph{Complement}: particle-hole symmetry in the multi-orbital Hubbard model}

Applying the particle-hole symmetry to the Kanamori Hamiltonian for two orbitals one obtains the value of the chemical potential yielding half-filling in cases where the hopping structure is particle-hole symmetric, as done for the single-band Hubbard model in complement \ref{compl:1band_PH}.

It can also be obtained by simply writing the Hamiltonian in a manifestly invariant form and then extract the chemical potential shift needed to obtain it from 
\be
\hat H_{int}-\mu\sum_{m\s}n_{m\s},
\label{eq:H_int_chem_shift}
\ee
where $\hat H_{int}$ is of the form eq. (\ref{eq:ham_kanamori}). Indeed we can write:
\be
\hat {\tilde {H}}_{int}\,=\,U\sum_m \tilde n_{m\up}\tilde n_{m\down}\,+\,U^\prime\sum_{m\neq\mp} \tilde n_{m\up}\tilde n_{m'\down}\,
+(U^\prime-J) \sum_{m<\mp,\sigma} \tilde n_{m\s}\tilde n_{m'\s}
\label{eq:Kanamori_Ising_tilde}
\ee
where all number operators have been shifted by $1/2$ using the notation $\tilde n_{m\s}\equiv n_{m\s}-1/2$, which means that we have only added one-body terms, and a constant energy shift.
We have also dropped the non density-density terms of (\ref{eq:ham_kanamori}) (the so-called  "spin-flip" and "pair-hopping" terms), which are manifestly particle-hole invariant and hence do not introduce any chemical potential shift under particle-hole transformation.

Developing the expression (\ref{eq:Kanamori_Ising_tilde}) and matching it with eq. (\ref{eq:H_int_chem_shift}) (dropping the non-diagonal terms there too) one finds that $\mu_{1/2filling}=\frac{U(2N-1)-5J(N-1)}{2}$.

\subsection{Extension of the Slave-spin formalism to multi-orbital Hubbard models}

In the multi-orbital context the Hubbard model becomes:
\be\label{eq:multiorb_Hubbard}
\hat H = \sum_{i\neq jmm'\s} t^{m m'}_{ij} d^\dag_{im\s}d_{jm'\s}+\sum_{im\s}\eps_mn^d_{im\s}+\hat {H}_{int}
\ee
where $\hat {H}_{int}$ is of the form eq. (\ref{eq:ham_kanamori}). We have also written explicitly the on-site one-body terms, that are the orbital energies $t^{mm}_{ii}\equiv \eps_m$, and we have made the hypothesis that there are no off-diagonal on-site hopping terms (local hybridizations), which is always the case in the Wannier function basis for a proper unit cell respecting the symmetries of the lattice. 

The Slave-spin formalism is easily extended to this multi-orbital case.
Indeed already in the one-band case every fermionic variable is supplemented with a corresponding auxiliary spin-1/2 variable, so that these spin variables were labeled, as the fermionic ones by site and spin indices. 

Now the fermions have also an orbital index $d_{im\s}$, and so will the auxiliary spins $S^z_{im\s}$ (and obviously the corresponding fermions of the enlarged Fock space $f_{im\s}$). The physical states are always individuated by the same constraint eq. (\ref{eq:constraint}), which is diagonal in all the indices so it is easily generalized:

\be
\langle n^f_{im\sigma}\rangle_{f}=\langle S^{z}_{im\sigma}\rangle_{s}+\frac{1}{2},
\label{eq:constraint_multiorb}
\ee
and is enforced at the mean-field level through orbital-dependent Lagrange multipliers $\lambda_m$.

The interaction has now further terms.
As already mentioned we will most often drop the spin-flip and pair-hopping terms, which necessitate by the way an approximate treatment at the present stage of development of the slave-spin formalism\cite{demedici_Slave-spins}\footnote{\label{foot:SF-PH}Where they have been used\cite{demedici_Slave-spins, demedici_3bandOSMT,demedici_Genesis,demedici_MottHund} the spin-flip and pair-hopping terms have been approximated using only slave-spin operators as $-J (S^+_{1\up}S^-_{1\down}S^+_{2\down}S^-_{2\up}+ S^+_{1\up}S^+_{1\down}S^-_{2\up}S^-_{2\down}+h.c.)$, which, even if they do not flip the fermionic part as they should, have the virtue of reproducing the atomic spectrum of the Kanamori hamiltonian. The results are satisfactory in the two-band case, whereas major discrepancies are found for $N\geq 3$ for the non-half-non-singly filled cases (the "Janus" cases detailed in sec. \ref{sec:Janus}).}. We will then use the interaction in the "Ising" form eq. (\ref{eq:Kanamori_Ising_tilde}), which can easily be expressed with the corresponding z-component of the slave-spins, as in the one-band case. 
That is eq. (\ref{eq:Kanamori_Ising_tilde}) is expressed as:
\be
\hat {\tilde {H}}_{int}[S]\,=\,U\sum_m S^z_{m\up} S^z_{m\down}\,+\,U^\prime\sum_{m\neq\mp} S^z_{m\up}\tilde S^z_{m'\down}\,
+(U^\prime-J) \sum_{m<\mp,\sigma}  S^z_{m\s} S^z_{m'\s}.
\label{eq:Kanamori_Ising_tilde_SS}
\ee

The hopping term remains of the form eq. (\ref{eq:SS_ham_oneband}) but now all terms carry orbital indices in which they can be off-diagonal.
The mean-field decoupling is strictly analogous, so that the final mean-field equations are:

\begin{align}
H_{f}=&\sum_{i\neq j, mm'\sigma}t^{mm'}_{ij}\sqrt{Z_mZ_{m'}}f^{\dag}_{im\sigma}f_{jm'\sigma} +\sum_{im\s}(\eps_m-\lambda_m-\mu)n^f_{im\s} \label{eq:Hf_multiorb},   \\
H_{s}=&\sum_{m,\sigma}\left[(h_mO^\dagger_{m\sigma} + H.c.) +\lambda_m(S^z_{m\s}+\frac{1}{2})\right]+\hat {\tilde {H}}_{int}[S], \label{eq:H_s_MF-multiOrb}\\
h_{m\sigma}=&\sum_{m'}\langle O_{m'\sigma}\rangle_s\sum_{j(\neq i)} t^{mm'}_{ij}\langle f^{\dag}_{im\sigma}f_{jm'\sigma}\rangle_f\\
Z_m=&\vert\langle O_{m\s}\rangle\vert^2
\end{align}
together with eq. (\ref{eq:constraint_multiorb}).

The gauge number $c_{im\s}$ of eq. (\ref{eq:gauge}) and treated for the single-band case in the complement  \ref{subsec:c_single-site}, is now orbital dependent, keeping the same dependence on the orbital density (provided, as already assumed here, the absence of local hybridizations), i.e.:
\be
c_{im\s}=c_{m}=(1/\sqrt{\langle n^f_{im\s}\rangle(1-\langle n^f_{im\s}\rangle)}-1)
\ee 
This choice, as again in the single-band case, leads to the correct values $Z_m=1$ in the noninteracting limit, and it is thus what we adopt.

One slight problem in this formulation is that it yields nonzero Lagrange multipliers in the noninteracting limit, when a crystal-field splitting is present, which is an annoying feature. However it is not fundamentally problematic.
We then compensate this by shifting the local energies $\eps_m$ to $\tilde \eps_m$ so that $\tilde \eps_m-\l_m(\{U,J\}=0,\{\tilde \eps_m\})=\eps_m$ in order to obtain the correct non-interacting population distribution among the orbitals at zero interaction strength.

Since we have used the particle-hole symmetric form of the interaction, here the chemical potential that leads to half-filling for a particle-hole symmetric DOS is $\mu=0$.


\subsection{The N-orbital Hubbard model}\label{sec:J0}

For the particular choice $J=0$ and for identical diagonal (in the orbital index) hopping integrals, the multi-band Hubbard model acquires the form:
\be\label{eq:Hubbard_Norb_deg}
\hat H = \sum_{ijm\s} t^m_{ij} d^\dag_{im\s}d_{jm\s}+\frac{U}{2}\sum_{i} \left(\sum_{m\s} (n^d_{im\s}-\frac{1}{2})\right)^2,
\ee
which is $SU(2N)$ symmetrical for rotations in the local spin-orbital space. Indeed the interaction term is nothing else than $U\sum_i (Q_i-N/2)^2$, where $Q_i$ is the total charge on site $i$, which is independent of the local basis chosen.

The Mott transition can intuitively happen for any integer number of electrons per atomic site, since what is needed is the blocking of the motion of the electrons due to their mutual repulsion, which can only work when there are no "spare" sites to move to.

Indeed this intuitive result is found in multi-orbital Hubbard models through several techniques. In Fig. \ref{fig:Lu_Gutz} the result from Ref. \cite{Lu_gutz_multiorb} in which the Gutzwiller approximation is used.
This approximation coincides with the Kotliar-Ruckenstein slave-bosons\cite{kotliar_ruckenstein}, which in many instances gives identical results with the slave-spin mean field. The results in Fig. \ref{fig:Lu_Gutz} can be considered representative of SSMF.

\begin{figure}
\begin{center}
\includegraphics[width=9cm, height=8cm]{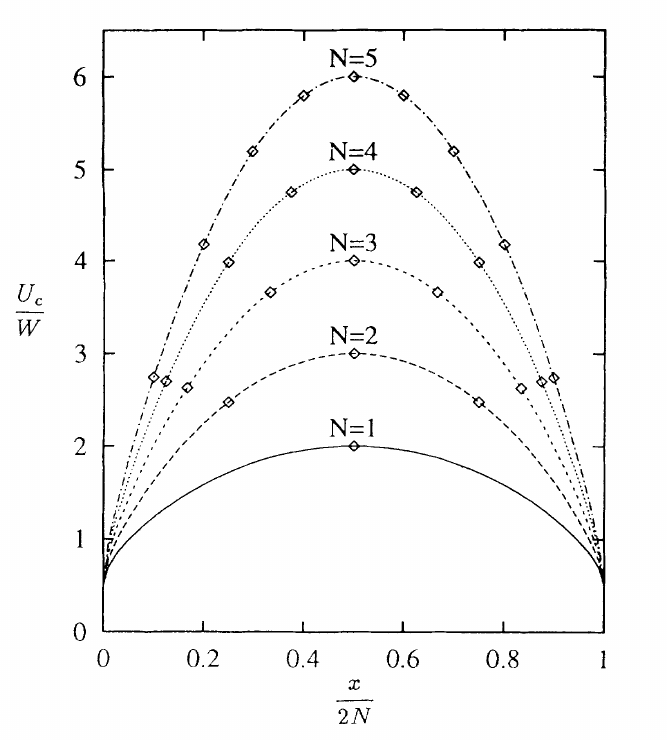}
\caption{Critical interaction strengths $U_c$ for the Mott transition in the SU(2N) symmetrical (i.e. with zero Hund's coupling) Hubbard model of bandwidth W, as a function of the filling (here x denotes the density per spin). Mott transitions (diamonds) happen for all the integer fillings allowed, and the $U_c$ grows in general with the number of orbitals and the proximity to half-filling, in this $J=0$ case. From Ref. \cite{Lu_gutz_multiorb}.}\label{fig:Lu_Gutz}
\end{center}
\end{figure}

The critical interaction strength for the Mott transition is in general a growing function of the atomic ground state degeneracy (as shown for example in complements \ref{comp:perturbative_Uc} and \ref{comp:perturbative_Uc_N2}), which implies, in presence of the simple atomic hamiltonian (\ref{eq:Hubbard_Norb_deg}) with $J=0$, a larger $U_c$ for a higher number of orbitals, and for fillings closer to half, as is clear from the figure.

Indeed these results are confirmed within other approximation schemes, like DMFT\cite{rozenberg_multiorb,Werner_3band}.

The exact large-N behaviour expected for the critical interaction $U_c$ at half-filling was calculated in Ref. \cite{florens_multiorb} and reads:
\be
\frac{U_c}{N}=8\vert\bar \eps\vert,\qquad (\mbox{at large N})
\ee

The same perturbative technique used in the complement \ref{comp:perturbative_Uc} can be used to determine the result from the Slave-spin technique (see complement \ref{comp:perturbative_Uc_N2}), that reads:
\be
U_c=8(N+1)\vert\bar \eps\vert
\ee
and is obviously confirmed by the numerical results reported in figure \ref{fig:Uc_vs_N}.

This shows then that the Slave-spin mean-field captures exactly the large-N critical interaction strength for the multi-orbital Hubbard model.

In Fig. \ref{fig:Uc_vs_N} the numerical calculation within SSMF of the quasiparticle weight as a function of the interaction strength U/D in this model is compared for different number of orbitals $N=1,2,3,4$.

\begin{figure}[h]
\includegraphics[width=12cm]{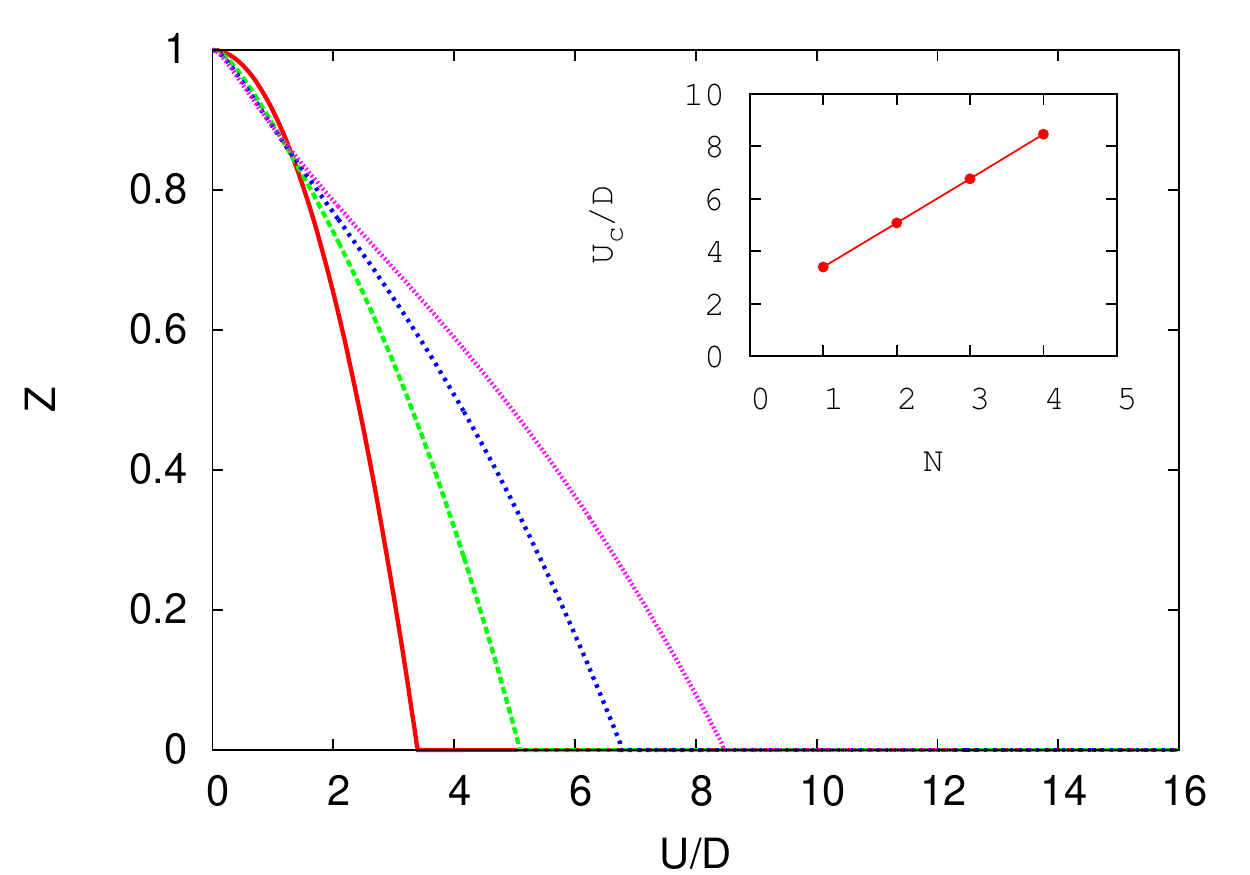}
\caption{\label{fig:Uc_vs_N} Quasiparticle weight, obtained from
slave-spin mean-field for the N-orbital Hubbard model at half-filling (with, from left to right:
$N=1,2,3,4$). The non-interacting density of states is a semi-circle with half bandwidth $D$
Inset: Dependence of the critical $U$ on $N$. The exact large- N behaviour\cite{florens_multiorb} is obtained (see text). From Ref. \cite{demedici_Slave-spins}.}
\end{figure}

\subsection{\emph{Complement}: critical U for the Mott transition in the 2-orbital Hubbard model}\label{comp:perturbative_Uc_N2}

Here we show how to obtain analytically within the slave-spin mean-field approximation the critical interaction strength $U_c$ for the Mott transition in the half-filled (particle-hole symmetric) 2-band Hubbard model in absence of Hund's coupling ($J=0$). The same technique can yield analogously the $U_c$ for any number N of orbitals, and reproduce exactly the numerical results shown in Fig. \ref{fig:Uc_vs_N}. The case for finite J is developed in complement \ref{comp:perturbative_Uc_N2_J}.

As done in the complement \ref{comp:perturbative_Uc}, the multi-orbital slave-spin equation (\ref{eq:H_s_MF-multiOrb}) can be developed around the insulating solution $h_{im\s}=0$ where $H_s=H_{at}$ with, for this multi-orbital case with $J=0$:
\be
H_{at}=\frac{U}{2}\left(\sum_{m\s} S^z_{m\s}\right)^2, 
\ee
of which we know the spectrum. This is: 

\be
\left\{
\begin{array}{cc}
|\up\up,\up\up\rangle |\down\down,\down\down\rangle & E=2U\\
&\\
\begin{array}{c}
 |\up\up,\up\down\rangle |\up\up,\down\up\rangle |\down\down,\up\down\rangle |\down\down,\down\up\rangle \\ 
|\up\down,\up\up\rangle |\down\up,\up\up\rangle |\up\down,\down\down\rangle |\down\up,\down\down\rangle 
\end{array}
& E=\frac{U}{2}\\
&\\
\begin{array}{c}
|\up\down,\up\down\rangle |\up\down,\down\up\rangle |\down\up,\up\down\rangle |\down\up,\down\up\rangle \\
|\up\up,\down\down\rangle |\down\down,\up\up\rangle
\end{array}
& E=0\\
\end{array}
\right.
\ee

We again calculate as a perturbation, the effect of the "kinetic" part of the slave-spin hamiltonian:
\be
H_{pert}=h\sum_{m\s} 2 S^x_{m\s}.
\ee
where $ O_{im\sigma} = 2 S^x_{m\s}$ in (\ref{eq:H_s_MF-multiOrb}) because both orbitals are half-filled and thus the gauge $c=1$ (eq. (\ref{eq:gauge}) for $n_m=1/2$), and we call $h\equiv h^*_{im\s}+h_{im\s}$. Also $\lambda_{m}=0$ because of particle-hole symmetry.

As for the single-band case we perform a perturbative expansion in $h$, which is small near the metal-insulator transition, if this is second-order.

In this $J=0$ case the ground state has degeneracy 6  (the subset above with $E=0$).  
Furthermore, as before, the perturbation $H_{pert}$ does not have any nonzero elements in the low-energy subspace one has to use second-order perturbation theory to find the right combination of states, in the ground state manifold, to which the perturbed state tends for $h\rightarrow 0$.
This is the lowest-lying eigenstate of the matrix $H'\equiv H_{pert}(E_0-H_{at})^{-1}H_{pert}$  in the degenerate subspace. In this case, since $H_{pert}$ couples the $E=0$ sector only with the $E=\frac{U}{2}$ sector, one can rewrite $H'=-\frac{2}{U}H_{pert}^2$ (where we have used the fact that $E_0-H_{at}$ is diagonal and $E_0=0$).

The complete 6x6 matrix for the restriction of $H_{pert}$ to the $E=0$ subspace to be diagonalized is reported in appendix \ref{comp:perturbative_Uc_t1t2_J0} (eq. (\ref{eq:Hpert_matrix}) with $h_1=h_2$). 
However by inspection one can easily convince himself that 
\be
|\phi_0\rangle\equiv \frac{1}{\sqrt{6}}(|\up\down,\up\down\rangle + |\up\down,\down\up\rangle + |\down\up,\up\down\rangle + |\down\up,\down\up\rangle + |\up\up,\down\down\rangle + |\down\down,\up\up\rangle) 
\ee
is the ground state to leading order in h. Indeed $H_{pert}^2$ has the effect of flipping any two spins (including twice the same), so that any state of the $E=0$ manifold is changed, when acted upon by $V^2$, into itself plus four of the others states of the manifold (only one is left out, since it needs 4 flips to be reached).
Then it is clear that the ground state is the symmetric sum of all the 6 states since it is obviously changed into itself by $H_{pert}^2$ and the plus signs in the sum add up all the $-\frac{2}{U}$ contributions, obtaining the lowest possible eigenvalue.

Once found the correct zeroth-order ground state $|\phi_0\rangle$ we need to find its ket correction to leading order in h.
 
The correction to the ket is nonzero at first order in $h$ and reads:
 \be
 |\phi_0^{(I)}\rangle=| \phi_0\rangle +\sum_{|s\rangle\neq |\phi_0\rangle}\frac{\langle s|H_{pert}|\phi_0\rangle}{E^0-E_s^0} |s\rangle=| \phi_0\rangle-\frac{6h}{\sqrt{6}U}\sum_{|s\rangle \subset \{E=\frac{U}{2}\}} |s\rangle,
 \ee
 because $\langle s|H_{pert}|\phi_0\rangle=\frac{3h}{\sqrt{6}}$ for any $|s\rangle \subset \{E=\frac{U}{2}\}$.
 
 Analogously to the 1-band case to calculate the critical interaction one uses the self-consistency condition $h=2 \bar \eps \langle 2 S_{m\s}^x\rangle$. 
 This condition has to be rigorously satisfied and since we can only calculate here a perturbative development for $\langle 2 S_{m\s}^x\rangle$, 
 which by definition is only exactly valid at the transition, (where $h\rightarrow 0$), the self-consistency condition becomes an equation determining the coupling U for which this condition can be satisfied rigorously, hence the critical coupling $U_c$.
 
Applying  $2 S_{m\s}^x$ to the ground state:  
\be
2 S_{1\up}^x|\phi_0^{(I)}\rangle=2 S_{1\up}^x(\frac{1}{\sqrt{6}}\sum_{{|s\rangle \subset \{E=0\}}} |s\rangle -\frac{\sqrt{6}h}{U}\sum_{{|s'\rangle \subset \{E=\frac{U}{2}\}}} |s'\rangle)
\ee
one obtains 6 of the 8 states of the $E=\frac{U}{2}$ manifold from the acting of $2 S_{1\up}^x$ on the $|s\rangle$ states and the 6 states of the $E=0$ manifold plus the two from the $E=2U$ one, from the acting on of $2 S_{1\up}^x$ on the $|s'\rangle$ states, which implies that
\be
\langle \phi_0^{(I)} |2 S_{1\up}^x|\phi_0^{(I)}\rangle = -\frac{12h}{U}.
 \ee
 
The self-consistency equation becomes then: 
 \be
 h=2 \bar \eps \langle 2 S_{m\s}^x\rangle=-\frac{24h}{U_c}\bar \eps \quad \Longrightarrow \quad U_c=-24\bar \eps.
 \ee
 For a semi-circular DOS of half-bandwidth D ($\mu=\lambda_{m}=0$ because of particle-hole symmetry),  $\bar \eps\simeq -0.2122 D$, which gives
 \be
 U_c\simeq 5.09
 \ee
 if $D=1$, in agreement with the numerical result fig. \ref{fig:Uc_vs_N}.

\subsection{Effect of Hund's coupling on the Mott transition}\label{sec:Hund_Mott}

We will now see what is the effect of a finite Hund's coupling $J\neq 0$ on the critical interaction strength needed to obtain a Mott insulator, and in general the influence that this has on the degree of correlation of the neighboring metallic phases.

Figure \ref{fig:2band_Hund-Kanamori-Ising} shows the quasiparticle weight as a function of the interaction strength, for various fixed values of $J/U$, in the 2-orbital Hubbard model at half-filling.
\begin{figure}[h!]
\begin{center}
\includegraphics[width=12cm]{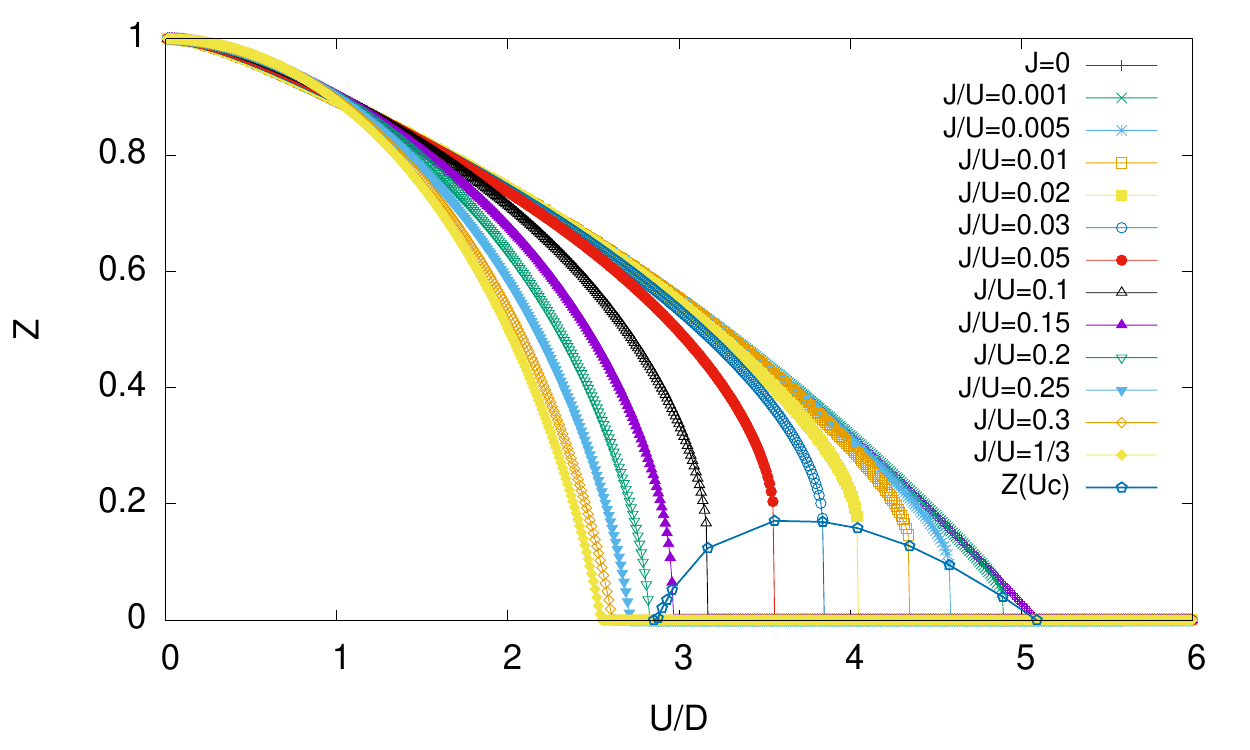}
\includegraphics[width=12cm]{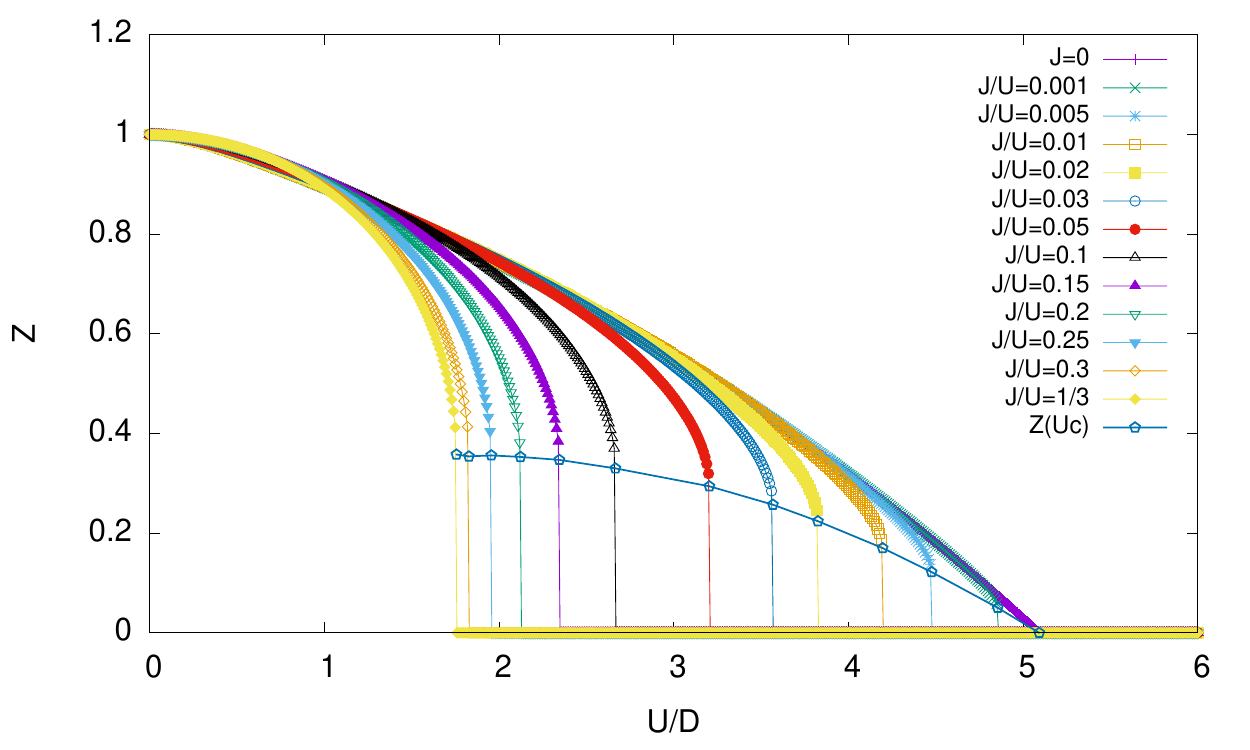}
\caption{Quasiparticle weight (also the inverse mass enhancement, in local schemes like SSMF) calculated for the half-filled 2-orbital Hubbard model as a function of the interaction strength U/D for various fixed values $J/U$. In this half-filled case one sees that the critical interaction strength for the Mott transition is reduced by an increasing Hund's coupling. Upper panel: Kanamori form of Hund's coupling. Lower panel: Ising form.}
\label{fig:2band_Hund-Kanamori-Ising}
\end{center}
\end{figure}
The two forms Hund's coupling are explored here: Kanamori (approximately treated, as indicated in footnote \ref{foot:SF-PH} on page \pageref{foot:SF-PH}) and Ising. In both cases Hund's coupling is seen to reduce the $U_c$ for the Mott transition. Besides a stronger effect from the Ising form (and consequently a more correlated metal at the same value of $J/U$ compared to the Kanamori form), the main difference that is noticed is on the order of the transition. 
Indeed the transition is second order at J=0. The onset of J in any form makes it immediately first order. However in the Kanamori form the second order of the transition is restored above $J/U\simeq 0.18$ (as signaled by the vanishing jump in Z at the transition). For the Ising form instead the first order character persists until the largest physical value $J/U=1/3$ and is always stronger for larger $J/U$.
Despite the approximate treatment of the Kanamori Hamiltonian this non-monotonous behaviour of the first-order jump is qualitatively well captured by SSMF, as can be shown by comparing to other techniques like rotationally-invariant slave bosons (RISB)\cite{Lechermann_RISB} or DMFT\cite{Ono_multiorb_linearizedDMFT}.

\begin{figure}
\begin{center}
\includegraphics[width=12cm]{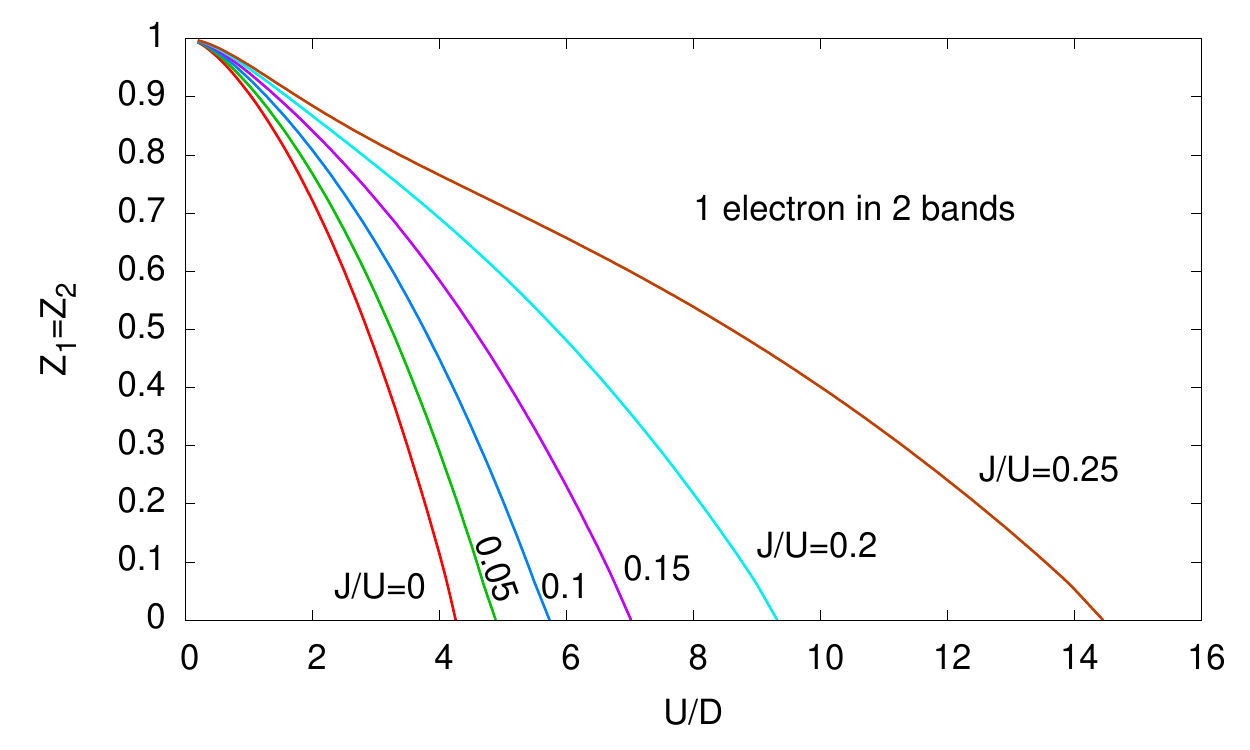}
\caption{Quasiparticle weight calculated in SSMF for a quarter-filled 2-orbital Hubbard model as a function of the interaction strength for various Hund's coupling strengths. From \cite{demedici_MottHund}.}
\label{fig:1el2bands}
\end{center}
\end{figure}

An opposite effect of Hund's coupling is found instead in the quarter-filled 2-orbital Hubbard model. Indeed as illustrated in Fig. \ref{fig:1el2bands}, a finite Hund's coupling decorrelates the system in this case, and pushes the $U_c$ for the Mott transition at very high interaction strengths. 

The main cause of this opposite behaviour in the half-filled and quarter-filled cases in the 2-band model is understood from the energetics of the atomic Mott gap.

\subsection{The physics of the Mott gap}\label{sec:MottGap}

Indeed the excitation spectrum of an idealized paramagnetic Mott insulator (such as the one obtained at large U in the Hubbard hamiltonian (\ref{eq:Hubbard_Norb_deg})) is easily put in correspondence with the atomic one. The so-called upper and lower Hubbard bands are the two main spectral features due to the addition and removal of a particle respectively, and correspond to the atomic transitions that one would obtain in the "atomic limit", i.e. putting all hoppings to zero in (\ref{eq:Hubbard_Norb_deg}). In such an isolated Hubbard atom the eigenstates are the local states with definite charge $n$ and the distance between the ground state and the excited states with charge $n \pm 1$ is $U/2$, in absence of Hund's coupling.
In the lattice model these excited states are as many as the lattice sites (only one site has a $\pm 1$ charge) times the local degeneracy due to orbital and spin indices, and are obviously degenerate. Reintroducing the hopping (for example as a perturbation) lifts this degeneracy and disperses the eigenstates on an energy range of the order of the bandwidth, thus forming the Hubbard bands.

From this point of view an intuitive criterion, due to J. Hubbard himself\cite{Hubbard_I}, for the breakdown of the Mott insulating state can easily be obtained.
Indeed the distance between the Hubbard bands is set by U (in the particle-hole symmetric case they are centered at energy $\pm U/2$) while their dispersion is set by the bandwidth W (and is $\simeq W$ in the single-band case, while it can be different, but always of order W, in multi-band cases, as it will be shown in the next section), so that at large $U/W$ an energy gap exists in both the addition and subtraction spectrum. This reflects the energy cost of moving a particle in the Mott insulator: it amounts to the energy of creating, on the ground state with all sites with $n$ particles, a site with an extra particle, thus leaving behind a site with an extra hole, minus the energy that can be gained with the delocalization of these particles. When this cost is zero, the Mott insulator cannot exist, and a metal is obtained (in absence of other symmetry breakings) which by definition is the state with delocalized excitations at vanishing energy.

Quantitatively this means that the Mott gap is the atomic excitation gap minus the bandwidth, i.e. (in the single-band case) $\D_{Mott}=\D_{at}-W$, and since $\D_{at}\equiv E_{at}(n+1)+E_{at}(n-1)-2E_{at}(n)=U$ in absence of Hund's coupling the Hubbard criterion for its closure reads:
\be\label{eq:Hubbard_crit}
\D_{Mott}=U-W=0 \Longrightarrow U_c=W
\ee
This reflect the natural physical expectation that a Mott insulator is obtained when the Coulomb repulsion between the electrons overcomes the delocalization energy due to the hopping.

Now, the atomic excitation spectrum is modified by a finite Hund's coupling J and so is the atomic Mott gap, thus implying that the Hubbard criterion yields $U_c=U_c(J)$.
Indeed when $H_{int}$ takes the form eq. (\ref{eq:ham_kanamori}) it is easily shown\cite{demedici_MottHund} that the atomic Mott gap becomes 
\be
\D^{at}_n=\begin{cases}
    U-3J, & \; \forall n\neq N \quad \text{(off half-filling)},\\
    U+(N-1)J,  & \; n=N  \quad \text{(half-filling)}.
  \end{cases}\label{eq:D_at_J}
 \ee
 
The atomic Mott gap is thus increased by J in the half-filled case, and reduced in all other cases. This implies that J favors the Mott insulator at half-filling, thus reducing the corresponding $U_c$, and disfavors it for all other fillings, moving thus the $U_c$ to very high values. Indeed in Fig. \ref{fig:Uc_vs_J_all_asynt} the numerical results from SSMF are reported for all integer fillings in both the 2-orbital and the 3-orbital Hubbard models.
It is clear that at large J these results are well captured by the simple Hubbard criterion with $\D^{at}$ from formulas \ref{eq:D_at_J}.

\begin{figure}
\begin{center}
\includegraphics[width=12cm]{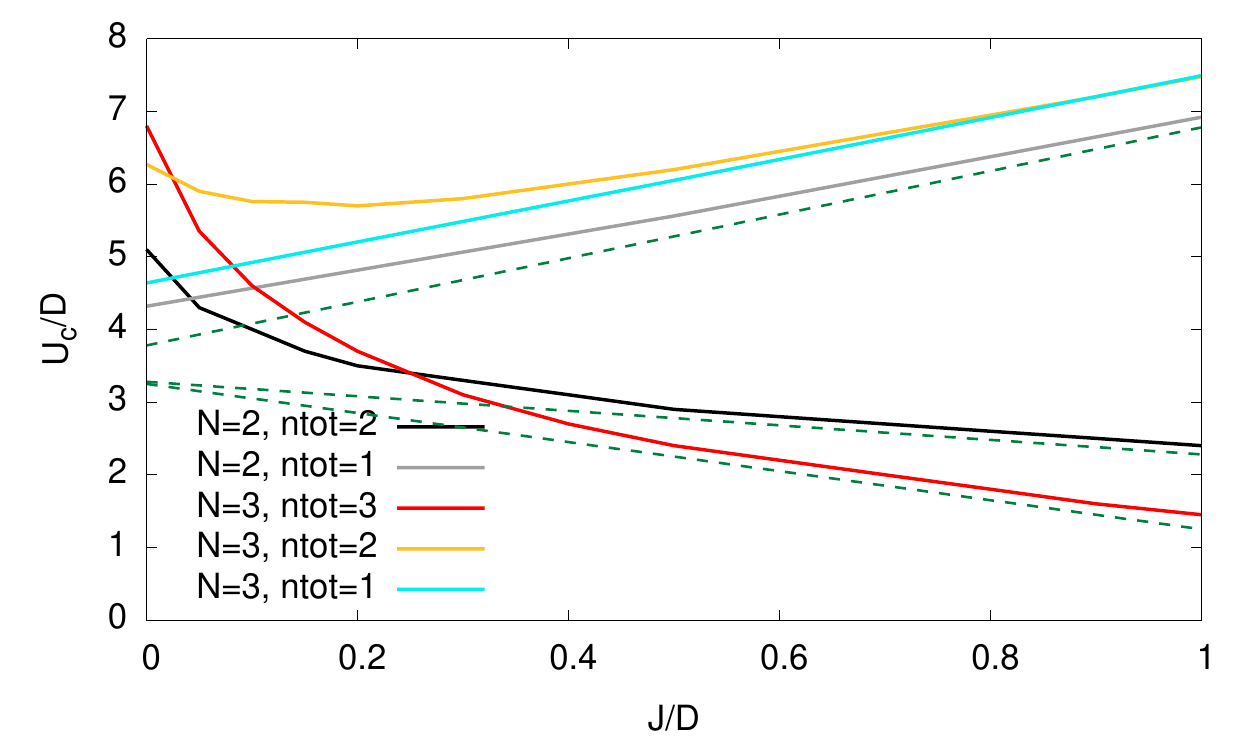}
\caption{Critical interaction strength for the Mott transition for all integer fillings (fillings for $n>N$ yield identical results because of the  particle-hole symmetry) in the 2-orbital and 3-orbital Hubbard model, as a function of Hund's coupling J. The dashed lines indicate the large-J asyntotics based on the dependence on J of the atomic gap, formulas (\ref{eq:D_at_J}) (from top to bottom $\d U_c\propto 3J, -J, -2J$). From \cite{demedici_MottHund}.}
\label{fig:Uc_vs_J_all_asynt}
\end{center}
\end{figure}

\subsection{Hund's effect on orbital fluctuations and the narrowing of the Hubbard bands}\label{sec:Janus}

The data plotted in Fig. \ref{fig:Uc_vs_J_all_asynt} show that at small J another effect of Hund's coupling is visible, that is a fast reduction of $U_c$ from the high values reached at $J=0$ in all cases but those in which the system is filled by one electron or one hole per site only. This reduction clearly adds up with the large-J behaviour, thus causing a superlinear reduction of $U_c$ in the half-filled cases and a non-monotonous $U_c(J)$ in the generic non-half-non-singly-filled cases (which are present in models from N=3 orbitals up).

In Fig.\ref{fig:Fanfarillo_comp-SSMF_DMFT} the calculation (performed in DMFT\cite{demedici_Janus} and SSMF\cite{Fanfarillo_Hund}) of the quasiparticle weight in the 3-orbital Hubbard model is reported, showing in detail the aforementioned two effects clearly at play. Indeed the central panel shows the non-half-non-singly-filled case which was dubbed "Janus-faced", because there the two effects of Hund's coupling play an opposite role\footnote{Janus, the ancient roman god of beginnings, transitions and endings and of time, gates, doors etc., was depicted with two faces looking in opposite directions.}: the shifting of $U_c$ towards high values prevails at large J, but an overall reduction of the quasiparticle weight and thus an increase in the correlation of the system with J happens for a large range of U and even causes the $U_c$ to decrease initially.
\begin{figure}[h!]
\begin{center}
\includegraphics[width=15cm]{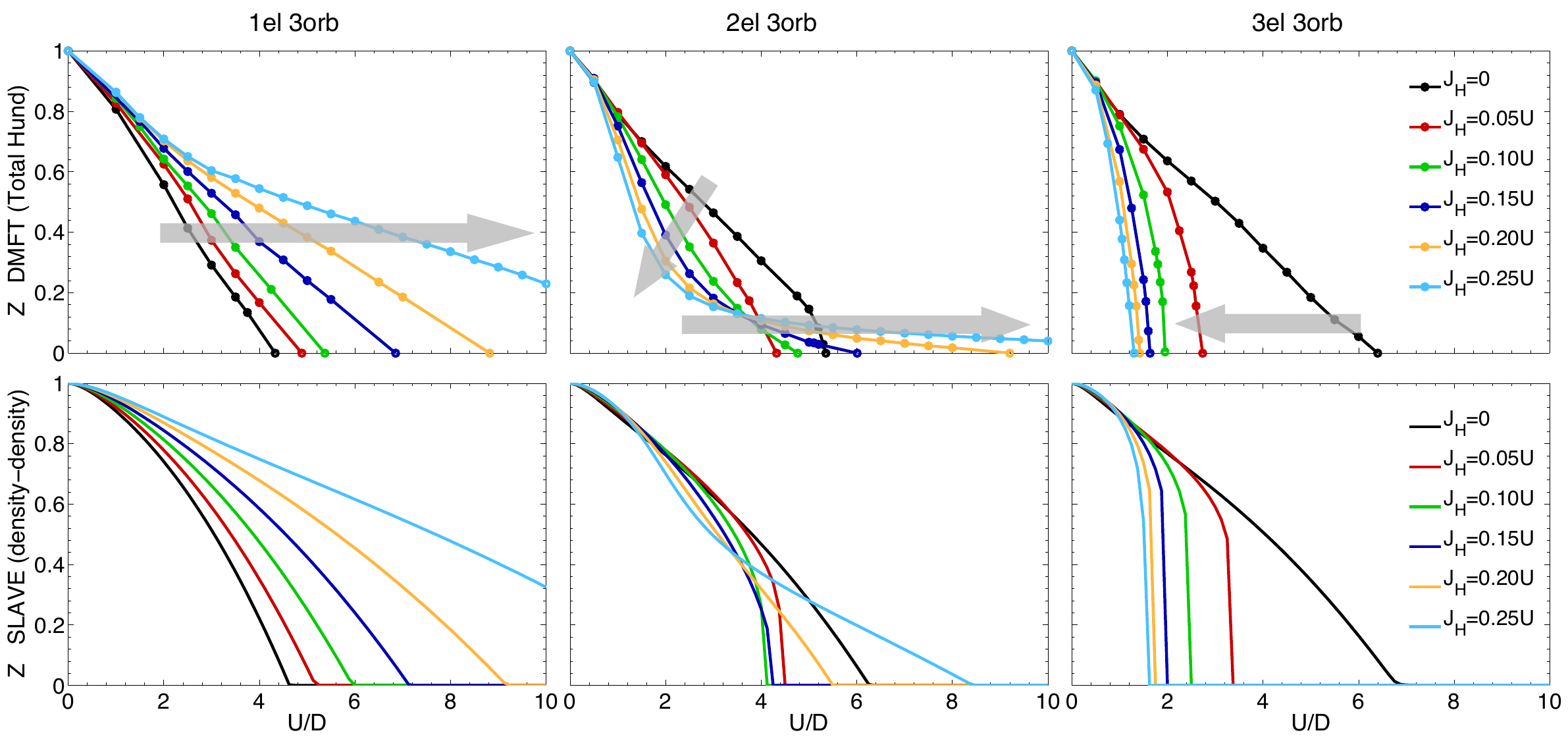}
\caption{Comparison of the quasiparticle weight and inverse mass enhancement Z calculated in DMFT (Kanamori Hamiltonian, top panels) and SSMF (Ising hamiltonian, bottom panels) in the 3-band Hubbard model, for various values of $J/U$. The grey arrows indicate schematically the effect of a growing Hund's coupling $J/U$. Adapted from Ref. \cite{Fanfarillo_Hund} (DMFT data from \cite{demedici_Janus}).}
\label{fig:Fanfarillo_comp-SSMF_DMFT}
\end{center}
\end{figure}

Intuitively this enhancement of correlation at small J can be viewed as the opposite of the effect reducing the correlation with the number of orbitals N at $J=0$, and ultimately leading to an increase in $U^N_c(J=0)$ with N. This effect was treated in section \ref{sec:J0} and stems from the systematic increase of the ground-state degeneracy and of the hopping channels with N, as also illustrated explicitly by the comparison of complements \ref{comp:perturbative_Uc} and \ref{comp:perturbative_Uc_N2}.
 A finite J, however, reduces the ground-state degeneracy again, lowering it to that of the high-spin multiplet (for a given fixed filling of n electrons in N orbitals). This in turn reduces also the effective hopping channels. As detailed in complement \ref{comp:perturbative_Uc_N2_J} for the 2-orbital Hubbard model, the ground state degeneracy is brought from 6 states for $J=0$ to 3 states at finite J (for the Kanamori form of the interaction Hamiltonian), and the perturbation theory leads to $U^{N=2}_c(J)=U^{N=1}_c-J$. This result is valid only at large J, but it captures the Mott gap linear scaling with J, and interestingly also shows that the intercept of this large-J behaviour at J=0 is  the critical interaction for the single-orbital case $U^{N=1}_c$. 
 
This fact can be interpreted in the simplified terms of the Hubbard criterion. Indeed the enhanced $U^N_c$ for growing N can be understood as a result of larger Hubbard bands, so that the criterion becomes:
\be\label{eq:Hubbard_crit_multiorb}
\D_{Mott}=\D_{at}-\tilde W
\ee
where $\tilde W$ is the width of the Hubbard bands, that was indeed found to increase with the number of orbitals in Ref. \cite{gunnarsson_fullerenes,gunnarsson_multiorb} as $\tilde W\simeq W\sqrt{N}$, due to the orbital quantum fluctuations in the ground state. The criterion for the closure of the gap $ \D_{Mott}=0$ then implies ($\D_{at}=U$ also in the multi-orbital case) that $U^N_c=\tilde W \simeq \sqrt{N} U^{N=1}_c$. 

This result is qualitatively in agreement with the increase of $U^N_c\propto N$ treated in section \ref{sec:J0}. 
The quantitative disagreement ($N$ vs $\sqrt{N}$) is explained by an increasingly larger coexistence zone between an insulator and a metal with a preformed gap. 
Indeed two spinodal lines (where the insulating and the metallic solutions disappear, respectively) are found corresponding to the Mott transition in DMFT. They give rise to a zone of coexistence of solutions, in which the actual transition is located exactly by comparing the energy of the two solutions\cite{georges_RMP_dmft}. This zone is seen to widen with N at $J=0$\cite{florens_multiorb}. 
However there are indications that this coexistence zone strongly narrows when J is nonzero\cite{Inaba_Koga_SecondOrder}. We will thus ignore to distinguish precisely between the two lines, even if conceptually we are making a leap, since the SSMF traces the disappearance of the metallic solution while the Hubbard criterion describes the closure of the insulating gap. Quantitatively however, at finite J, this discrepancy is a small one.

Thus by making use, for the finite-J case, of the second of eqs. (\ref{eq:D_at_J}) (relevant for this half-filled case at $N=2$) $\D^{at}=U+J$, in (\ref{eq:Hubbard_crit_multiorb}) one finds, for the closure condition of the Mott gap,  $U_c=\tilde W -J$. This  compared with  large-J behaviour  $U^{N=2}_c(J)=U^{N=1}_c-J$ derived for the 2-orbital Hubbard model in complement \ref{comp:perturbative_Uc_N2_J}, signals then that the width of the Hubbard bands is reduced by Hund's coupling back to a value of the order of the one-band case $\tilde W=W$.
\footnote{The precise result $\tilde W=W$ should be taken with some care however, since our treatment of the Kanamori hamiltonian is approximate. Indeed when deriving the same perturbative result with the Ising form of Hund's coupling one finds $U^{N=2}_c(J)=U^{N=1}_c/2-J$. The Ising hamiltonian is treated exactly, however the Mott transition is always first order, which invalidates the perturbative expansion. We take nevertheless these results together as indicating the large-J behaviour for $U^N_c$ having a $J=0$ intercept of the order of the single-band $U_c$. See also the discussion at the end of complement \ref{comp:perturbative_Uc_N2_J}.}

This result is confirmed in DMFT. Indeed Fig. \ref{fig:SSMF_DMFT_rescaled} shows the $U_c$ for all the Mott transitions in the 2-band and 3-band degenerate Hubbard models as a function of J, and compares the result obtained by SSMF and DMFT, with the Kanamori Hamiltonian. In the plotted $U_c(J)$ a linear contribution in J is added or subtracted, following eqs. (\ref{eq:D_at_J}), to compensate the effect due to the J-dependence of the Mott gap, and thus highlighting the remaining effect of reduction by J of the orbital fluctuations in the ground state, and of the width of the Hubbard bands.

Indeed the very argument used in  \cite{gunnarsson_fullerenes,gunnarsson_multiorb} to support the enhanced width of the Hubbard bands at J=0, was shown to break down by Koga et al. in Ref. \cite{Koga_OSMT}. The argument is essentially that Hund's coupling suppresses the orbital fluctuations responsible for the enhancement.

\begin{figure}
\begin{center}
\includegraphics[width=12cm]{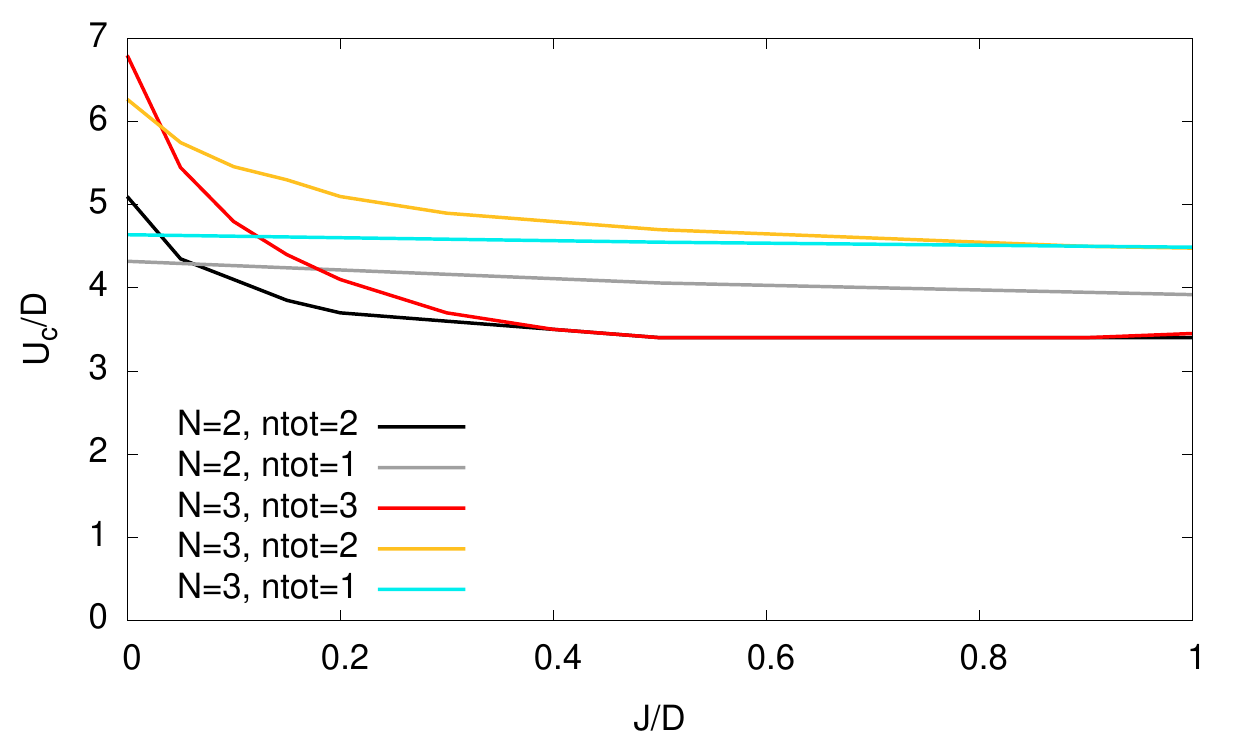}
\includegraphics[width=12cm]{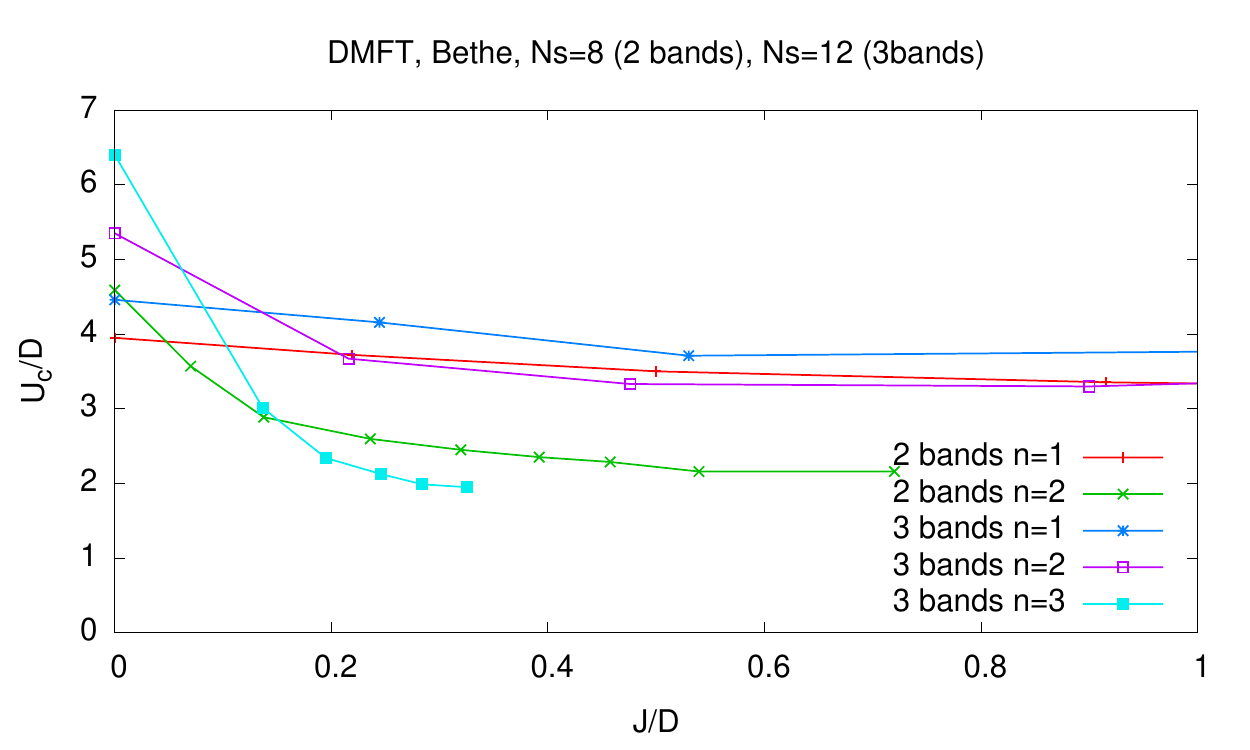}
\caption{Critical interactions for the Mott transition in the 2-band and 3-band Hubbard models compensated for the J-dependence of the Mott gap. Specifically J and 2J have been added to the N=2 and N=3 half-filled cases, respectively, while 3J has been subtracted for all the other cases, following eqs. (\ref{eq:D_at_J}). Upper panel: SSMF. Lower panel: DMFT. In the half-filled cases this rescaled $U_c$, indicative of the width of the Hubbard bands (see text), is of the order of the bandwidth at finite J, while it is strongly enhanced for vanishing J.}
\label{fig:SSMF_DMFT_rescaled}
\end{center}
\end{figure}
 
The width of the Hubbard bands can be also assessed in a direct manner within DMFT, where an accurate local spectral function can be calculated. This result is reported in Fig. \ref{fig:Hubbard_band_narrowing} for a 2-orbital Hubbard model, and confirms the conclusion that we have anticipated. Indeed the width of the Hubbard bands diminishes upon the onset of J, and it goes roughly from a value of order $\sim \sqrt{2} W$ at $J=0$ to a value roughly of order $\sim W$  at large J/U.\footnote{The uncertainty on the numerical result is due to the fact that exact diagonalization of a finite bath in the auxiliary impurity system used in the DMFT scheme is used for the calculation of the local spectral function. The discretization implied by this impurity solver ($N_s=10$ total sites between bath sites and the two-orbital impurity are used) yields a spiky spectral function that needs to be evaluated slightly away from the real axis in order to appear continuous. This broadening preserves the spectral weight distribution however, letting us conclude reliably on the width of the Hubbard bands.}

In agreement with this result another complementary analysis can be done in this framework. Indeed starting from a half-filled Mott insulator at large U, the chemical potential can be shifted in the gap until the edge of the gap itself. When the chemical potential is shifted further one recovers a metal with a finite density of states at the Fermi level. By moving then the chemical potential in the opposite direction there is another value of the chemical potential where this density of states vanishes and the chemical potential is within the gap again. This cycle defines the border of the Hubbard band within an error bar mainly due to the coexistence zone found in DMFT. 
The result of this procedure as a function of J (not shown, due to large error bars caused by the extremely low coherence scale of the metal in this regime that makes the convergence of the algorithm very hard at zero temperature) confirms that the Hubbard bands are roughly narrowed down by J from a value $\sim\sqrt{N}W$ to one of $\sim W$, and that the coexistence zone shrinks sizably.

\begin{figure}	
\begin{center}
\includegraphics[width=12cm]{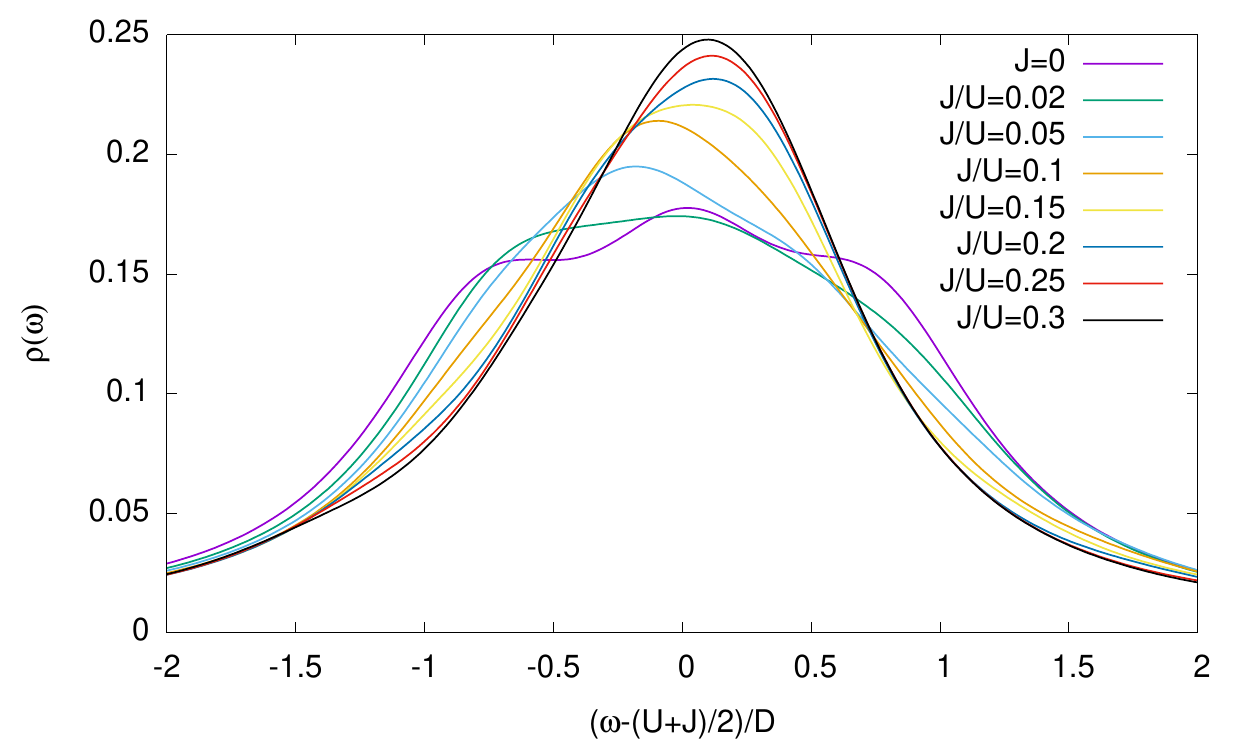}
\caption{Hubbard bands in the local spectral function $\rho(\omega)$ for the half-filled 2-orbital Hubbard model as a function of J (Kanamori Hamiltonian) in DMFT. The interaction is $U=7D$, so that at all $J/U$ the system is in a Mott insulating phase, and a gap is open in the spectrum. As for Fig. \ref{fig:SSMF_DMFT_rescaled}, following eqs. (\ref{eq:D_at_J}), a scaling of $(U+J)/2$ is performed (here on the abscissas), that brings the position of the atomic-limit excitation (roughly the barycenter of the Hubbard bands in the lattice model) to coincide for all $J/U$. Here we show thus the Hubbard band of the addition spectrum(positive frequencies). The result shows directly that the Hubbard bands have a width $\sim W=2D$ at large $J/U$, reduced from the $J=0$ value $\sim \sqrt{2}W$.}
\label{fig:Hubbard_band_narrowing}
\end{center}
\end{figure}
  
\subsection{\emph{Complement}: comparison with DMFT}

In order to assess the accuracy of the SSMF a more detailed comparison is in order with results from the state-of-the-art technique for the analysis of local correlations in such Hubbard-like models, Dynamical Mean-Field Theory.

The upper panel of figure \ref{fig:compDMFT-SSMF} shows such a comparison for the $U_c$ for the Mott transition in the half-filled 2-orbital Hubbard model with both Ising and Kanamori interaction hamiltonians.
It can be notices that SSMF slightly but systematically overestimates the $U_c$ compared to DMFT. DMFT also confirms the stronger correlations induced by the Ising interaction, as highlighted in Fig. \ref{fig:2band_Hund-Kanamori-Ising}.

\begin{figure}[h!]
\begin{center}
\includegraphics[width=11cm]{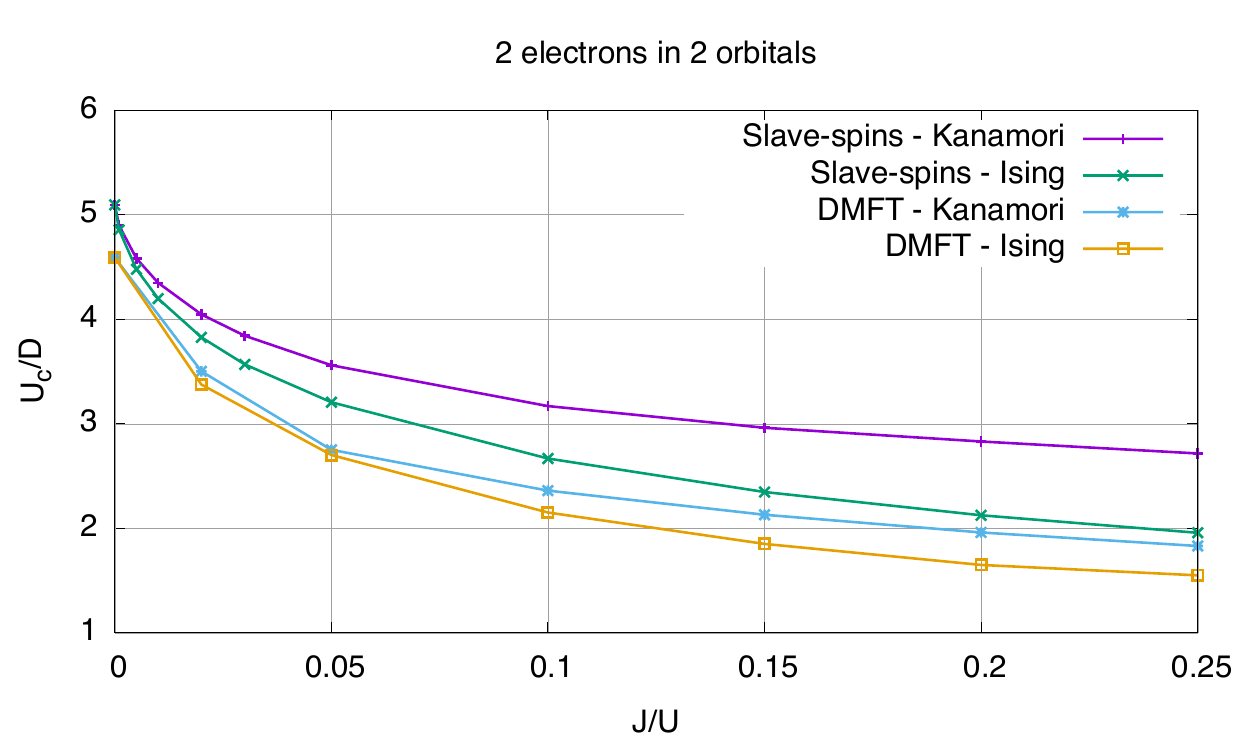}
\includegraphics[width=11cm]{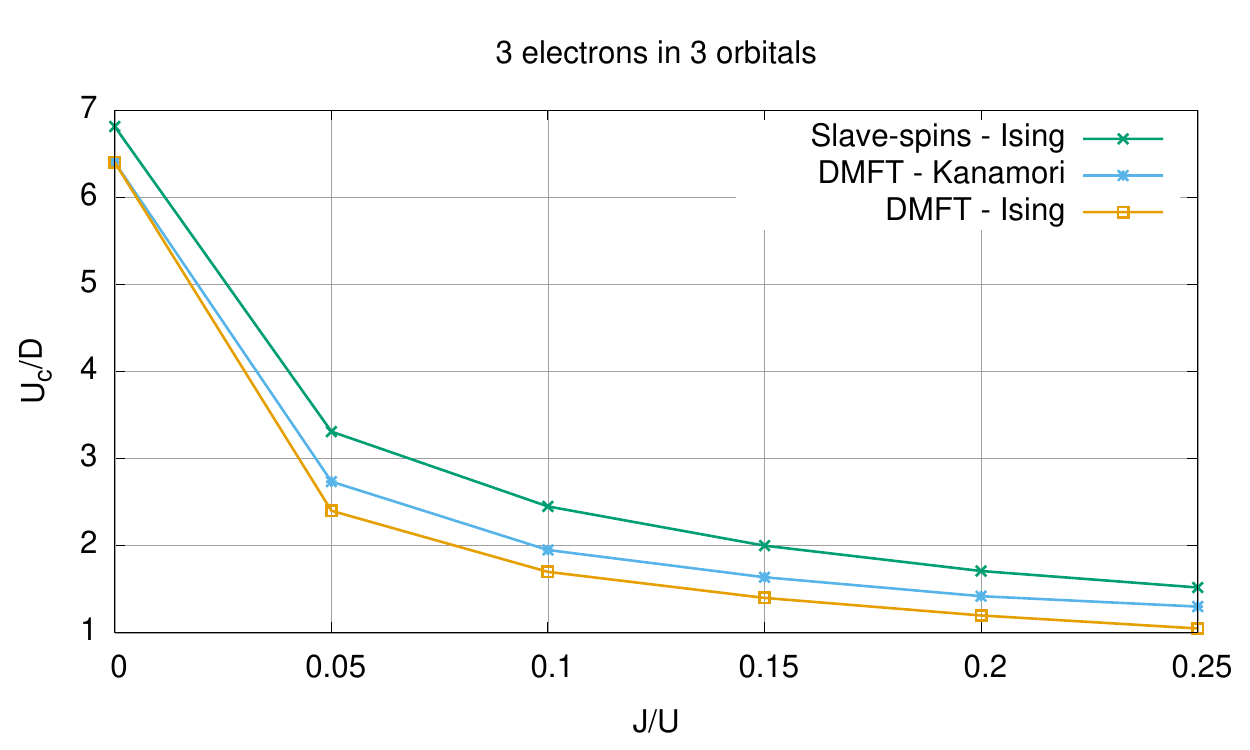}
\caption{Comparison of the SSMF critical interaction strength for the Mott transition in the half-filled 2-band (upper panel) and 3-band (lower panel) Hubbard models with both Ising and Kanamori interactions with the analogous results in DMFT. A similar comparison for the Mott transition in the quarter-filled 2-band Hubabrd model with Kanamori interaction is reported in Ref. \cite{demedici_MottHund} (figure 6, lower panel).}
\label{fig:compDMFT-SSMF}
\end{center}
\end{figure}
It is to be noticed that the SSMF treatment of the Ising hamiltonian (of which the mapping in the auxiliary Hilbert space is exact) gives results reasonably close those for the Kanamori hamiltonian treated in DMFT. Indeed in both the 2-band and the 3-band half-filled models results for typical physical values of $J/U\simeq 0.12\div 0.15$ for the DMFT treatment of the Kanamori hamiltonian are reproduced by the SSMF treatment of the Ising hamiltonian with $J/U\simeq 0.2\div 0.25$.

This estimate is confirmed in more detail in the comparison performed in Ref. \cite{Fanfarillo_Hund} and reported in Fig.\ref{fig:Fanfarillo_comp-SSMF_DMFT}.

\subsection{\emph{Complement}: critical U for the Mott transition in the 2-orbital Hubbard model for large Hund's coupling J, and orbital decoupling}\label{comp:perturbative_Uc_N2_J}

Here we show how to obtain analytically within the slave-spin mean-field approximation the critical interaction strength $U_c$ for the Mott transition in the (particle-hole symmetric) half-filled 2-band Hubbard model in presence of Hund's coupling ($J\neq 0$). As shown by the numerical results (Fig.\ref{fig:2band_Hund-Kanamori-Ising}) the Mott transition in presence of Hund's coupling (in the Kanamori form) is second order only at large J. Thus the present results bear validity only at large J (which is why they do not connect with the analytical results obtained for $J=0$ in Section \ref{comp:perturbative_Uc_N2}.

We also show explicitly that in this large J limit the equations determining $U_c$ for each band decouple, so that when the bandwidth of the two bands differs the $U_c$ for the two band differs too, and a zone where an orbital-selective Mott phase is realized opens.

As done previously we develop the multi-orbital slave-spin equation (\ref{eq:H_s_MF-multiOrb}) around the insulating solution $h_{im\s}=0$ where $H_s=H_{at}$ with, now that $J\neq 0$:
\begin{align}\label{eq:Kanamori_SS}
\hat {\tilde {H}}_{int}[S]\,&=\,U\sum_m S^z_{m\up} S^z_{m\down}\,+\,U^\prime\sum_{m\neq\mp} S^z_{m\up}\tilde S^z_{m'\down}\,
+(U^\prime-J) \sum_{m<\mp,\sigma}  S^z_{m\s} S^z_{m'\s}\nonumber\\
&-J (S^+_{1\up}S^-_{1\down}S^+_{2\down}S^-_{2\up}+ S^+_{1\up}S^+_{1\down}S^-_{2\up}S^-_{2\down}+h.c.)
\end{align}
of which we know the spectrum. This is \footnote{a constant energy shift (U+J)/2 was added everywhere to position the ground state on the zero of energy.}: 

\be
\left\{
\begin{array}{cc}
|\up\up,\up\up\rangle |\down\down,\down\down\rangle & E=2U-2J\\
&\\
\begin{array}{c}
 |\up\up,\up\down\rangle \quad |\up\up,\down\up\rangle \quad |\down\down,\up\down\rangle \quad |\down\down,\down\up\rangle \\ 
|\up\down,\up\up\rangle \quad |\down\up,\up\up\rangle \quad |\up\down,\down\down\rangle \quad |\down\up,\down\down\rangle 
\end{array}
& E=\frac{U+J}{2}\\
&\\
\begin{array}{c}
\frac{1}{\sqrt{2}}(|\up\up,\down\down\rangle -|\down\down,\up\up\rangle)
\end{array}
& E=4J\\
\begin{array}{c}
\frac{1}{\sqrt{2}}(|\up\down,\down\up\rangle+|\down\up,\up\down\rangle)  \qquad
\frac{1}{\sqrt{2}}(|\up\up,\down\down\rangle +|\down\down,\up\up\rangle)
\end{array}
& E=2J\\
\begin{array}{c}
|\up\down,\up\down\rangle  \qquad \frac{1}{\sqrt{2}}(|\up\down,\down\up\rangle+|\down\up,\up\down\rangle)  \qquad |\down\up,\down\up\rangle \\
\end{array}
& E=0\\
\end{array}
\right.
\ee

We again calculate as a perturbation, the effect of the "kinetic" part of the slave-spin hamiltonian:
\be
H_{pert}=\sum_{m\s} h_m 2 S^x_{m\s}.
\ee
where $ O_{im\sigma} = 2 S^x_{m\s}$ because both orbitals are half-filled and thus the gauge $c=1$ (eq. (\ref{eq:gauge}) for $n_m=1/2$), and we call $h_m\equiv h^*_{im\s}+h_{im\s}$. Also $\lambda_m=0$ at half-filling because of particle-hole symmetry.

As done previously we perform a perturbative expansion in $h_m$, which is small near the metal-insulator transition, if this is second-order.

In this $J\neq0$ case the ground state has degeneracy 3  (the subset above with $E=0$). 
Furthermore, as before, the perturbation $H_{pert}$ does not have any nonzero elements in the low-energy subspace one has to use second-order perturbation theory to find the right combination of states, in the ground state manifold, to which the perturbed state tends for $h_m\rightarrow 0$.
This is the ground state of the matrix $H'\equiv H_{pert}(E_0-H_{at})^{-1}H_{pert}$  in the degenerate subspace. In this case, since $H_{pert}$ couples the $E=0$ sector only with the $E=\frac{U+J}{2}$ sector, one can rewrite $H'=-\frac{2}{U+J}H_{pert}^2$ (where we have used the fact that $E_0-H_{at}$ is diagonal and $E_0=0$).

By applying explicitly $H'$ to the ground state triplet of states one finds the following 3x3 matrix (we keep here the $h_m$ distinct):
\be\label{eq:Hpert_matrix_largeJ}
H'=-\frac{4(h_1^2+h_2^2)}{U+J}\left[ \begin{array}{ccc}
1 & \frac{1}{\sqrt{2}} & 0 \\
\frac{1}{\sqrt{2}} & 1 & \frac{1}{\sqrt{2}}  \\
 0  & \frac{1}{\sqrt{2}}  & 1 \\
\end{array}
\right]
\ee
of which the ground state to zero order in $h_m$ is: 
\be
|\phi_0\rangle\equiv \frac{1}{2}(|\up\down,\up\down\rangle + |\up\down,\down\up\rangle + |\down\up,\up\down\rangle + |\down\up,\down\up\rangle).
\ee

 The correction to the ground state to first order in $h_m$ reads:
 \bea
 |\phi_0^{(I)}\rangle=\@&&\@ | \phi_0\rangle +\sum_{|s\rangle\neq |\phi_0\rangle}\frac{\langle s|H_{pert}|\phi_0\rangle}{E^0-E_s^0} |s\rangle\nonumber \\
 =\@&&\@ | \phi_0\rangle-\frac{2}{U+J}\left[h_1(|\up\up,\up\down\rangle+|\up\up,\down\up\rangle+|\down\down,\up\down\rangle+|\down\down,\down\up\rangle)\right.\nonumber\\
 \@&&\@\qquad\qquad\quad + \left. h_2(|\up\down,\up\up\rangle+|\down\up,\up\up\rangle+|\up\down,\down\down\rangle+|\down\up,\down\down\rangle) \right]
 \eea
 This gives:
  \be
  \langle\phi_0^{(I)}| 2 S_{m\s}^x|\phi_0^{(I)}\rangle=-\frac{8}{U+J}h_m.
  \ee
 Analogously to the 1-band case to calculate the critical interaction one uses the self-consistency condition $h_m=2 \bar \eps_m \langle 2 S_{m\s}^x\rangle$. 
 This linearized condition determines the critical coupling $U_c$.

The self-consistency equations for each orbital then read: 
 \be
 h_m=2 \bar \eps_m \langle 2 S_{1\s}^x\rangle=-\frac{16h_m}{U+J}\bar \eps_m \quad \Longrightarrow \quad U^m_c=-16\bar \eps_m-J.
 \ee

This equations show two important results:
\begin{itemize}
\item In the half-filled Hubbard model with two identical bands (i.e. $\bar\eps_1=\bar\eps_2$ here), at large J the two bands undergo a common Mott transition of which the $U_c$ follows the Mott gap asymptotics (as described in section \ref{sec:MottGap}). It is interesting however to notice that the intercept at $J=0$ is the $U_c$ for the Mott transition in the 1-band Hubbard model $U_c(N=1)=-16\bar \eps$ (section \ref{comp:perturbative_Uc}), so that the result for the half-filled 2-band model can be written:
\be
U_c(N=2, \mbox{large}\; J)=U_c(N=1)-J.
\ee 

\item the equations for the two orbitals are \emph{decoupled}, contrary to what happens at $J=0$ (see eq. \ref{eq:selfcons_linear_J0_h1h2}). This is due to the fact that when starting from the triplet ground state at finite J, the second order perturbation theory that leads to the linearized self-consistency equation necessarily implies two hopping processes (two slave-spin flips, in the context of the slave-spin mapping) in the same orbital, or one would end up in one of the states of the $E=2J$ or $E=4J$ groups, and thus outside the degenerate ground state manifold. This is not the case at $J=0$ because all these states are degenerate within the ground state manifold of dimension 6. This is the low-energy \emph{orbital decoupling} induced by Hund's coupling J, and it is materialized - in this large J limit\footnote{In this sense the limit is large-J: the states outside the lowest triplet are excluded altogether from the perturbation theory, which would happen rigorously only in the large-J limit.} - in the fact that for $\bar\eps_1\neq\bar\eps_2$ each band undergoes a separate Mott transition at a critical interaction proportional to the respective kinetic energy $\bar\eps_m$. Thus the Mott transition becomes \emph{orbital-selective} (OSMT) and generates the phase diagram drawn with dashed lines in Fig. \ref{fig:orb_dec}. 

\end{itemize}

One last remark is in order.
The approximate treatment of the spin-flip and pair-hopping terms of the Kanamori hamiltonian in SSMF might lead to consider a more physical result the one extracted from the Ising form of the interaction. Indeed this is the present standard of use in realistic calculations based on SSMF.

However the perturbative expansion performed in this section is justified if the transition is second order, since only in that case it makes sense to perform a calculation for vanishing $h_m$. This indeed happens for the Kanamori hamiltonian at large J (see Fig. \ref{fig:2band_Hund-Kanamori-Ising} upper panel). In the case of the Ising form of Hund's coupling instead the transition is always first order at finite J (Fig. \ref{fig:2band_Hund-Kanamori-Ising} lower panel), thus the perturbative treatment is not justified.
It can nevertheless be performed (the treatment is strictly analogous to the one performed in this complement for the Kanamori case) and it yields the result:
\be
U_c(N=2, \mbox{large} \; J)=U_c(N=1)/2-J \qquad \mbox{("Ising" interaction)},
\ee
(this smaller value for $U_c$ compared to the value found for the Kanamori hamiltonian is indeed in line with the fact that the ground state degeneracy is brought down to a doublet in the Ising hamiltonian, instead than a triplet like in the Kanamori case).

This results signals that indeed a self-consistent solution exists for vanishing $h_m$, and it is thus a metastable one, which is pre-empted by the first-order transition (as studied in \cite{Ono_multiorb_linearizedDMFT,Lanata-Efficient_Gutzwiller}). 

In any case these results collectively point to a $J=0$ intercept of the large-J result for $U_c$ of the order of the critical coupling for the single-band Hubbard model.


\subsection{Hund's-induced orbital decoupling}\label{sec:orb_decoupling}

As suggested in complement \ref{comp:perturbative_Uc_N2_J}, another striking effect of Hund's coupling is to favor orbital-selective Mott transitions (OSMT)\cite{Anisimov_OSMT}. This was first shown in \cite{Koga_OSMT} and studied in more detail in \cite{demedici_Slave-spins, Ferrero_OSMT}. 
The main outcome of these studies on the 2-band Hubbard model with different bandwidths (OSMT in this model has been intensively studied in the period 2004-2005, a quite complete list of references can be found in \cite{Inaba_Koga_SecondOrder}) is summarized in Fig. \ref{fig:orb_dec}.
\begin{figure}
\begin{center}
\includegraphics[width=12cm]{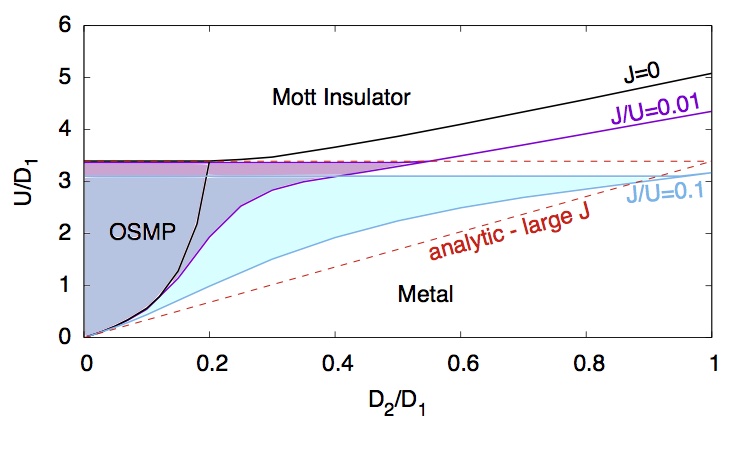}
\caption{Phase diagram of the 2-band Hubbard model with different bandwidths, in the plane bandwidth ratio ($D_2/D_1$) - interaction strength ($U/D_1$)), for several values of Hund's coupling $J/U$ and for $J=0$ (adapted from \cite{demedici_Slave-spins}).The analytic result showing how in the large-J limit the orbitals get decoupled is obtained in complement \ref{comp:perturbative_Uc_N2_J}.}
\label{fig:orb_dec}
\end{center}
\end{figure}
There, the colored areas represent - for different Hund's coupling strength J/U - an orbital-selective Mott phase (OSMP) arising in between a metal at low interaction strength U and a Mott insulator at high U. 

A difference in bandwidth between the two bands (which have here semi-circular densities of states, respectively half-bandwidth $D_1$ and $D_2$, and do not hybridize between them) is necessary to trigger the OSMT, however at J=0 a quite extreme bandwidth ratio $D_2/D_1=0.2$ is needed. Indeed when a band is more than five times wider than the other, orbital fluctuations that promote a unique Mott transition (as investigated in Appendix \ref{comp:perturbative_Uc_t1t2_J0}) cannot make up for the strong difference in bare kinetic energy between the two bands and an OSMT takes place. 

Hund's coupling however is seen to promote the OSMT and quickly shift this critical bandwidth ratio to moderate values, up to a point in which, already for moderate Hund's coupling $J/U=0.1$, a little difference in bandwidth is enough to separate the transitions between the bands. Complement \ref{comp:perturbative_Uc_N2_J}, in which a large J is assumed, shows that the same mechanism leading to the reduction of the width of the Hubbard bands highlighted in the previous section \ref{sec:Janus} is the one that leads to decoupled equations for the Mott transition in the two bands (these equations yield the phase diagram shown in Fig. \ref{fig:orb_dec} by dashed lines).
Indeed complement \ref{comp:perturbative_Uc_N2_J} is the low-energy counterpart in SSMF of the mechanism yielding decoupled Hubbard bands pointed out by Koga et al. in Ref. \cite{Koga_OSMT} (and illustrated in Ref. \cite{Georges_annrev}, Fig. 9), which ultimately motivates also their reduction in width (since it counteracts the enhancement mechanism\cite{gunnarsson_fullerenes} mentioned in section \ref{sec:Janus}).

Indeed numerical evidence that the core of all OSMT is an \emph{orbital decoupling} caused by Hund's J is provided in Ref. \cite{demedici_MottHund}, where several multi-band models were analyzed with bands of different width or bands of identical width but finite crystal-field splitting.  There, it is pointed out that Hund's coupling causes a change from a collective to an individual orbital behaviour in which then it is the individual bandstructure and/or population in each orbital (and correspondingly in each band, if there is no hybridization) that determines the correlation strength of the electrons with that orbital character. 
This results does not rely necessarily on the bands having different width, but only on the independence of the gap to charge excitations in each orbital. The correlation between charge excitations can be measured in these models and it is strongly suppressed in the vicinity of the half-filled Mott insulator \cite{demedici_MottHund,demedici-SpringerBook}, where orbital-selective Mott transitions are realized\footnote{Even for OSMT with 4 electrons in 3 bands like in \cite{demedici_3bandOSMT} or 6 electrons in 5 bands\cite{demedici_Genesis,Lauchli_Werner_Krylov_5band} - seemingly far from half-filling - the orbital-selective insulating gap is connected with the gap of the half-filled Mott insulator.}.
Indeed the striking result that bands of identical width could undergo separate Mott transitions was indeed remarked in \cite{Werner_high-low_Hund} and studied in\cite{demedici_3bandOSMT} and \cite{Werner_3band,Huang_XiDai_hi-lo_spin_3band_cfs}. In these cases it is the final orbital population that sets the correlation strength and when an orbital is individually half-filled it can become insulating independently of the others and an orbital-selective Mott phase can be realized.

In general, all the OSMT at finite J can be rationalized within the Hubbard criterion as a result of decoupled Hubbard bands \cite{demedici-SpringerBook}. This is illustrated in Fig. \ref{fig:OrbDec_cartoon}.

\begin{figure}[h]
\begin{center}
\includegraphics[width=8cm]{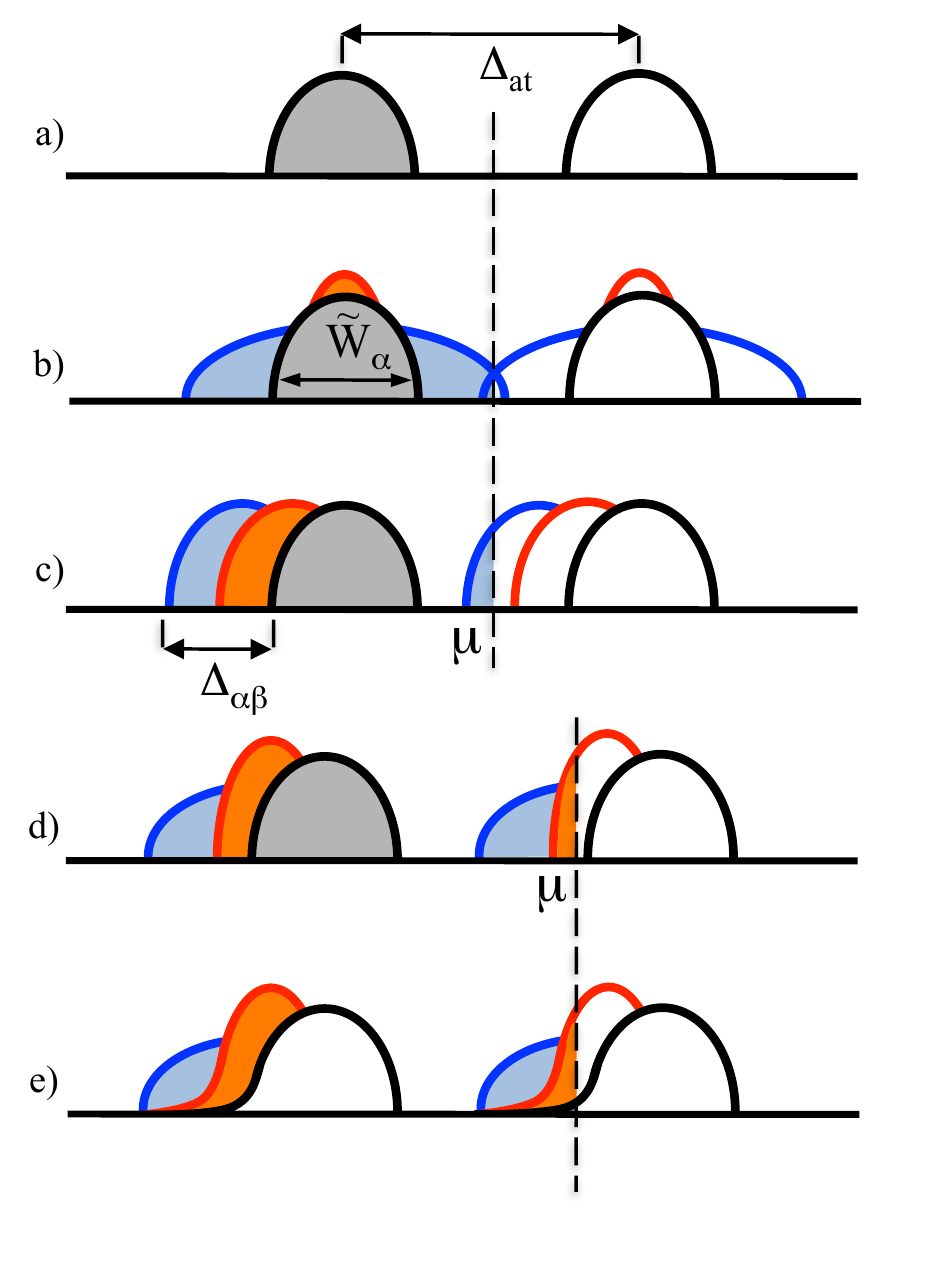}
\caption{Cartoon of a general mechanism for Hund's promoted orbital-selective Mott transitions and selective Mottness (see text). a) Spectrum of a degenerate half-filled Hubbard model; b) Orbital selective Mott phase triggered by a difference in bandwidths; c) Orbital selective Mott phase triggered by a lifting of degeneracy of the orbital energies due to crystal-field splittings; d) General situation with both the previous effects combined, OSM phase due to doping; e) Onset of hybridization compared to d), transforming an orbital-selective Mott phase in a metallic phase with orbital-selective Mottness. From \cite{demedici-SpringerBook}.}
\label{fig:OrbDec_cartoon}   
\end{center}
\end{figure}

A finite inter-orbital hopping or local hybridization acts against Hund's coupling, as also does the crystal-field splitting. However as long as Hund's coupling remains the dominant scale determining the atomic multiplet and is able to quench orbital fluctuations, these can be treated as perturbations and the physics of orbital decoupling is robust against them.
Hybridization however ultimately can win over Hund's coupling at low energy \cite{Pavarini_Hund_vs_hyb,Winograd_pseudogap} (however also competing non-local energy scales can prevail\cite{Pepin_OSMT, DeLeo_OSMT}) turning the localized component of the OSMP into a very heavy delocalized state. Thus the OSMP, in presence of hybridization is typically turned into a phase with  coexistence of strongly and weakly correlated electrons in the conduction bands. This is what happens in Fe-based superconductors, and it is the subject of the next section.


\section{Slave-spin modeling of Fe-based superconductors}\label{sec:ironbased}

As anticipated in the introduction, SSMF has turned out very profitable in the investigation of the normal phase of Fe-based superconductors (FeSC).
Indeed these are Fermi-liquid metals - apart from the phases where they turn superconductors - albeit with strongly renormalized band structures. Also the conduction bands are inherently multi-orbital in nature, and Hund's coupling is believed to play a major role. The large number of different materials belonging to this class and the large number of parameters determining the physics in these materials (interaction strength, pressure, doping, ...) make them a large playground, where an agile, albeit simplified, method for the realistic modeling of such phases like DFT+SSMF, can extract a lot of physical information and provide guidance in the understanding and predicting the physics of the presently known and unknown similar materials.

Most of the compounds from the different families have been studied with DFT+SSMF: LaFeAsO\cite{YuSi_LDA-SlaveSpins_LaFeAsO,  demedici_OSM_FeSC}, AFe$_2$As$_2$ (with A=Ba,K,Rb,Cs)\cite{demedici_OSM_FeSC,Eilers_Qcrit,Hardy_122_SlaveSpin_exp}, intercalated chalcogenides\cite{Yu_Si_KFeSe,Yi_Shen_ARPES_OSMT_KFeSe} FeTeSe and the FeSe monolayer \cite{Yi_Universal_OSM_Chalcogenides}, La$_2$O$_3$Fe$_2$Se$_2$\cite{Giovannetti-La2O3Fe2Se2}.
\begin{figure}[h!]
\begin{center}
\includegraphics[width=11cm]{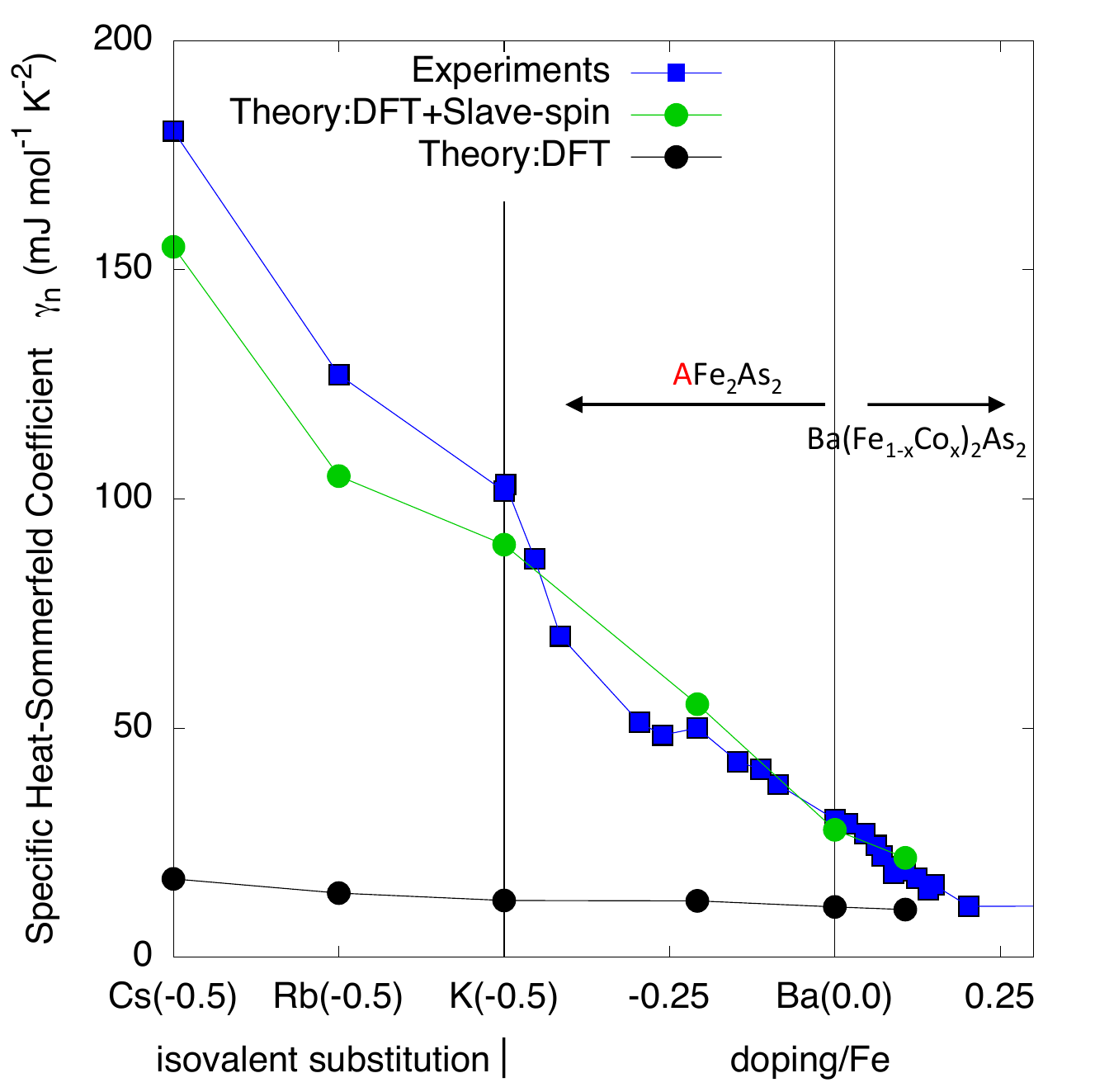}
\caption{Normal-phase Sommerfeld coefficient of the electronic specific heat in throughout the whole 122 family. Comparison between experiments (blue squares), DFT (black circles), DFT+Slave spins with a single set of interaction parameters (U=2.7eV, J/U=0.25) for all the compounds (green circles).  Adapted from Hardy et al., Ref. \cite{Hardy_122_SlaveSpin_exp}.}
\label{fig:Sommerfeld122}
\end{center}
\end{figure}

The procedure for this realistic DFT+SSMF modeling is the following\footnote{This is indeed completely analogous to the procedures used in the framework of DFT+DMFT calculations.}.  A tight binding fit (using a basis of localized wavefunctions, for example maximally localized Wannier functions\cite{Marzari_RMP}) of the conduction bands issued from the material-specific DFT simulations is performed. The local intra- and inter-orbital interactions between electrons on the orbitals of this local basis are included explicitly in this low-energy Hamiltonian, yielding thus a multi-orbital Hubbard model eq. (\ref{eq:multiorb_Hubbard}), which can be treated within SSMF as explained in the previous sections\footnote{If the low-energy tight-binding model includes local orbitals other than those in which the interaction is treated explicitly (e.g. typically if ligand-centered orbitals are also used in the local basis, for which the treatment of interaction is left at the DFT level, yielding a so-called "p-d model") then a relative energy shift between the correlated and uncorrelated orbitals has to be included. This "double-counting" term is of difficult assessment and many different recipes have been proposed\cite{Czyzyk_Sawatzky_DoubleCounting,Karolak-DoubleCounting,Park-UprimeDoubleCounting,Haule_ExactDoubleCounting}. The problem of having this, basically uncontrolled, extra parameter can be avoided altogether by limiting the low-energy tight-binding fit to the "correlated" orbitals (i.e. the five 3d Wannier functions centered on each Fe atom, in the present case of FeSC, yielding the so-called "d-model"). In this case the double counting term is absorbed in the chemical potential, which is fixed by the total electron population. The price to pay for this advantage is to have broader d-like Wannier functions, for which the Hubbard-like treatment is less justified. This is nevertheless the strategy that has been successfully followed thus far in all the published SSMF studies.\label{foot:d-model}}.

All of the physics described in the previous sections and due to Hund's coupling in the multiorbital Hubbard model has been found to play an important role in these realistic investigations of Fe-superconductors. In particular an influence of the half-filled Mott insulator has been clearly found in these simulations.

Indeed one striking feature of these calculations is a steady increase of the normal-phase Sommerfeld coefficient (the coefficient of the low-temperature linear contribution to the specific heat) in the 122 FeSC family. This is illustrated in Fig. \ref{fig:Sommerfeld122}.
The plotted experimental data (blue squares) concern BaFe$_2$As$_2$ throughout the full electron-doped (with partial Fe $\rightarrow$ Co substitution) and hole-doped (with Ba $\rightarrow$ K substitution) range, until the stoichiometric end-member KFe$_2$As$_2$, and further isovalent substitutions of the cations in this last compound giving stoichiometric RbFe$_2$As$_2$ and CsFe$_2$As$_2$.
The measures are taken in high-purity single-crystals and the modeling with standard DFT (black points in Fig. \ref{fig:Sommerfeld122}) is clearly incapable to describe correctly this increase.

Some coupling to low-energy bosons (e.g. phonons or spin-fluctuations) or to critical fluctuations due to the proximity to a quantum critical point\cite{Eilers_Qcrit} can be invoked to try to explain this physics.
However it appears simpler the explanation based on the fact that DFT modeling neglects local correlations, and these are particularly important for FeSC since Mott physics plays a major role.  

Indeed the realistic modeling based on SSMF, which includes the local correlations as detailed in the previous sections and can describe the Mott transition, predicts a Sommerfeld coefficient in very good agreement with experimental data (green points in Fig. \ref{fig:Sommerfeld122}), with only one set of interaction parameters throughout all the compounds (U=2.7eV, J/U=0.25 \footnote{This values depend on the type of Wannier orbitals used in the modeling of the conduction bands. These values are appropriate for the "d-model" scheme. See also footnote \ref{foot:d-model}.\label{foot:interactions_d-model}}).

It can be shown that in the investigated low-energy models for FeSC, for realistic values of the interaction parameters\cite{miyake_interactions_jpsj_2010}, a Mott insulator is always realized at half-filling (i.e. for an electronic density of 5 electrons per Fe atom, of which all the five 3d orbitals contribute predominantly to the conduction bands). At the density of 6 electrons/Fe, corresponding to that of the stoichiometric parent FeSC of all families (LaFeAsO, BaFe$_2$As$_2$, FeSe, etc.), no Mott transition is found until very high interaction strengths\cite{YuSi_LDA-SlaveSpins_LaFeAsO,  demedici_OSM_FeSC, Yi_Universal_OSM_Chalcogenides} (with the exception of the intercalated chalcogenides in the special Fe-vacancy-ordered phase\cite{Yu_Si_KFeSe} and of La$_2$O$_3$Fe$_2$Se$_2$, this being however a very different family, with very large crystal-field splitting strongly contrasting the effect of Hund's coupling on the Mott transitions\cite{Giovannetti-La2O3Fe2Se2}).

This can be shown in general, as indeed Mott physics depends very little on the details of the k-space band dispersions and mostly on k-integrated quantities, like the total kinetic energy of the DFT band structure.
Indeed when normalizing the energy scale in a way to equate the bare kinetic energy of the realistic low-energy models for the various compounds to that of a simple 5-band Hubbard model with semicircular densities of states, one can show that the critical interaction strength for the half-filled Mott transitions are essentially coincident, and below the range of typical interaction values estimated for these compounds. 
The same phase diagram suggests that the electron densities relevant for FeSC (5.5$\div$ 6.3 electrons/Fe) fall in general in the zone of influence of this half-filled Mott Insulator. 
Both these relevant ranges in the interaction/density plane are indicated by the grey box in Fig. \ref{fig:5OrbModel_BaFe2As2_SSMF} (upper panels). 

\begin{figure}[h!]
\begin{center}
\includegraphics[width=15cm]{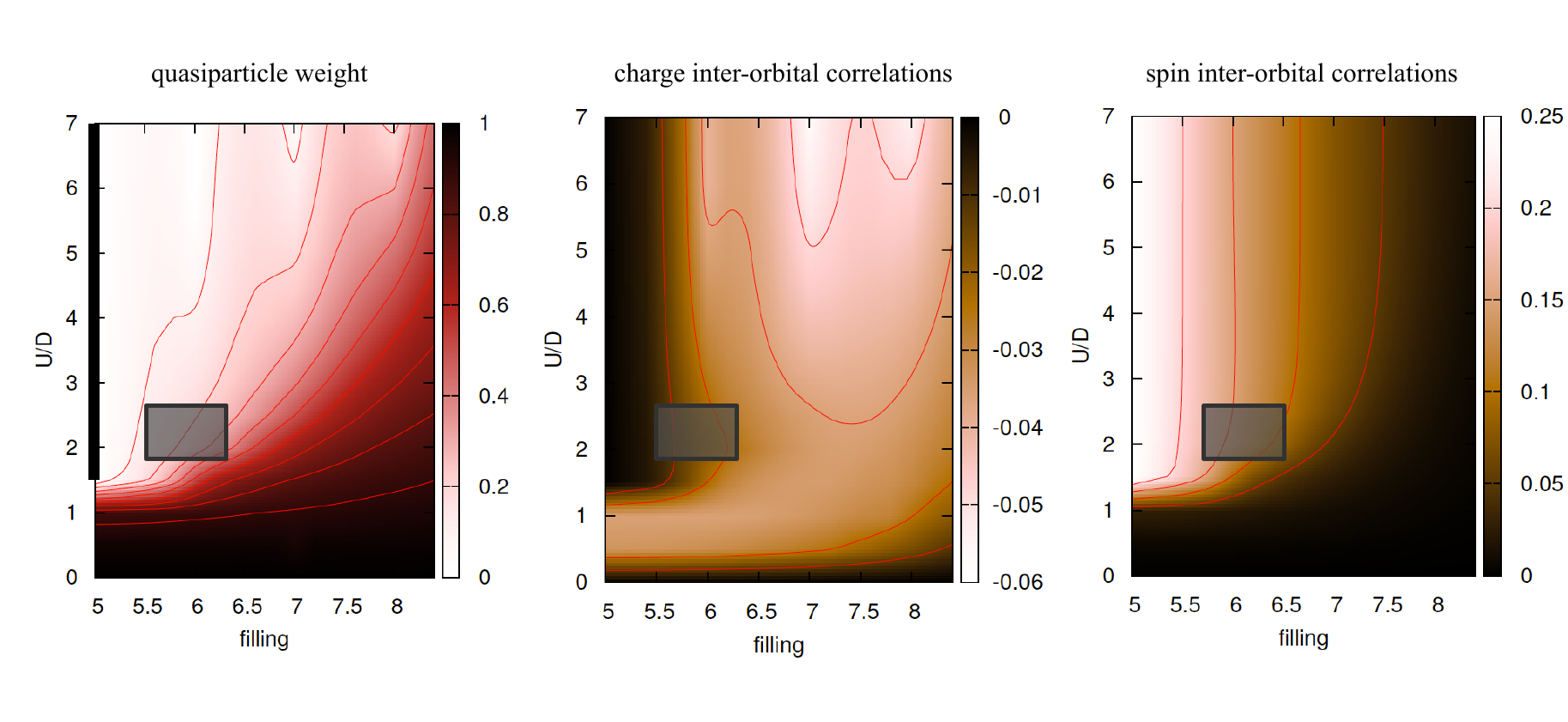}
\includegraphics[width=5cm]{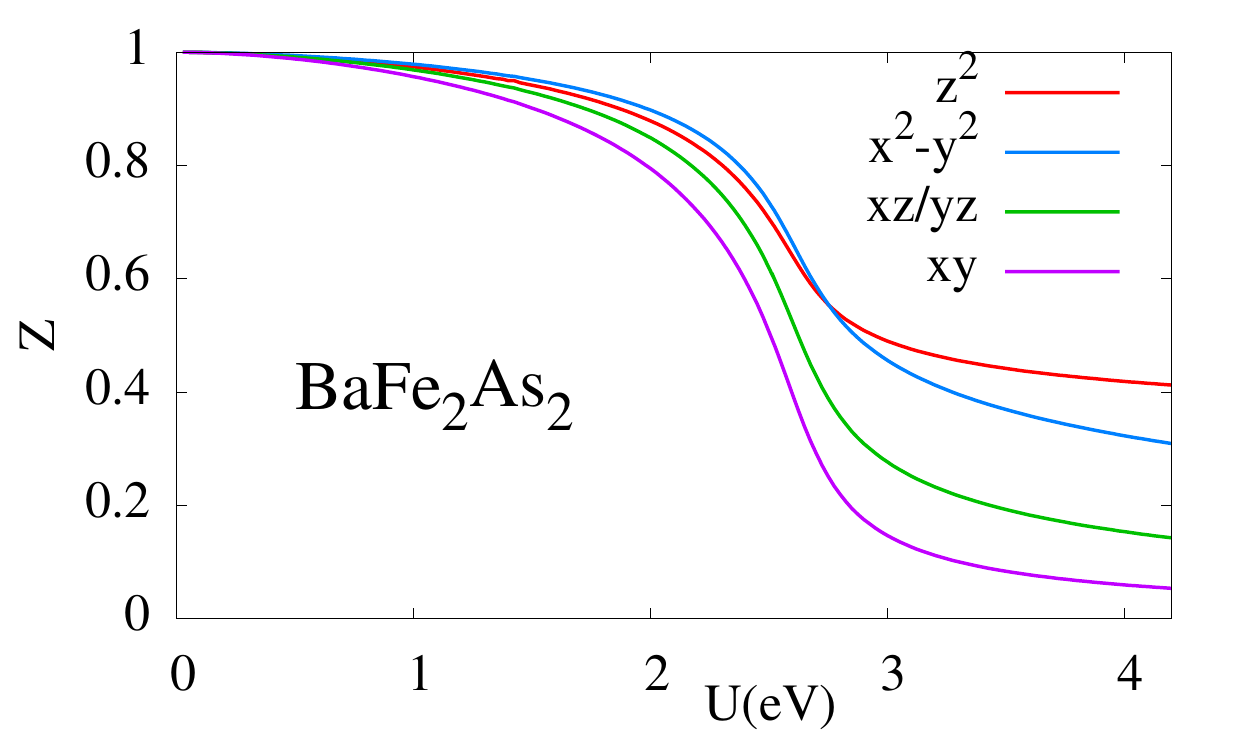}
\includegraphics[width=5cm]{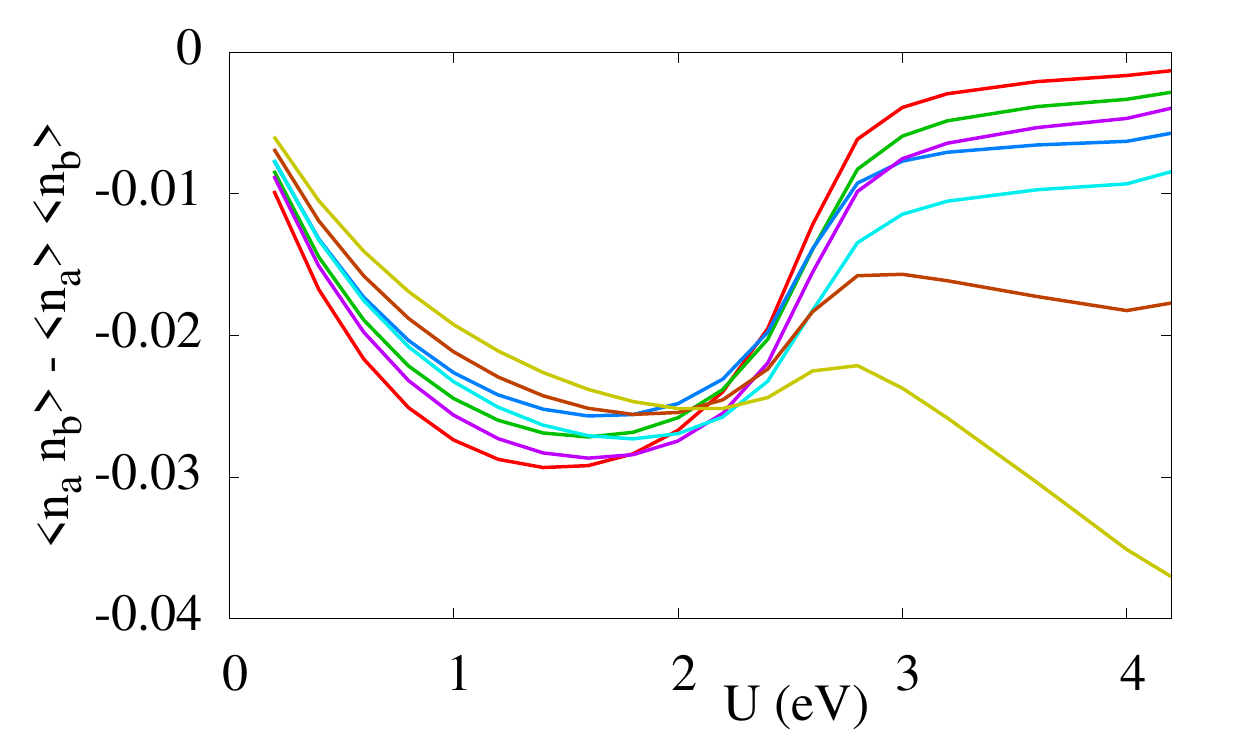}
\includegraphics[width=5cm]{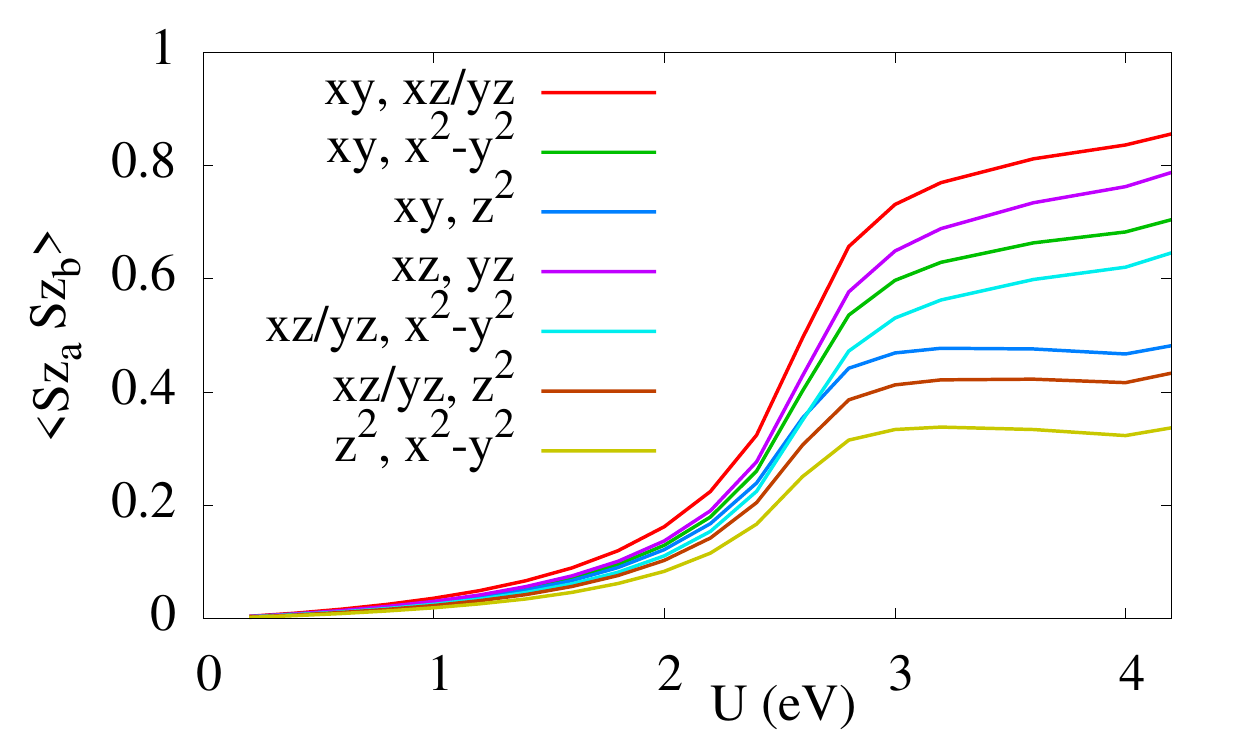}
\caption{{\bf Upper panels} Quasiparticle weight (left), inter-orbital charge (center) and spin (right) correlations for the 5-orbital Hubbard model (with semi-circular DOS of half-bandwidth D) for J/U=0.2, in the plane electron density(filling)/interaction strength (reproduced from \cite{demedici-SpringerBook}). The black bar on the left graph marks the Mott insulating phase at half-filling. The gray box locates the relevant zone for FeSC (see text). {\bf Lower panels}. Ab-initio modeling for BaFe$_2$As$_2$ within SSMF,  from \cite{demedici_OSM_FeSC} (Supplementary material). Same quantities as above at fixed filling n=6, as a function of interaction strength for J/U=0.25. The estimated value of U is 2.7 - 2.8 eV \cite{miyake_interactions_jpsj_2010} (see footnote \ref{foot:interactions_d-model}).}
\label{fig:5OrbModel_BaFe2As2_SSMF}
\end{center}
\end{figure}

Thus the experimentally measured strong enhancement of the Sommerfeld coefficient is simply explained with the materials approaching a Mott insulator which would be realized at half filling\footnote{The realization of a Mott insulator for half-filled FeSC is hard to test directly. Indeed the end member of the family KFe$_2$As$_2$ has a density of 5.5 electrons/Fe and no further hole-doped material has been synthesized in the family. A similar material with 5 electrons in the 5 conduction bands is BaMn$_2$As$_2$ however, which is found insulating\cite{An_Sefat-BaMn2As2_ins} indeed, albeit the nature of this insulating state is still being clarified\cite{McNally_BaMn2As2_Mott_Hund,Zing-BaMn2As2_LaMnAsO}.}. Indeed the Sommerfeld coefficient\cite{ashcroft,nozieres} is 
\be
\gamma=\frac{\pi^2k_B^2}{3}N^*(\eps_F) 
\ee
with $N^*(\eps_F)$ the total quasiparticle density of states at the Fermi energy $\eps_F$. In our SSMF framework this reduces to the DFT density of states at $\eps_F$ for $U=J=0$, while its value is enhanced at finite interaction strength. This is an outcome of the quasiparticle mass enhancement induced by the interactions (in our method the renormalization of the hopping integrals), which produces bands which are less dispersive, thus denser in energy.

But besides the overall enhancement of the quasiparticle masses mainly set by the distance in doping from the half-filled Mott insulator, the influence of the proximity to this Mott phase in FeSC shows up in the several other aspects. These are seen in all realistic DFT+SSMF calculations (Fig. \ref{fig:5OrbModel_BaFe2As2_SSMF}, lower panels) and are the main outcomes of the Hund's physics illustrated in the previous section, thus already present at the model level (Fig. \ref{fig:5OrbModel_BaFe2As2_SSMF}, upper panels).
Indeed ferromagnetic spin correlations between electrons in different orbitals are favored by Hund's coupling, thus leading to the formation of a large local moment in the paramagnetic metallic phase (this aspect received recently support from X-ray emission experiments, Ref. \cite{Lafuerza_XES_122}). Also inter-orbital charge correlations are suppressed, due to the Hund's induced orbital-decoupling (the mechanism detailed in section \ref{sec:orb_decoupling}).
Overall these features develop after a rapid crossover as a function of the interaction strength (clearly identifiable in the lower panels of Fig. \ref{fig:5OrbModel_BaFe2As2_SSMF} by the sharp changes around U=2.5$\div$2.8eV in all the quantities calculated for BaFe$_2$As$_2$, and by the accumulation of the contour lines in the model calculation in the upper panels), defining a broad frontier departing from the half-filled Mott transition\cite{Ishida_Mott_d5_nFL_Fe-SC,Fanfarillo_Hund}. 

The successful denomination of "Hund's metal" \cite{Yin_FeSC_kinetic_frustration} might be used for the zone beyond this frontier, where the Hund's coupling is essential in shaping the aforementioned phenomenology of the doped half-filled Mott insulator. This is more of a matter of taste however, for the present lack of a precise definition of a Hund's metal.
 
Another remarkable feature of this zone is that for temperatures higher than the Fermi-liquid coherence scale, the metal develops a finite spin-spin correlation at long times (and a sublinear frequency dependence of the self-energy at low frequency) which has been dubbed "spin-freezing" in the model context\cite{Werner_SpinFreezing} and has been found in realistic simulations for FeSC\cite{Liebsch_FeSe_spinfreezing,Werner_dynU_122}.
The coherence temperature is strongly suppressed, when approaching the half-filled Mott insulator (see discussion in Ref. \cite{Georges_annrev} and references therein).
  
Predictably from the model studies (see section \ref{sec:orb_decoupling}), the features of the realistic band structure that break the orbital symmetry (most notably the crystal-field splitting of the orbital energies) induce an orbital differentiation of the correlation strength, as is clear from the lower-left panel in Fig. \ref{fig:5OrbModel_BaFe2As2_SSMF} in which one sees that for BaFe$_2$As$_2$, in the strongly correlated region just after the crossover, mass enhancements range from 2 to 10 depending on the orbital.
This kind of orbital differentiation has been found in all the abovementioned simulations of FeSC within DFT+SSMF, and within many other methods:
 DMFT
\cite{Haule_pnictides_NJP,Shorikov_LaFeAsO_OSMT,Laad_SusceptibilityPnictides_OSM,Craco_FeSe,Aichhorn_FeSe,Yin_kinetic_frustration_allFeSC,Ishida_Mott_d5_nFL_Fe-SC, Liebsch_FeSe_spinfreezing},
variational Montecarlo\cite{Misawa_d5-proximity_magnetic,Misawa_LaFeAsO},
Slave-spins \cite{YuSi_LDA-SlaveSpins_LaFeAsO,Yu_Si_KFeSe},
Hartree-Fock mean-field\cite{Bascones_OSMT_Gap_halffilling},
fluctuation-exchange approximation\cite{Ikeda_pnictides_FLEX},
Gutzwiller approximation\cite{Lanata_FeSe_LDA+Gutz}.

The actual degree of orbital differentiation in the material depend on the realistic value of the interaction strength, which is estimated U=2.7-2.8eV in BaFe$_2$As$_2$\cite{miyake_interactions_jpsj_2010}\footnote{A sizable error bar is actually to be considered around these c-RPA estimated values, U in particuler. A small 5 \% adjustment in LaFeAsO for instance was found necessary in Ref.\cite{Misawa_LaFeAsO}, but the error on U can in principle be substantially higher.}, yielding mass enhancements ranging between 1.5 and 3, consistently with the experimental estimate.

A very detailed comparison between theoretical predictions and experimental estimates of the mass enhancement was done in Ref. \cite{demedici_OSM_FeSC}, clearly establishing the orbital selectivity of electronic correlations in the whole phase diagram of the 122 family and assessing the degree of selectivity.
Indeed (as shown in the upper panel of Fig. \ref{fig:OSM_FeSF}) a substantial discrepancy among the various experimental data quantifying the correlation strength is found, when trying to uniquely define if electrons are strongly \emph{or} weakly correlated in these materials. As already mentioned the Sommerfeld coefficient raises uniformly, indicating a strong mass enhancement, from the electron-overdoped (where the electrons are essentially non correlated) to the stoichiometric compounds (where the mass enhancement is $\sim$3), all the way to the hole-overdoped family end-members (where the mass enhancement is $\sim$9).
On the other hand other probes like low-frequency optical conductivity (of which the integrated spectral weight is lowered by correlations) or ARPES (that can directly measure the reduction in band dispersion due to the mass enhancement) report much more moderate correlation strengths over the whole phase diagram (topping at $\sim$3 for KFe$_2$As$_2$).
\begin{figure}[h!]
\begin{center}
\includegraphics[width=12cm]{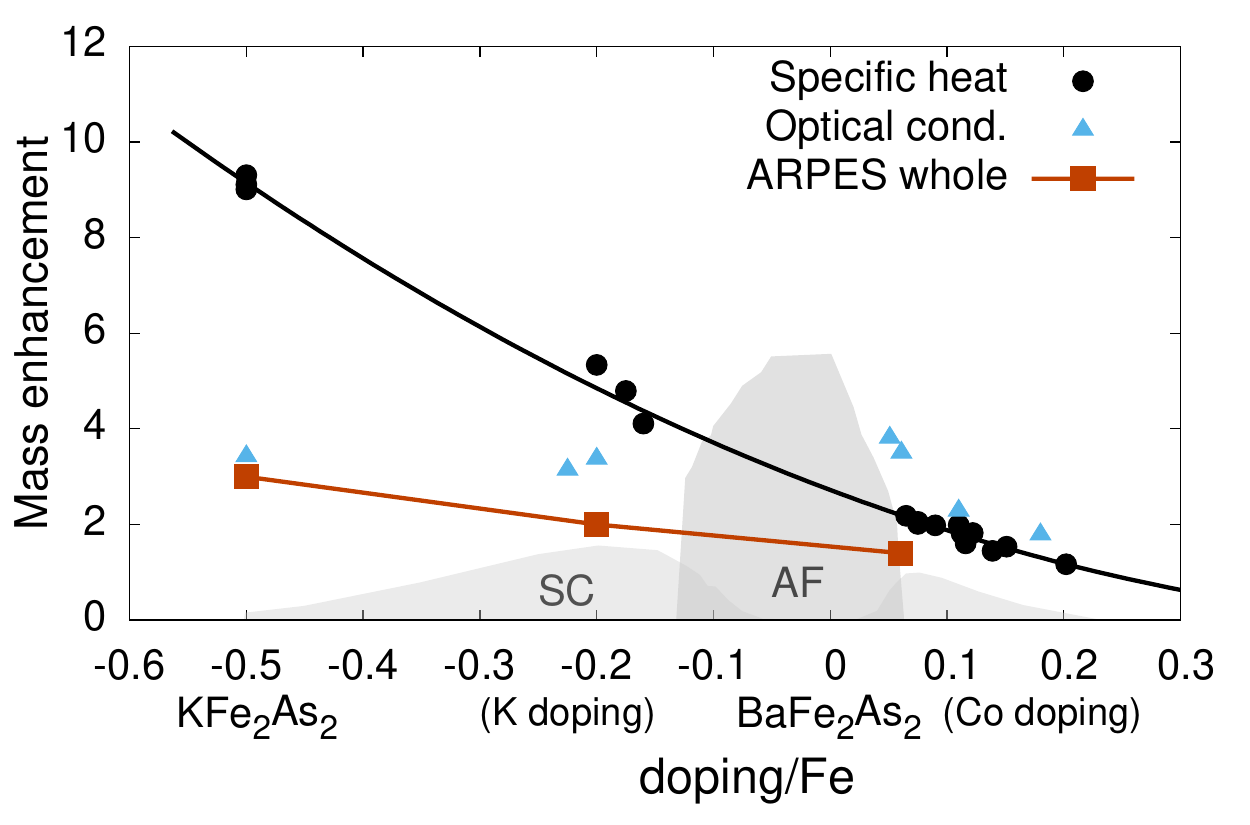}
\includegraphics[width=12cm]{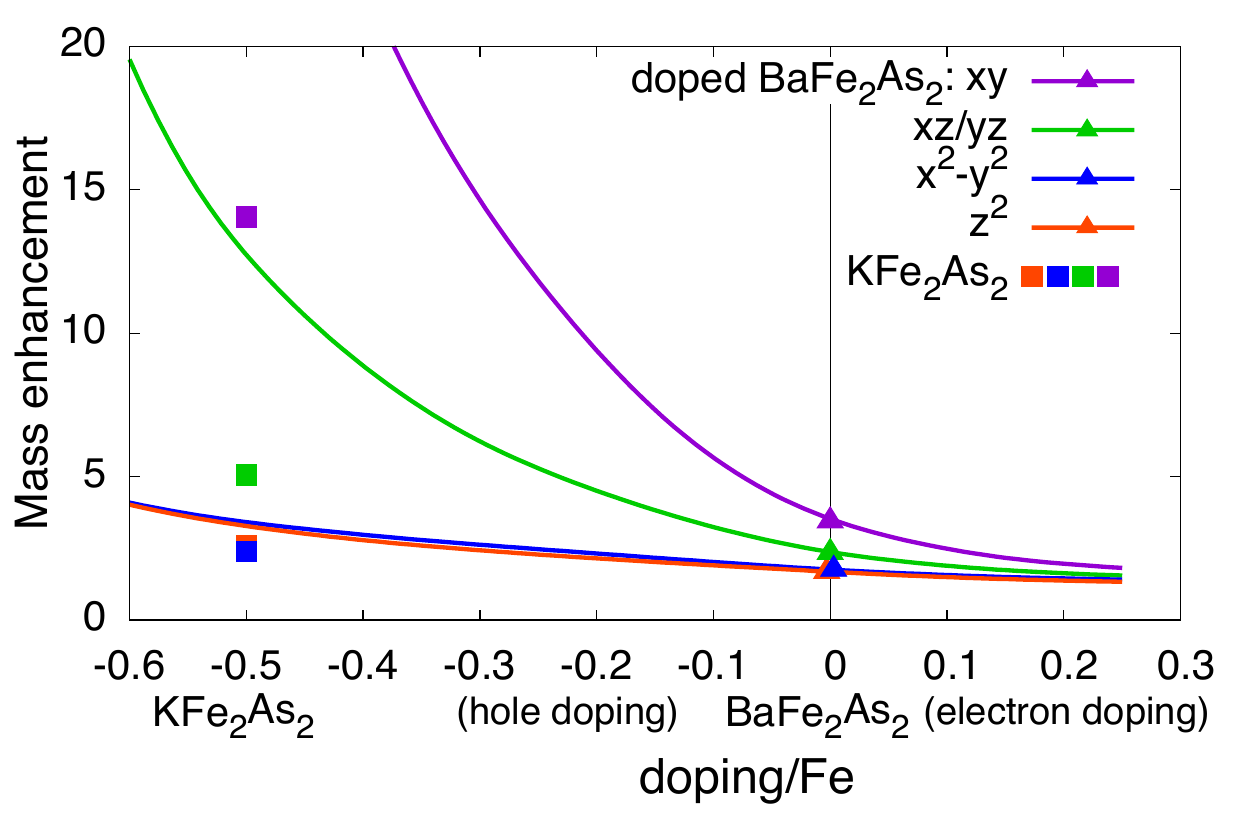}
\caption{Mass enhancement for doped BaFe$_2$As$_2$ and KFe$_2$As$_2$. Experiments (upper panel, experimental references in Ref. \cite{demedici_OSM_FeSC} and supplementary material) and theory within SSMF (U=2.7eV, J/U=0.25). Triangles (stoichiometric) and solid lines (doped) correspond to calculations performed using  the DFT bandstructure for BaFe$_2$As$_2$, while the squares correspond to the calculation for stoichiometric KFe$_2$As$_2$. The different colors indicate electrons of different orbital character (see legend).}
\label{fig:OSM_FeSF}
\end{center}
\end{figure}

However this discrepancy is readily interpreted as a \emph{coexistence of weakly and strongly correlated electrons}. Indeed when the mass enhancement is not equal for all the electrons the different experimental quantities can be renormalized in very different ways. For instance the heaviest electrons tend to strongly enhance the Sommerfeld coefficient, which is a sum of contributions from each orbital character, each weighed by the respective mass enhancement, to the total density of states. On the other hand the Drude weight (the integrated low-frequency spectral weight of the optical conductivity) is a sum of contributions from each orbital each weighed by its \emph{inverse} mass enhancement. Thus the contribution from the most renormalized electrons tends to be suppressed, and the renormalization of the lighter electrons will tend to dominate the renormalization of the total weight.
Analogously, the bandstructure, which is an intricate manifold of conduction bands that can be seen as the hybridization due to inter-orbital hoppings of five Fe 3d bands (one for each orbital), will still disperse as the more dispersive of these bands, even if selective correlations strongly reduce the dispersion corresponding to some orbitals. 

Thus, naturally, estimates of the correlation strength from optics or ARPES whole bandstructures will yield much smaller values compared to specific heat estimates, if weakly and strongly correlated electrons coexist in the conduction bands.

ARPES and quantum oscillations can however also probe the renormalization of each band separately.
Indeed they yield values for the mass enhancement in the different sheets of the Fermi surface (see these values overlaid to Fig. \ref{fig:OSM_FeSF} in Ref. \cite{demedici_OSM_FeSC}) which are concentrated around $\sim 2\div3$ in the electron doped side of the phase diagram, spread progressively with decreasing filling, and range between 2 and 20 for KFe$_2$As$_2$. This is in agreement with the theoretical estimates (reported in the lower panel of Fig. \ref{fig:OSM_FeSF}). 

As predicted by the theory (already in Refs. \cite{demedici_3bandOSMT,demedici_Genesis,demedici_MottHund} and then in most of the aforementioned works) the strongest renormalization always happen for the electrons of the $xy$ orbital characterAfter being theoretically predicted in  the stronger correlations in $xy$ electrons \footnote{One should mention that in some works an alternative convention is used and the orbital label $xy$ is switched with $x^2-y^2$. It is just a question of nomenclature however, the most correlated orbital being the planar one of $t_{2g}$ symmetry, that we call $xy$ following the most used convention.} . 
The main reason for the $xy$ orbital being the most correlated of all is its final population, which is the closest to half-filling among the five orbitals. Indeed in the framework of Hund's-induced orbital-decoupling detailed in section \ref{sec:orb_decoupling}, each orbital is distinctly influenced by Mott physics ("selective Mottness"\cite{demedici_OSM_FeSC}). Indeed in this regime the enhancement of the correlation among electrons in each orbital depends on the proximity to the Mott state that each orbital can reach separately once individually half-filled. A more thorough discussion on this mechanism and a specific analysis of the stronger correlations in the $xy$ orbital in FeSC can be found in the last sections and appendix of Ref.  \cite{demedici-SpringerBook}.

\begin{figure}[h!]
\begin{center}
\includegraphics[width=14cm]{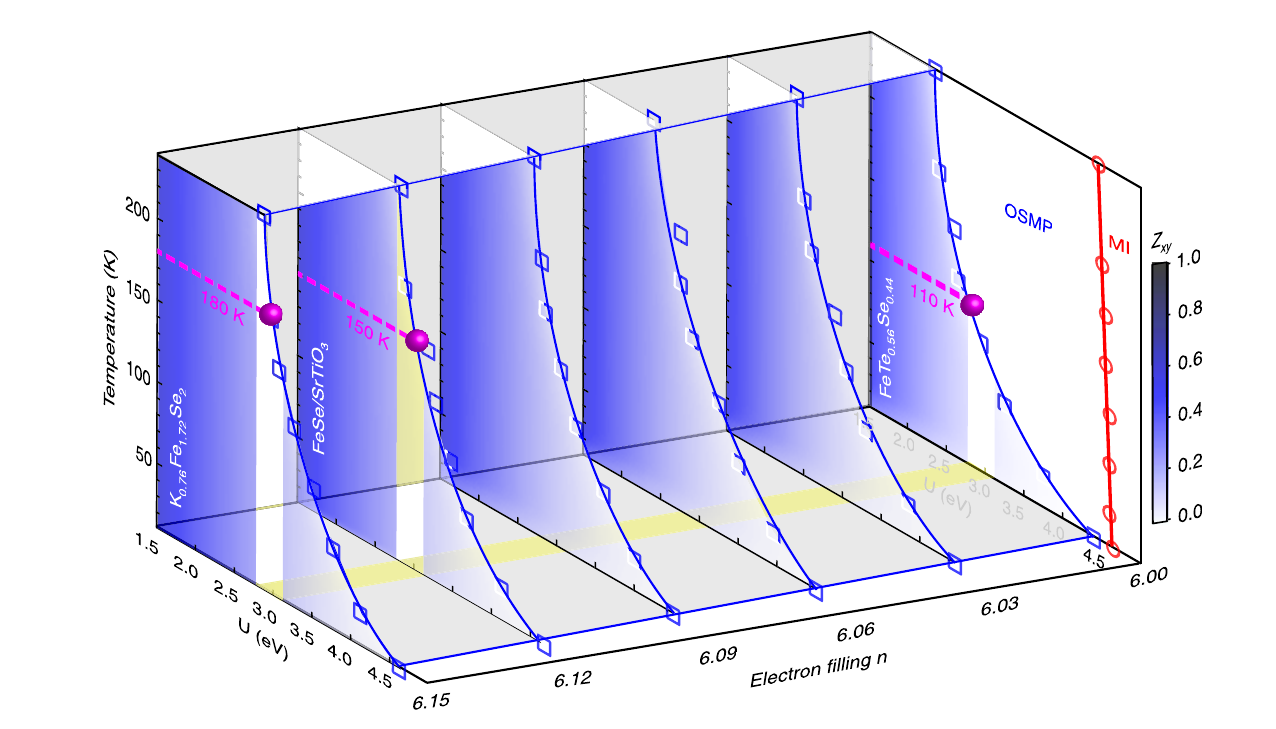}
\caption{Coherence scale (blue lines) for the $xy$ electrons in several Fe-chalcogenides within SSMF. Hund's coupling is kept to a fixed value throughout these calculations, thus causing the Mott transition for 6 electrons in this 5-orbital model to happen at a lower $U_c$ than in the modeling at fixed J/U. The experimentally determined values are marked by pink lines. From. Ref. \cite{Yi_Universal_OSM_Chalcogenides}.}
\label{fig:Yi_OSMT_ARPES}
\end{center}
\end{figure}

The stronger correlations of the $xy$ electrons imply a lower coherence temperature for the corresponding quasiparticles. When this temperature is crossed indeed a quick depletion of the spectral weight associated to $xy$ quasiparticles has been directly observed in ARPES, in Refs. \cite{Yi_Shen_ARPES_OSMT_KFeSe, Yi_Universal_OSM_Chalcogenides} in most of the Iron chalcogenides. This was successfully modeled within SSMF as shown in Fig. \ref{fig:Yi_OSMT_ARPES}.

Another useful application of SSMF has been the determination of the contribution of quasiparticle interband transitions to the infrared optical conductivity of the 122 family\cite{Calderon-Optics_Pnictides}. Indeed the renormalization of the bandstructure induces a shift of these transitions to lower frequency compared to the DFT estimate, and in this work it is predicted that a substantial contribution from these transitions extends down to the whole range of the mid-infrared.

\section{Concluding remarks}\label{sec:conclusions}
In this chapter we gave a pedagogical introduction to the slave-spin technique. After an introduction on the physics of strongly correlated materials, and of the Fermi-liquid metallic phases, alongside with the details of the technique we have seen the main aspects of the influence of Mott and Hund physics in the correlated metallic phases. We have finally applied these known concepts issued of both old and recent research to the physics of the recently discovered Fe-based superconductors. Some new concepts are also introduced here, that are useful to complete our view on the normal phases of these superconductors: the shrinking of the Hubbard bands of the half-filled multi-orbital Mott insulator and the low-energy description of the "orbital-decoupling", both induced by a sizable Hund's coupling.

Obviously this manuscript is a path through many topics, bridging several theoretical and experimental areas in the field of strongly correlated materials, that is exquisitely the result of the taste of the authors, and no claim of completeness could be associated to it, in any of the respective areas. 
Thus, for example, it cannot be a complete treatment of the physics of Iron-based superconductors, far from it actually, which receive a much more complete and equilibrate treatment in many of the recent and less recent reviews available in the literature (a non-exhaustive list being the following Refs.\cite{Johnston_Review_FeSC,Stewart_RMP,Chubukov_Hirschfeld-PhysToday,Springer_Book_FeSC,Alloul_Cano_ComptesRendus} and references therein). 
Nor it is a complete treatment of Mott and Hund's physics, of which the interested reader can deepen his knowledge in another series of available reviews (a very incomplete list being the following Refs. \cite{mottRMP,imada_mit_review,Vollhardt_RMP_He3,georges_RMP_dmft}and \cite{Georges_annrev}).
Finally, even on the slave-spin technique the focus is put primarily on the flavor of this technique used by the authors, and the reader is redirected to Ref. \cite{YuSi_LDA-SlaveSpins_LaFeAsO} for the main alternative - albeit very similar both in the spirit and in the results on FeSC - take, besides the various publications mentioned in section \ref{sec:ironbased}.

However we believe to have given our best possible intersection of the three previous subjects, and in this sense this manuscript probably represent a rather thorough coverage. 

It is worth mentioning two natural extensions of the SSMF that have not been worked out yet, still they are very much needed: the formalism for symmetry-broken (magnetic, superconducting etc) phases, and the rotationally invariant formulation, capable to treat the Kanamori-Hund interaction on the same level of approximation than the Ising form.

\appendix

\section{Appendix: Mott transition in the 2-orbital Hubbard model with different bandwidths at $J=0$}\label{comp:perturbative_Uc_t1t2_J0}

Here we show analytically that in the half-filled degenerate 2-band Hubbard model with different bandwidths (due to different diagonal hopping integrals $t_1$ and $t_2$), the Mott transition at $J=0$ remains common between the two bands even when $t_1\neq t_2$, until quite large bandwitdth rations. However, in accordance with the numerical evidence the largest bandwidth ratio where this is possible is $t_1/t_2=5$. For larger ratios the $U_c$ for the Mott transition differs between the bands, and one opens the $J=0$ orbital-selective Mott phase seen in Fig. \ref{fig:orb_dec}.

As done previously we develop the multi-orbital slave-spin equation (\ref{eq:H_s_MF-multiOrb}) around the insulating solution $h_{im\s}=0$ where $H_s=H_{at}$ with $J=0$:
\be
H_{at}=\frac{U}{2}\left(\sum_{m\s} S^z_{m\s}\right)^2, 
\ee
of which we know the spectrum. This is: 

\be
\left\{
\begin{array}{cc}
|\up\up,\up\up\rangle |\down\down,\down\down\rangle & E=2U\\
&\\
\begin{array}{c}
 |\up\up,\up\down\rangle |\up\up,\down\up\rangle |\down\down,\up\down\rangle |\down\down,\down\up\rangle \\ 
|\up\down,\up\up\rangle |\down\up,\up\up\rangle |\up\down,\down\down\rangle |\down\up,\down\down\rangle 
\end{array}
& E=\frac{U}{2}\\
&\\
\begin{array}{c}
|\up\down,\up\down\rangle |\up\down,\down\up\rangle |\down\up,\up\down\rangle |\down\up,\down\up\rangle \\
|\up\up,\down\down\rangle |\down\down,\up\up\rangle
\end{array}
& E=0\\
\end{array}
\right.
\ee

We again calculate as a perturbation, the effect of the "kinetic" part of the slave-spin hamiltonian:
\be
H_{pert}=\sum_{m\s}h_m 2 S^x_{m\s}.
\ee
where $ O_{im\sigma} = 2 S^x_{m\s}$ because both orbitals are half-filled and thus the gauge $c=1$ (eq. (\ref{eq:gauge}) for $n_m=1/2$), and we call $h_m\equiv h^*_{im\s}+h_{im\s}$. Also $\lambda_m=0$ at half-filling because of particle-hole symmetry. Notice that here we let $h_1$ and $h_2$ differ.

We perform as in the other cases a perturbative expansion in $h_1$ and $h_2$, which are small near the metal-insulator transition, if this is second-order.

In this $J=0$ case the ground state has degeneracy 6  (the subset above with $E=0$).  
Furthermore, as before, the perturbation $H_{pert}$ does not have any nonzero elements in the low-energy subspace one has to use second-order perturbation theory to find the right combination of states, in the ground state manifold, to which the perturbed state tends for $h_m\rightarrow 0$.
This is the ground state of the matrix $H'\equiv H_{pert}(E_0-H_{at})^{-1}H_{pert}$  in the degenerate subspace. In this case, since $H_{pert}$ couples the $E=0$ sector only with the $E=\frac{U}{2}$ sector, one can rewrite $H'=-\frac{2}{U}H_{pert}^2$ (where we have used the fact that $E_0-H_{at}$ is diagonal and $E_0=0$).

By applying $H'$ to one of the states with orbital polarization (i.e. $ |\up\up,\down\down\rangle$ and $|\down\down,\up\up\rangle$ that corresponds to the original states in which the two electrons occupy both the same orbital), one finds that both are only coupled with one combination of the four non-polarized states, which is
\be
|+,+\rangle=\frac{1}{2}(|\up\down,\up\down\rangle + |\up\down,\down\up\rangle + |\down\up,\up\down\rangle + |\down\up,\down\up\rangle) 
\ee
In detail:
\bea
\@&&\@ H'|\up\up,\down\down\rangle= 
\left[
2(h_1^2+h_2^2) |\up\up,\down\down\rangle
+2h_1^2\down\down,\down\down\rangle +2h_2^2 |\up\up,\up\up\rangle 
+2h_1h_2(|++\rangle)
\right]\nonumber \\
\@&&\@ H'|\down\down,\up\up\rangle = [\qquad\qquad\qquad " \qquad\qquad  h_1 \longleftrightarrow h_2 \qquad\qquad " \qquad\qquad ]
\eea
Indeed $V^2$ has the effect of flipping any two spins (including twice the same), so that any state of the $E=0$ manifold is changed, when acted upon by $V^2$, into itself plus four of the others states of the manifold (only one is left out, since it needs 4 flips to be reached), plus two states of the $E=2U$ manifold.

This suggest to pick the following three other orthogonal states\footnote{This notation is easily understood as follows: the state for the two-orbital system is obtained by tensor product of two one-orbital states, which are chosen either the following triplet state $|+\rangle\equiv (|\up\down\rangle+|\down\up\rangle)/\sqrt{2}$, or the singlet state $|-\rangle\equiv (|\up\down\rangle-|\down\up\rangle)/\sqrt{2}$. Thus for example the state $|+,-\rangle$ is the tensor product of a triplet state in orbital 1 and the singlet state in orbital two, yielding the first formula in \ref{eq:basis}.} as basis states for writing $H'$:
\bea\label{eq:basis}
\@&&\@ 
|+,-\rangle=\frac{1}{2}(|\up\down,\up\down\rangle - |\up\down,\down\up\rangle + |\down\up,\up\down\rangle - |\down\up,\down\up\rangle) \nonumber \\
\@&&\@ 
|-,+\rangle=\frac{1}{2}(|\up\down,\up\down\rangle + |\up\down,\down\up\rangle - |\down\up,\up\down\rangle - |\down\up,\down\up\rangle) \nonumber \\
\@&&\@ 
|-,-\rangle=\frac{1}{2}(|\up\down,\up\down\rangle - |\up\down,\down\up\rangle - |\down\up,\up\down\rangle + |\down\up,\down\up\rangle) 
\eea

It is straightforward to verify that the restriction of $H'$ to the $E=0$ manifold in this basis is given by the following 6x6 matrix:
\be\label{eq:Hpert_matrix}
H'=-\frac{2}{U}\left[ \begin{array}{cccccc}
2h_1^2+2h_2^2 & 0 & 4h_1h_2 & 0 & 0 & 0 \\
0 & 2h_1^2+2h_2^2 & 4h_1h_2 & 0 & 0 & 0 \\
 4h_1h_2 &  4h_1h_2 & 4h_1^2+4h_2^2&  0 & 0 & 0 \\
0 & 0 & 0 & 4h_1^2 & 0 & 0\\
0 & 0 & 0 & 0 & 4h_2^2 & 0\\
0 & 0 & 0 & 0 & 0 & 0\\
\end{array}
\right]
\ee

The diagonalization of this matrix yields the ground state energy at the first order in the perturbation i.e. $E_0^I= (-6(h_1^2+h_2^2)-2\sqrt{(h_1^2+h_2^2)^2+32h_1^2h_2^2})/U$
and the correct zeroth order ground state, that is 
\be
|\phi_0\rangle = X (|\up\up,\down\down\rangle+ |\down\down,\up\up\rangle) + Y|+,+\rangle
\ee
with $X=\frac{4 h_1 h_2}{\sqrt{{\cal N}}}$, $Y=\frac{h_1^2+h_2^2+\sqrt{(h_1^2+h_2^2)^2+32h_1^2h_2^2}}{\sqrt{{\cal N}}}$, and 
$ {\cal N} = 64h_1^2h_2^2+2(h_1^2+h_2^2)^2+2(h_1^2+h_2^2)\sqrt{(h_1^2+h_2^2)^2+32h_1^2h_2^2}$.

To first order in the perturbation the ground state becomes
\footnote{Here, as specified previously, $|+,\down\down\rangle$ is intended as the tensor product of the triplet state $|+\rangle\equiv (|\up\down\rangle+|\down\up\rangle)/\sqrt{2}$ for orbital 1 with the state $|\down\down\rangle$ for orbital 2, i.e.$|+,\down\down\rangle\equiv (|\up\down,\down\down\rangle+|\down\up,\down\down\rangle)/\sqrt{2}$}:
\bea
|\phi_0^{(I)}\rangle=
\@&&\@
|\phi_0\rangle+\sum_{|s\rangle\neq |\phi_0\rangle}\frac{\langle s|H_{pert}|\phi_0\rangle}{E_0-E_s} |s\rangle\nonumber\\
=\@&&\@|\phi_0\rangle
-\frac{2}{U}X[\sqrt{2}h_1(|+,\down\down\rangle + |+,\up\up\rangle)+\sqrt{2}h_2(|\up\up,+\rangle + |\down\down,+\rangle)]\nonumber\\
\@&&\@ 
-\frac{2}{U}Y[\sqrt{2}h_1(|\up\up,+\rangle + |\down\down,+\rangle)+\sqrt{2}h_2(|+,\down\down\rangle + |+,\up\up\rangle)]
\eea

We now use the self-consistency conditions $h_m=2 \bar \eps_m \langle\phi_0^{(I)}| 2 S_{m\s}^x|\phi_0^{(I)}\rangle$ to obtain, as in the complements \ref{comp:perturbative_Uc} and \ref{comp:perturbative_Uc_N2} the linearized equations for the $h_m$ which determine the critical coupling for the Mott transition. By calculating $\langle 2 S_{m\s}^x\rangle$ for both orbitals we get to the following coupled equations:
\bea\label{eq:selfcons_linear_J0_h1h2}
\@&&\@ h_1=2\bar \eps_1(-\frac{4}{U})\left[ (2 X^2 + 2Y^2) h_1 + 4XYh_2 \right] \nonumber \\
\@&&\@ h_2=2\bar \eps_2(-\frac{4}{U})\left[ (2 X^2 + 2Y^2) h_2 + 4XYh_1 \right] 
\eea
 
These equations as usual determine, for $(h_1,h_2)\rightarrow 0$, the U for the Mott transition, as long as $h_1$ and $h_2$ vanish together. But the equations are two because, since $h_1$ and $h_2$ are determined selfconsistently they vanish with a given ratio $\a\equiv h_2/h_1$, which is fixed by the asymmetry between the bands, namely by $\bar\eps_2/\bar\eps_1$.

By substituting $h_2=\a h_1$ in (\ref{eq:selfcons_linear_J0_h1h2}) and in the expressions for X and Y, and simplifying $h_1$ everywhere one gets two equations determining $U_c$ and $\a$ for given $\bar\eps_1$ and $\bar\eps_2$. U can be eliminated to get the equation determining implicitly $\a$ for each $\bar\eps_2/\bar\eps_1$ that reads:
\be\label{eq:eps1eps2_vs_alpha}
\frac{\bar\eps_2}{\bar\eps_1}=\frac{X^2+Y^2+2\a XY}{X^2+Y^2+\frac{2}{\a} XY}
\ee
with $X=\frac{4 \a}{\sqrt{{\cal N_\a}}}$, $Y=\frac{1+\a^2+\sqrt{(1+\a^2)^2+32\a^2}}{\sqrt{{\cal N_\a}}}$, and 
$ {\cal N_\a} = 64\a^2+2(1+\a^2)^2+2(1+\a^2)\sqrt{(1+\a^2)^2+32\a^2}$.
This formula gives the dependence shown in Fig. \ref{fig:eps1eps2_vs_alpha}.
\begin{figure}[h]
\begin{center}
\includegraphics[width=12cm]{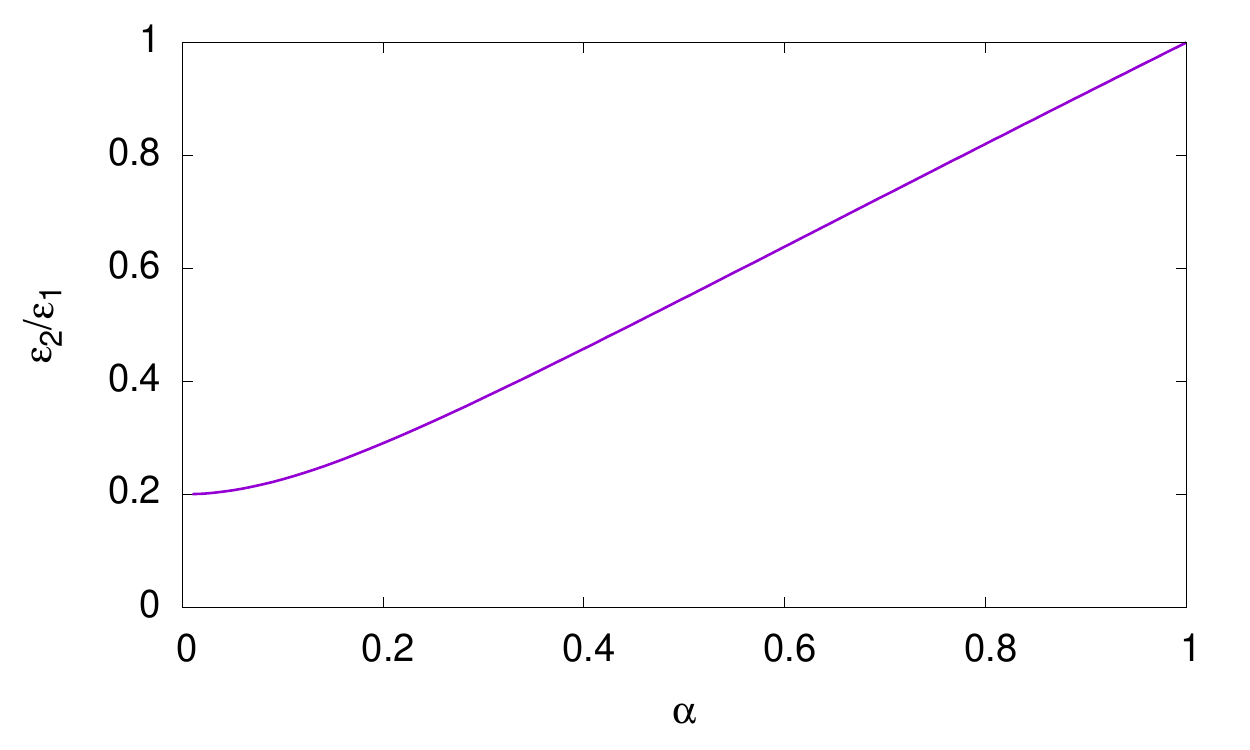}
\caption{\label{fig:eps1eps2_vs_alpha} The relation between $\a\equiv h_2/h_1$ and $\bar\eps_2/\bar\eps_1$ implied in formula (\ref{eq:eps1eps2_vs_alpha})}.
\end{center}
\end{figure}
This shows that a common Mott transition within the two bands can be found for any bandwidth ratio between 1 and 0.2, which is the limiting ratio. 
This can also be calculated by expanding for small $\a$. One has $X^2+Y^2\simeq 1$ and $XY\simeq 2\a$ at the leading order, giving:
\be
U_c=-16\bar\eps_1, \qquad \bar\eps_2/\bar\eps_1=0.2.
\ee
which locates the point where the three transition lines merge as found in the numerical calculation shown in Fig. \ref{fig:orb_dec} for $J=0$\cite{demedici_Slave-spins,Ferrero_OSMT}.

For $\bar\eps_2/\bar\eps_1=1$ instead, one has $\a=1$ by symmetry and recovers the result of the complement \ref{comp:perturbative_Uc_N2} for the $U_c$ in the degenerate 2-band Hubbard model at $J=0$ by evaluating $X=1/\sqrt{6}$, $Y=2/\sqrt{6}$, which yield indeed $U_c=-24\bar\eps_1=-24\bar\eps_2$.


\end{document}